\documentclass{aa}  

\usepackage{booktabs}
\usepackage{graphicx}
\usepackage{txfonts}
\usepackage{natbib}
\usepackage{placeins}
\bibpunct{(}{)}{;}{a}{}{,} 

\begin{document} 

\title{The traveling-PWN modeling attempt on the enigmatic LHAASO dumbbell-like structure}

\author{Caijin Xie
        \inst{1}
        \and 
        Yihan Liu
        \inst{1}
        \and
        Chengyu Shao\inst{1}
        \and
        Yudong Cui\inst{1}\fnmsep\thanks{cuiyd@mail.sysu.edu.cn}
        \and
        Lili Yang\inst{1,2}\fnmsep\thanks{yanglli5@mail.sysu.edu.cn}
}

\institute{School of Physics and Astronomy, Sun Yat-Sen University, No. 2 Daxue Road, 519082, Zhuhai China
\and
Center for Astro-Particle Physics, University of Johannesburg,
P.O. Box 524, Auckland Park 2006, South Africa
}
 
\abstract
   {The first Large High Altitude Air Shower Observatory (LHAASO) catalog presents six enigmatic ultra-high-energy (UHE) $\gamma$-ray sources with lonely $>$ 25 TeV emission being detected, which are indicated as 1LHAASO: J0007$+$5659u, J0206$+$4302u, J0212$+$4254u, J0216$+$4237u, J1740$+$0948u, and J1959$+$1129u. No counterparts of the six sources have been observed, except two energetic pulsars, PSR J0218$+$4232 and PSR J1740$+$1000. Among them, 1LHAASO: J0206$+$4302u, J0212$+$4254u, and J0216$+$4237u are connected on the significance map and constituted a dumbbell-like structure. They are close in position and show a similar spectral shape, suggesting a physical association among them.}
   {To explain the origin of the six LHAASO sources, especially the intriguing dumbbell-like structure, we conducted the leptonic and hadronic modeling research according to our multiwavelength and multimessenger study. For the dumbbell-like structure, models with traveling-PWNe were considered.}
   {The multiwavelength and multimessenger study was based on the \textit{Fermi}-LAT, \textit{Swift}-XRT, \textit{Planck}, CfA $^{12}$CO survey, and IceCube neutrino datasets. In the traveling-PWN modeling research, we assumed an isotropic and homogeneous diffusion condition and discussed the influence of diffusion coefficient, distance, and proper motion velocity.}
   {No counterparts are discovered in our multiwavelength and multimessenger study, except the two known pulsars. The traveling-PWN modeling attempt with a single PWN appears implausible to explain the dumbbell-like structure, as the diffusion coefficient needs to be much lower than the Bohm limit. A double traveling-PWNe model is also explored and can account for the results of LHAASO-KM2A observation. However, the probability of occurrence of this explanation is significantly lower than that of a conventional triple PWNe explanation. Furthermore, according to our model, the only known pulsar PSR J0218$+$4232 is unlikely to be associated with the dumbbell-like structure in an isotropic and homogeneous diffusion environment. The expected TeV emission of this pulsar is far from explaining even the eastern part of the structure.}
   {}


\maketitle

\section{Introduction} \label{sec:intro}

\begin{table*}
  \centering
  \caption{\label{tab:1}Overview of the six UHE $\gamma$-ray sources detected solely by LHAASO-KM2A \citep{cao2024first}, which are the focus of this study.}
  \begin{tabular}{cccccc}
  \hline
    Source & Galactic longitude & Galactic latitude & TS$_{100}^{\rm{a}}$ & Spectral index & Energetic pulsars nearby \\ \hline
    J0007$+$5659u & 116\fdg94 & $-$5\fdg36 & 43.6 & $3.10\pm0.20$ & / \\
    J0206$+$4302u & 137\fdg27 & $-$17\fdg71 & 82.8 & $2.62\pm0.16$ & / \\
    J0212$+$4254u & 138\fdg28 & $-$17\fdg55 & 30.2 & $2.45\pm0.23$ & / \\
    J0216$+$4237u & 139\fdg17 & $-$17\fdg55 & 65.6 & $2.58\pm0.17$ & PSR J0218$+$4232 \\
    J1740$+$0948u & 33\fdg79 & 20\fdg26 & 37.2 & $3.13\pm0.15$ & PSR J1740$+$1000 \\
    J1959$+$1129u & 51\fdg10 & $-$9\fdg42 & 60.8 & $2.69\pm0.17$ & / \\
    \hline
  \end{tabular}
  \tablefoot{${\rm{a}}$. Test statistic value at energies exceeding 100 TeV.}
\end{table*}

As one of the state-of-the-art $\gamma$-ray and cosmic ray (CR) detector, the Large High Altitude Air Shower Observatory (LHAASO) located at Daocheng site, Sichuan province, with altitude of 4410 m, possesses an excellent detection ability in ultra-high-energy (UHE) band and has expanded the way for discovering new UHE $\gamma$-ray sources. LHAASO covers energies from 100 GeV to 1 EeV \citep{di2016lhaaso} and achieves sensitivity below 1 $\times$ 10$^{-12}$ erg cm$^{-2}$ s$^{-1}$ sr$^{-1}$ at TeV energy \citep{neronov2020lhaaso}. It consists of three main components: The Water Cherenkov Detector Array (WCDA) for TeV $\gamma$-ray detection, The Kilometer Squared Array (KM2A) for detecting $\gamma$-ray above 10 TeV, and a Wide Field-of-view atmospheric Cherenkov Telescope Array (WFCTA) mainly for CR detection \citep{cao2025data}. Recently, the LHAASO Collaboration has presented its first catalog of $\gamma$-ray sources, reporting 10 sources without 1-25 TeV $\gamma$-photons detected by WCDA, but with emissions ranging from 25 TeV to above 100 TeV observed by KM2A \citep{cao2024first}. The experimental data are fitted by a power-law function, with the fixed reference energy set at 50 TeV \citep{cao2024first}. To avoid the complex background, our work focus on the six lonely $>$ 25 TeV point-like sources located outside the Galactic plane, which are 1LHAASO: J0007$+$5659u, J0206$+$4302u, J0212$+$4254u, J0216$+$4237u, J1740$+$0948u, and J1959$+$1129u (the prefix 1LHAASO is omitted hereafter). For an overview, the position and spectral information of them are presented in Table \ref{tab:1}. If these lonely $>$ 25 TeV sources are Geminga-like pulsar wind nebula (PWN) halos, their electron cutoff energies will be far beyond 100 TeV. An even more intriguing aspect is that the sources J0206$+$4302u, J0212$+$4254u, and J0216$+$4237u appear to be spatially linked, forming an extended, dumbbell-like structure on the significance map, and exhibit a similar spectral shape with an index of approximately 2.5. So far, only two energetic pulsars, PSR J0218$+$4232 and PSR J1740$+$1000, have been observed in the vicinity of the six LHAASO sources. Here, PSR J0218$+$4232 is a millisecond pulsar located near J0216$+$4237u\textemdash the eastern source constituting the dumbbell-like structure. However, no physical associations between the structure and the pulsar can be confirmed, due to the relatively large position offset \citep{cao2024first} and the low predicted TeV emission according to the observation of \textit{Fermi Large Area Telescope} (\textit{Fermi}-LAT) and Major Atmospheric Gamma Imaging Cherenkov (MAGIC) \citep{acciari2021search}. PSR J1740$+$1000 is a middle-aged radio pulsar spatially close to J1740$+$0948u, with a position offset of approximately 0\fdg2. Due to the scarcity of observational data, the physical association between J1740$+$0948u and PSR J1740$+$1000 is still unclear. These mysteries on the six LHAASO UHE sources motivate us to investigate their potential origins, which may suggest the existence of new physical mechanisms.

Among the six LHAASO sources, the intriguing dumbbell-like structure comprising J0206$+$4302u, J0212$+$4254u, and J0216$+$4237u has attracted much attention. We attempt to use the proper motion of a central pulsar powering a PWN to explain its origin and study the traveling-PWN model under the isotropic and homogeneous diffusion condition. The central pulsar can either be a known pulsar or a hypothetical one. \citet{zhang2021morphology} has discussed the impact of a pulsar's proper motion on the morphology of its $\gamma$-ray halo. The results infer that it is difficult to use the proper motion of a middle-aged pulsar with common distance and proper motion velocity to account for an extended structure. As a further research, we have explored a broader parameter space for the model with a single traveling-PWN and studied the multiple traveling-PWNe model, as well as the probabilities of occurrence for different scenarios. 

In this work, to discover the origin of the six LHAASO sources listed in Table \ref{tab:1}, we conduct a multiwavelength and multimessenger study, aiming to check if there are any possible counterparts of them observed in other energy bands. The results are presented in Sect. \ref{sec:multi} and some details about data analysis are provided in Appendix \ref{sec:ana}. Based on the constraints set by the multiwavelength and multimessenger study, we perform the leptonic and hadronic modeling research on the six sources. The best-fit models are presented in Sect. \ref{sec:Mod}, while the corner plots illustrating the models' ability to explain the data are given in Appendix \ref{sec:cor}. The expected neutrino flux in the hadronic scenario compared with the sensitivity of next generation neutrino observatory is shown in Appendix \ref{sec:neu}. The traveling-PWN modeling attempt within an isotropic and homogeneous diffusion environment, aiming to explain the intriguing dumbbell-like structure, is discussed in Sect. \ref{sec:tra}. We have explored the influence of diffusion coefficient, distance, and proper motion velocity on the final morphology of $\gamma$-ray emission. Using appropriate parameters, models involving a single traveling-PWN, two traveling-PWNe, and three PWNe are studied, along with their respective probabilities of occurrence. Finally, the conclusion and discussion are given in Sect. \ref{sec:dis}.

\section{Multiwavelength and multimessenger study}\label{sec:multi}

\subsection{\textit{Fermi}-LAT dataset}\label{subsec:fermi}

\textit{Fermi}-LAT was launched into space on June 11, 2008. It is an all-sky $\gamma$-ray spatial observatory covering energies from 20 MeV to above 300 GeV \citep{abdo2009fermi}, with a typical sensitivity of $\sim$ 10$^{-12}$ erg cm$^{-2}$ s$^{-1}$ for observation of 10 years \citep{funk2013comparison}. To investigate the morphology of the six LHAASO UHE sources in the lower energy band, we analyzed 10 years of data recorded by \textit{Fermi}-LAT from 2014 January 1 to 2024 January 1 with energies ranging from 300 MeV to 300 GeV. As a result, at the exact locations of the six LHAASO sources, no abundant observational data was obtained and the $\gamma$-ray fluxes of the six sources acquired by maximum likelihood analysis had error bars larger than the true values. The analysis was conducted by using the publicly available toolkit Fermitools. Since no accurate $\gamma$-ray fluxes of the six sources could be extracted from \textit{Fermi}-LAT dataset, the upper limits of spectral energy distribution (SED) were calculated and used to constrain the leptonic and hadronic modeling fits discussed in Sect. \ref{sec:Mod}. 

\begin{figure*}[!h]
    \centering
    \includegraphics[width=1.0\textwidth]{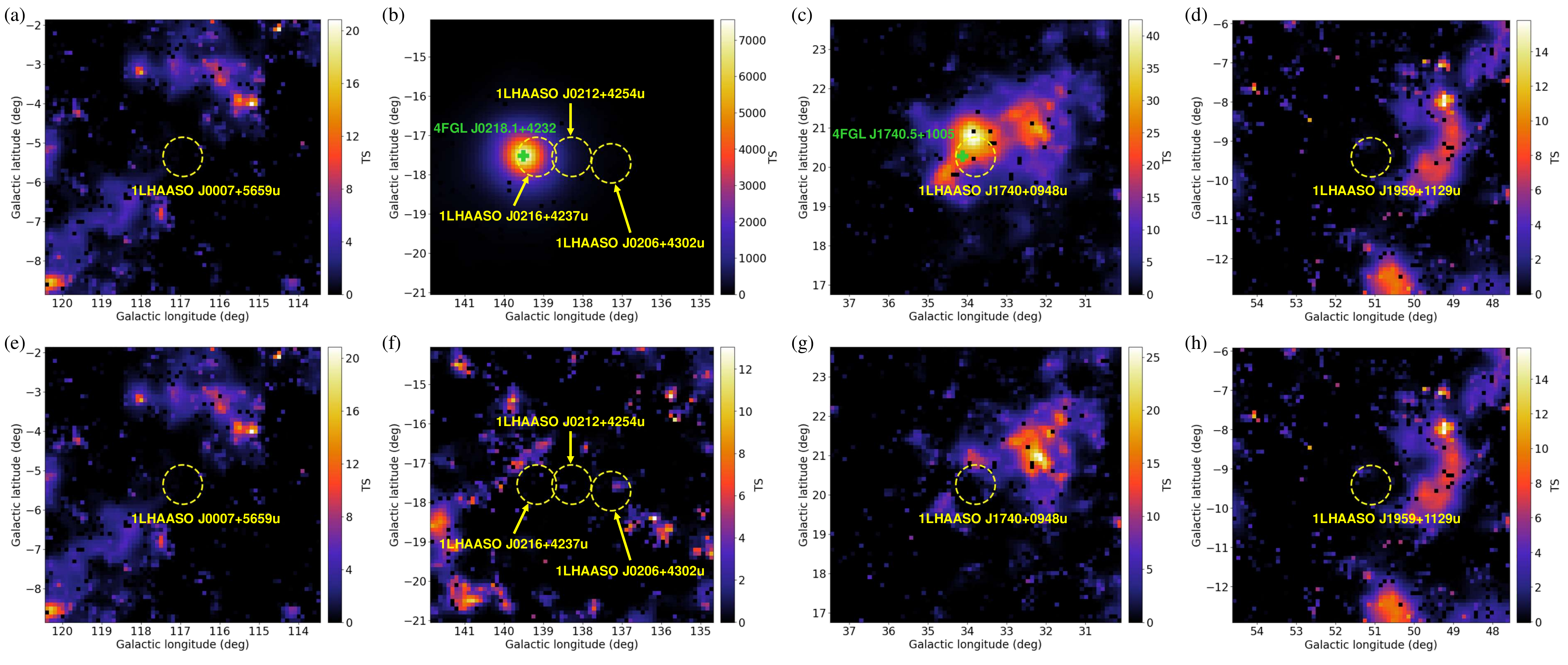}
    \caption{\label{fig:1}\textit{Fermi} TS maps within the 5$^{\circ}$ radius ROIs around the six LHAASO UHE sources. Panels (a)-(d) show TS maps where the nearby 4FGL sources associated with the LHAASO sources are retained. Yellow dashed circles with a radius of 0\fdg5 are centered on the labeled LHAASO sources, indicating the regions used to filter the 4FGL sources by position. Panels (e)-(h) display the residual TS maps after excluding all 4FGL sources.}
\end{figure*}

We also calculated the test statistic (TS) maps of the six LHAASO sources with the gttsmap function, to examine the $\gamma$-ray sources observed by \textit{Fermi}-LAT. Figure \ref{fig:1} (a)-(d) present TS maps corresponding to all the fourth \textit{Fermi}-LAT catalog sources, the so-called 4FGL sources \citep{abdollahi2020fermi}, in the vicinity of the six LHAASO sources. Two 4FGL sources are found within the 0\fdg5 radius region centered on each LHAASO source, and can be confirmed as the two known pulsars PSR J0218$+$4232 and PSR J1740$+$1000 respectively.\footnote{For more details, visit the Public List of LAT-Detected Gamma-Ray Pulsars at \url{https://confluence.slac.stanford.edu/display/GLAMCOG/Public+List+of+LAT-Detected+Gamma-Ray+Pulsars}.} Figure \ref{fig:1} (e)-(h) show the residual TS maps with all 4FGL sources removed, indicating that no other significant sources appear to be associated with the six LHAASO sources in the energy range of 300 MeV to 300 GeV. Details regarding the maximum likelihood analysis and TS map calculations are provided in Appendix \ref{subsec:fermi ana}.

\subsection{\textit{Swift}-XRT dataset}\label{subsec:swift}

The \textit{Swift Gamma-Ray Explorer} is a spatial observatory designed to make prompt multiwavelength observations of gamma-ray bursts (GRBs) and GRB afterglows. The integrated X-ray telescope on the \textit{Swift} satellite (\textit{Swift}-XRT) enables it to precisely locate the positions of GRBs with a few arcsecond accuracy within 100 s of the burst onset \citep{burrows2005swift}. The \textit{Swift}-XRT covers an energy range from 0.2 keV to 10 keV, and its sensitivity is 2 $\times$ 10$^{-14}$ erg cm$^{-2}$ s$^{-1}$ in 10$^{4}$ s \citep{burrows2005swift}. We have searched for X-ray sources spatially close to the six LHAASO UHE sources (position offset $<$ 0\fdg3) within the \textit{Swift}-XRT dataset. As a result, no such source was observed.\footnote{The \textit{Swift}-XRT 2SXPS catalog at \url{https://www.swift.ac.uk/2SXPS/} has been checked.} Furthermore, among the six sources, \textit{Swift}-XRT observations were exclusively available within the 0\fdg3 radius region of interest (ROI) centered on J0206$+$4302u.\footnote{The data were acquired from \url{https://heasarc.gsfc.nasa.gov/cgi-bin/W3Browse/swift.pl}.} These data with exposure time of 1577.1 s were used to set the constraint on SEDs of the six sources in the keV band as no abundant efficient photons were detected. The upper limit results were 7.0 $\times$ 10$^{-14}$ erg cm$^{-2}$ s$^{-1}$ at 1.5 keV and 6.1 $\times$ 10$^{-13}$ erg cm$^{-2}$ s$^{-1}$ at 8.1 keV. Appendix \ref{subsec:swift ana} presents more details about the data processing. 

\subsection{\textit{Planck} dataset}\label{subsec:planck}

The \textit{Planck} satellite, launched on 2009 May 14, is a third-generation space experiment aiming to study the cosmic microwave background (CMB). It covers the observation wavelength from 350 $\mu$m to 1 cm and has a good angular resolution of $\sim$ 5 arcmin \citep{tauber2010planck}. 

\begin{figure*}[!h]
    \centering
    \includegraphics[width=0.875\textwidth]{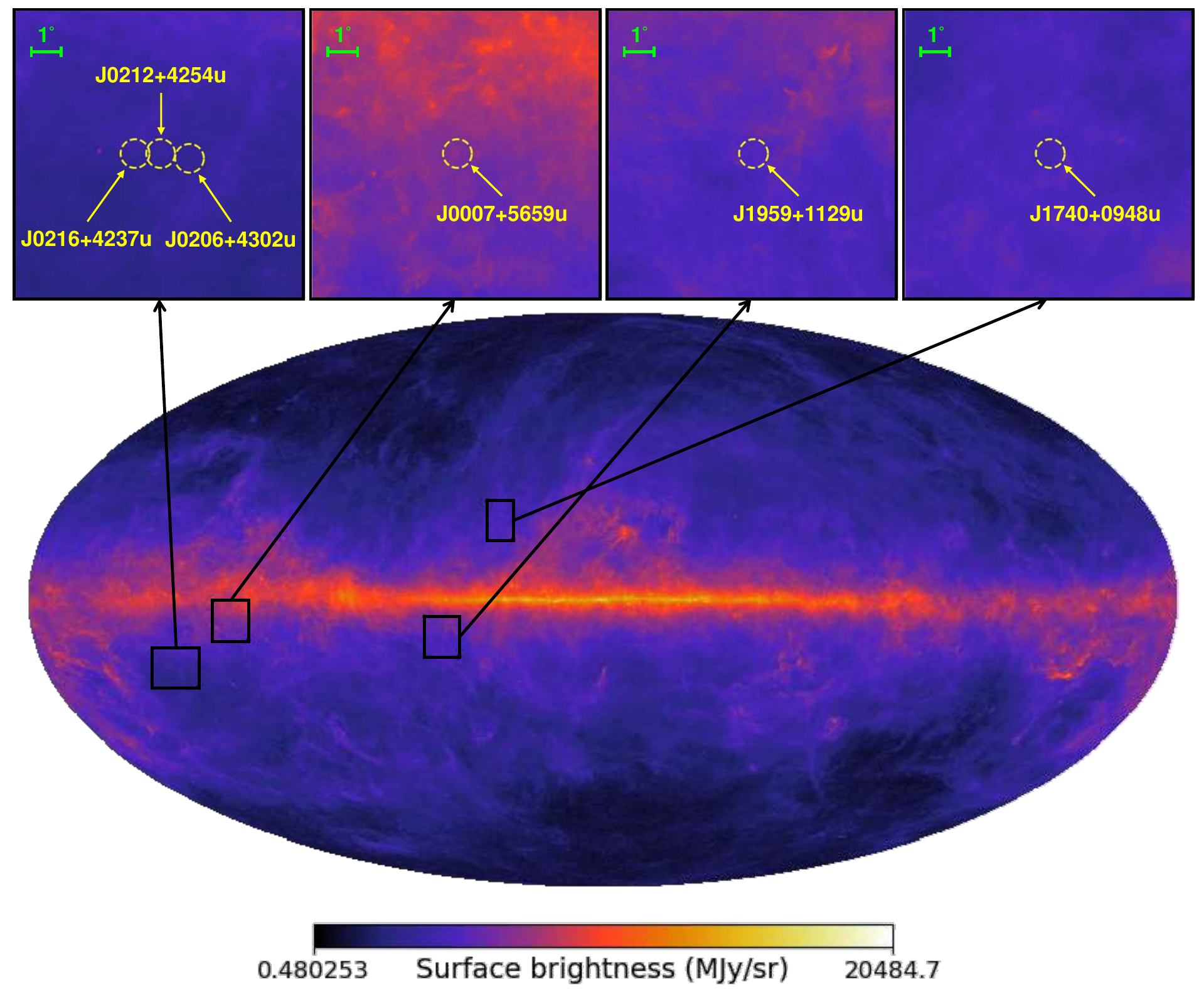}
    \caption{\label{fig:2}\textit{Planck} all-sky map at 857 GHz. The map is presented in Galactic coordinate and the color-bar is plotted in logarithmic scale. The zoomed-in views of 10$^{\circ}$ $\times$ 10$^{\circ}$ ROIs centered on the six sources are presented at the top of the figure. The yellow dashed circles indicates the positions of labeled sources.}
\end{figure*}

We hoped to find counterparts, especially molecular clouds (MCs), of the six LHAASO UHE sources in the far infrared band with \textit{Planck} observation. The up-to-date \textit{Planck} all-sky map at 857 GHz, obtained from the \textit{Planck} Public Data Release 3 is shown in Fig. \ref{fig:2}. The original data are presented in the hierarchical equal-area isolatitude pixelization (HEALPix) fits table. This pixelization produces a subdivision of a spherical surface in which each pixel covers the same surface area as every other pixel. The raw HEALPix fits table was reprojected into a standard fits table in the world coordinate system (WCS) using the astronomical image toolkit Montage with conventional processing \citep{jacob2009montage}, and the data within 10$^{\circ}$ $\times$ 10$^{\circ}$ ROIs centered on the six sources were extracted for more detailed analyzes.\footnote{More details about the HEALPix fits table and Montage processing can be found at \url{http://montage.ipac.caltech.edu/docs/HEALPix/}.} As indicated by the yellow dashed circles in Fig. \ref{fig:2}, no significant evidence of counterparts in proximity to these sources can be found, except for J0007$+$5659u. The exact environment of this source needs to be further identified with carbon monoxide (CO) observation.

The \textit{Planck} catalog of compact sources has also been checked. However, no source with position offset to the six sources less than 0\fdg3 was found. The catalog can be accessed via the NASA/IPAC Infrared Science Archive (IRSA) official website.\footnote{\url{https://irsa.ipac.caltech.edu/frontpage/}}

\subsection{CfA $^{12}$CO survey dataset}\label{subsec:co}

MCs can provide abundant protons for the inelastic interaction with CR protons, the so-called proton-proton (p-p) interactions, which can result in high-energy $\gamma$-ray and neutrino emission \citep{de2022exploring}. The results of \textit{Planck} observation can be a reference of whether there exist MCs. However, a more accurate and efficient way to trace MCs is to observe the $^{12}$CO J $=$ 1-0 transition line at 115 GHz \citep{su2019milky}. We examined the moment-masked \citep{dame2011optimization} data of the recent $^{12}$CO survey covering the entire northern sky using the CfA 1.2 m telescope and its twin instrument in Chile \citep{dame2022co}.\footnote{The data is available at \url{https://lweb.cfa.harvard.edu/rtdc/CO/NorthernSkySurvey/}.} Since the six LHAASO UHE sources are located far from the Galactic plane, this $^{12}$CO survey was the only one with publicly available data we could find that directly observed the 115 GHz $^{12}$CO line, meanwhile, covering all these six sources. The coverage of the local standard of rest (LSR) velocity is $\pm47.1$ km s$^{-1}$ and the velocity resolution is 0.65 km s$^{-1}$. The angular resolution and typical noise fluctuation are 0\fdg25 and 0.18 K, respectively \citep{dame2022co}.

\begin{figure*}[!h]
    \centering
    \includegraphics[width=0.85\textwidth]{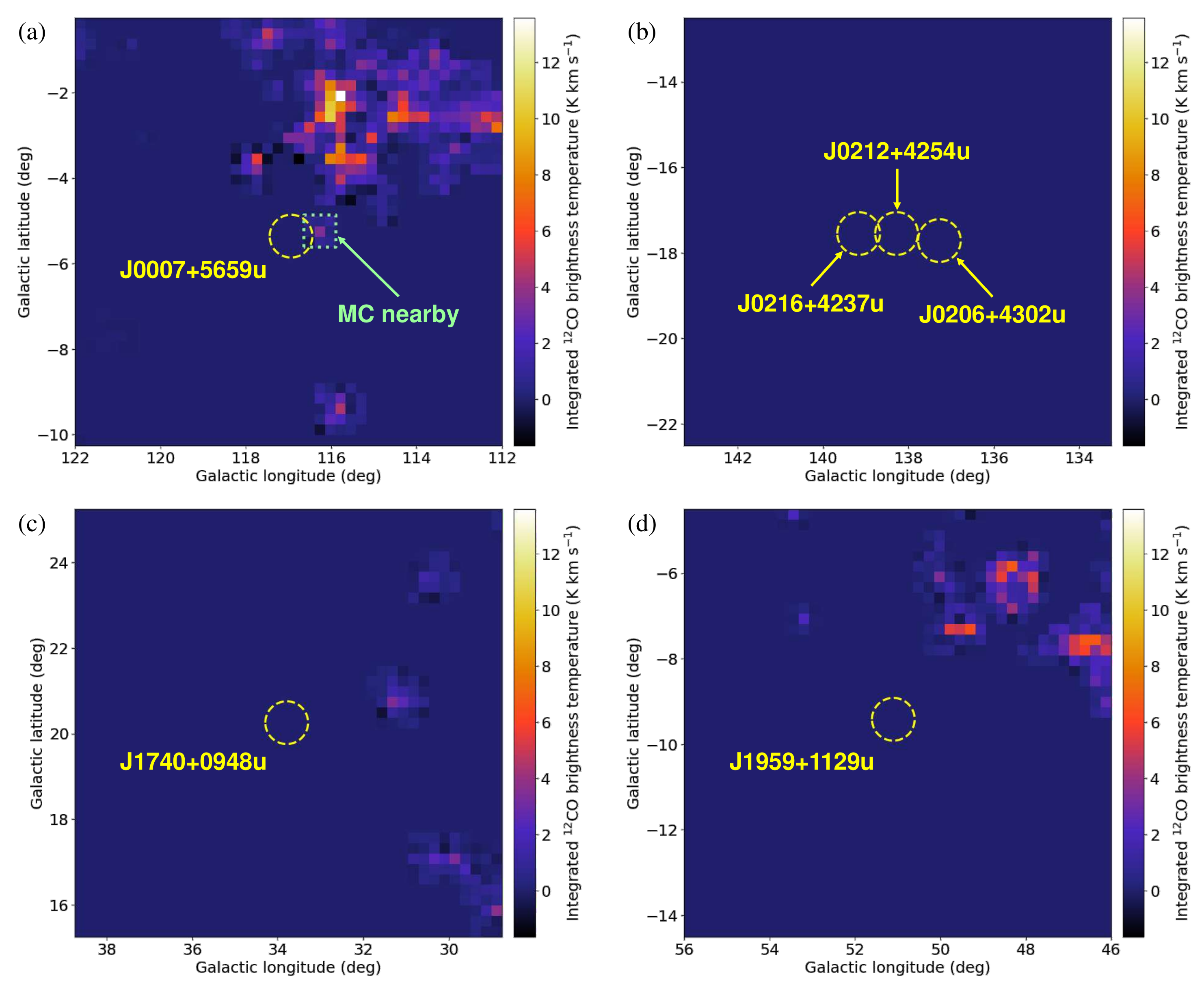}
    \caption{\label{fig:3}Results of $^{12}$CO survey around the six LHAASO UHE sources. The data in panel (a) was integrated with a LSR velocity coverage from $-4.9$ km s$^{-1}$ to 1.6 km s$^{-1}$, covering the MC in proximity to J0007$+$5659u. For the data in panels (b)-(d), the LSR velocity integrating range was $\pm47.1$ km s$^{-1}$, reaching the limitation of this survey \citep{dame2022co}. The positions of the six LHAASO sources are marked by yellow dashed circles with radius of 0\fdg5. The MC proposed to be associated with J0007$+$5659u is indicated by the dashed box in light green.}
\end{figure*}

The moment-masked data can be divided into 146 $^{12}$CO brightness temperature maps with different LSR velocities. We firstly checked each $^{12}$CO brightness temperature map and found that only the LHAASO source J0007$+$5659u appeared to have a thin MC nearby. The other five sources with larger galactic latitudes were unlikely to be associated with significant MCs. Figure \ref{fig:3} presents the LSR velocity integrated $^{12}$CO brightness temperature maps in the vicinity of the six LHAASO sources to clearly show the result, which is consistent with the \textit{Planck} observation. More details on data analysis are provided in Appendix \ref{subsec:co ana}.

The MC proposed to be associated with J0007$+$5659u covers a sky region of approximately 0\fdg75 $\times$ 0\fdg75. The $^{12}$CO spectrum of this MC is shown in Fig. \ref{fig:A1} of Appendix \ref{subsec:co ana}. With a Gaussian fitting, the central LSR velocity obtained is $-1.83\pm0.02$ km s$^{-1}$. Based on this result, the expected distance of the MC can be calculated online with a Bayesian approach in which sources are assigned to spiral arms according to their galactic longitudes, latitudes, and LSR velocities with respect to arm signatures seen in CO and H$_{{\rm{I}}}$ surveys \citep{reid2016parallax, reid2019trigonometric}.\footnote{\url{http://bessel.vlbi-astrometry.org/node/378}} The result is $0.70\pm0.04$ kpc with a probability of 72\%. The mean H$_{2}$ column density $\overline{N}({\rm{H_{2}}})$ of the MC calculated by adopting the $^{12}$CO-to-H$_{2}$ conversion factor of 2 $\times$ 10$^{20}$ cm$^{-2}$ (K km s$^{-1}$)$^{-1}$ \citep{bolatto2013co} is $1.98\pm0.57$ $\times$ 10$^{20}$ cm$^{-2}$. Assuming a symmetric spatial distribution of the MC, the mean particle number density $\overline{n}({\rm{H_{2}}})$ of the MC can be calculated with \citep{rani2023identification}
\begin{equation}\label{eq-nmc}
	\overline{n}({\rm{H_{2}}})=\frac{3}{4\pi}\frac{M({\rm{H_{2}}})}{\mu_{p}m_{p}R^{3}_{eq}},
\end{equation}
where $M({\rm{H_{2}}})$ and $R^{3}_{eq}$ are the mass and equivalent radius of the MC, respectively. $\mu_{p}m_{p}=2.72m_{p}$ stands for the mean molecular weight, while $m_{p}$ represents the mass of proton. Using the area, distance, and mean H$_{2}$ column density $\overline{N}({\rm{H_{2}}})$ of the MC previously obtained, $\overline{n}({\rm{H_{2}}})$ calculated from Eq. (\ref{eq-nmc}) is $10.50\pm3.08$ cm$^{-3}$.

\subsection{IceCube neutrino dataset}\label{subsec:icecube}

High-energy neutrinos are the key to answer whether the $\gamma$-ray sources are hadronic or leptonic origins. Together with the multiwavelength observations, from radio to $\gamma$-rays, they may shed light on the intrinsic properties of the Galactic sources and their environments. 

IceCube neutrino observatory is a kilometer-scale neutrino experiment located at the south pole with 5160 digital optical modules deployed under ice. It was fully constructed on December 18, 2010, and has been collecting data ever since \citep{aartsen2017icecube}. For the first time, IceCube detected high-energy neutrinos coming from extraterrestrial origins \citep{aartsen2013first, icecube2013evidence}. Recent years, IceCube has found evidence for neutrino sources, such as TXS 0506+056, NGC1068, and Galactic plane \citep{IceCube:2018dTXS, IceCube:2018TXSnu, IceCube:2022NGC, IceCube:2023Galactic}. Some correlation studies of IceCube neutrino and LHAASO sources have been performed, like the works done by \citet{IceCube:2022heu}, \citet{Fang:2024LHAASOnu}, and \citet{Li:2024gnb}. 

We examined the IceCube high-energy starting events (HESE) data release of 12 years. The dataset contains 164 events that are reconstructed based on the latest ice model via DirectFit \citep{abbasi2023updated}. No significant association with an offset angle of 2$^{\circ}$ has been found for these six LHAASO sources. For the public data release of neutrino track-like events between April 6, 2008 and July 8, 2018, no significant excess was found. Therefore, the upper limits are obtained based on the information of exposure time, effective area, and atmospheric background rate at the level of 1 $\times$ 10$^{-9}$ erg cm$^{-2}$ s$^{-1}$ sr$^{-1}$ for the dumbbell-like structure \citep{IceCube:10yeardata, IceCube:10year}.

\section{Modeling research}\label{sec:Mod}

\subsection{Leptonic modeling}\label{subsec:lep}

As observed by the High-Altitude Water Cherenkov Observatory (HAWC) in 2017, the Geminga pulsar (PSR J0633$+$1746) is surrounded by a prominent TeV $\gamma$-ray halo, characterized by a hard spectral index of approximately 2.34 \citep{abeysekara2017extended}. Notably, the energy range and spectral index of the $\gamma$-ray halo around Geminga closely resemble those of the six enigmatic UHE sources detected by LHAASO. This similarity provides a compelling basis to consider Geminga as a reference leptonic model for interpreting the six LHAASO sources.

The relativistic wind generated by a rapid rotation pulsar like Geminga can accelerate the ambient particles, including electrons, and finally forms a bubble of shocked relativistic particles, the so-called PWN \citep{gaensler2006evolution}. For relativistic electrons, there are three typical cooling mechanisms\textemdash synchrotron radiation, inverse Compton (IC) scattering, and bremsstrahlung \citep{blumenthal1970bremsstrahlung,ghisellini1988synchrotron,baring1999radio}. All of these cooling mechanisms can lead to the emission of photons with various wavelengths. Most of the photons in the keV band are generated by synchrotron radiation, which originates from relativistic electrons propagating through the ambient magnetic field. Furthermore, with UHE electrons and a strong magnetic field, the energy of photons emitted from this mechanism can be up to GeV \citep{bednarek2003gamma}. On the other hand, IC scattering between relativistic electrons and the ambient radiation field is a major contributor to high-energy $\gamma$-ray emission. The radiation field in IC scattering can be CMB, far-infrared or near-infrared radiation, and visible light \citep{popescu2017radiation, zhang2021morphology}. The energy spectrum of this radiation mechanism can be extended up to the TeV range. Also, the bremsstrahlung caused by the collision between relativistic electrons and other particles can contribute to the emission of TeV $\gamma$-photons, while in general its $\gamma$-photon flux is significantly lower than that of IC scattering \citep{bednarek2003gamma}.

Considering all of the three electron cooling mechanisms mentioned above and assuming Geminga PWN as a reference counterpart, we study the leptonic origins of the six LHAASO sources. The $\gamma$-ray source modeling package GAMERA \citep{hahn2015gamera}, developed by the Max Planck Institute for Nuclear Physics in Heidelberg (MPIK), was used to calculate the leptonic spectra. The injection electrons of PWN were assumed to satisfy a power-law spectrum with an exponential cutoff. The function can be written as
\begin{equation}\label{eq-plcu}
	\frac{dN}{dE_{e}}=N_{e}\left(\frac{E_{e}}{1\,\rm{TeV}}\right)^{-\alpha_{e}}{\rm{exp}}\left(-\frac{E_{e}}{E_{c}}\right),
\end{equation}
where $E_{e}$ denotes the energy of injection electron. $\alpha_{e}$ and $E_{c}$ are the spectral index and cutoff energy, which are free parameters in our leptonic modeling research. $N_{e}$ is the normalization factor depending on the total energy $W_{e}$ of PWN electrons. The relationship between $N_{e}$ and $W_{e}$ is 
\begin{equation}\label{eq-re}
	W_{e}=N_{e}\int_{E^{e}_{\rm{min}}}^{E^{e}_{\rm{max}}}E_{e}\left(\frac{E_{e}}{1\,\rm{TeV}}\right)^{-\alpha_{e}}{\rm{exp}}\left(-\frac{E_{e}}{E_{c}}\right)dE_{e}.
\end{equation}

Here, $W_{e}$ was assumed to be 1.23 $\times$ 10$^{45}$ erg, corresponding to 0.01\% of the total spin-down energy of Geminga \citep{fang2018two}. The minimum and maximum energy of the PWN electrons $E^{e}_{\rm{min}}$ and $E^{e}_{\rm{max}}$ were set to be 0.1 GeV and 10 PeV to match the KM2A data. Consulting the approach in work \citet{lhaaso2025ultra}, the magnetic field strength associated with the six sources was set to 1 $\mu$G for calculating the intensity of synchrotron radiation, aligning with the X-ray observation results from \textit{Swift}-XRT. This was a reasonable value in Galactic halo \citep{jansson2012new}. Following \citet{popescu2017radiation} and \citet{zhang2021morphology}, the radiation fields in IC scattering were assumed to be CMB (with temperature of 2.73 K and energy density of 0.25 eV cm$^{-3}$), far-infrared radiation (with temperature of 40 K and energy density of 1 eV cm$^{-3}$), near-infrared radiation (with temperature of 500 K and energy density of 0.4 eV cm$^{-3}$), and visible light (with temperature of 3500 K and energy density of 1.9 eV cm$^{-3}$). The ambient particle number density around the PWN for bremsstrahlung calculation was assumed to be 0.1 cm$^{-3}$ \citep{de2022exploring}. The distance $d$ from the PWN to Earth was left free, as it determines the total flux intensity when $W_{e}$ was kept fixed. 

In summary, the free parameters to fit in our model were $\alpha_{e}$, $E_{c}$, and $d$. The Markov chain Monte Carlo (MCMC) method was used for fitting \citep{geyer1992practical}. In this procedure, the input and output in our model were the energy of $\gamma$-photons and the corresponding SED, respectively. The experimental data to fit were the fitting KM2A SEDs of the six sources presented by the LHAASO catalog \citep{cao2024first}. For leptonic modeling, the upper limits of SEDs in different energy bands based on \textit{Fermi}-LAT (see Sect. \ref{subsec:fermi}), \textit{Swift}-XRT (see Sect. \ref{subsec:swift}), and WCDA \citep{cao2024first} datasets were adopted to constrain the fitting. A punishment on likelihood while the output of model exceeding the upper limit was added to the likelihood function of the MCMC fitting. Before running the fitting, the scientific calculation toolkit Scipy was used to obtain the start point of the Markov chain. Each chain contained 2000 steps while the first 200 steps were regarded as burn-in steps. The entire calculation and fitting were performed under Python3 environment.

\begin{figure*}[!h]
    \centering
    \includegraphics[width=1.0\textwidth]{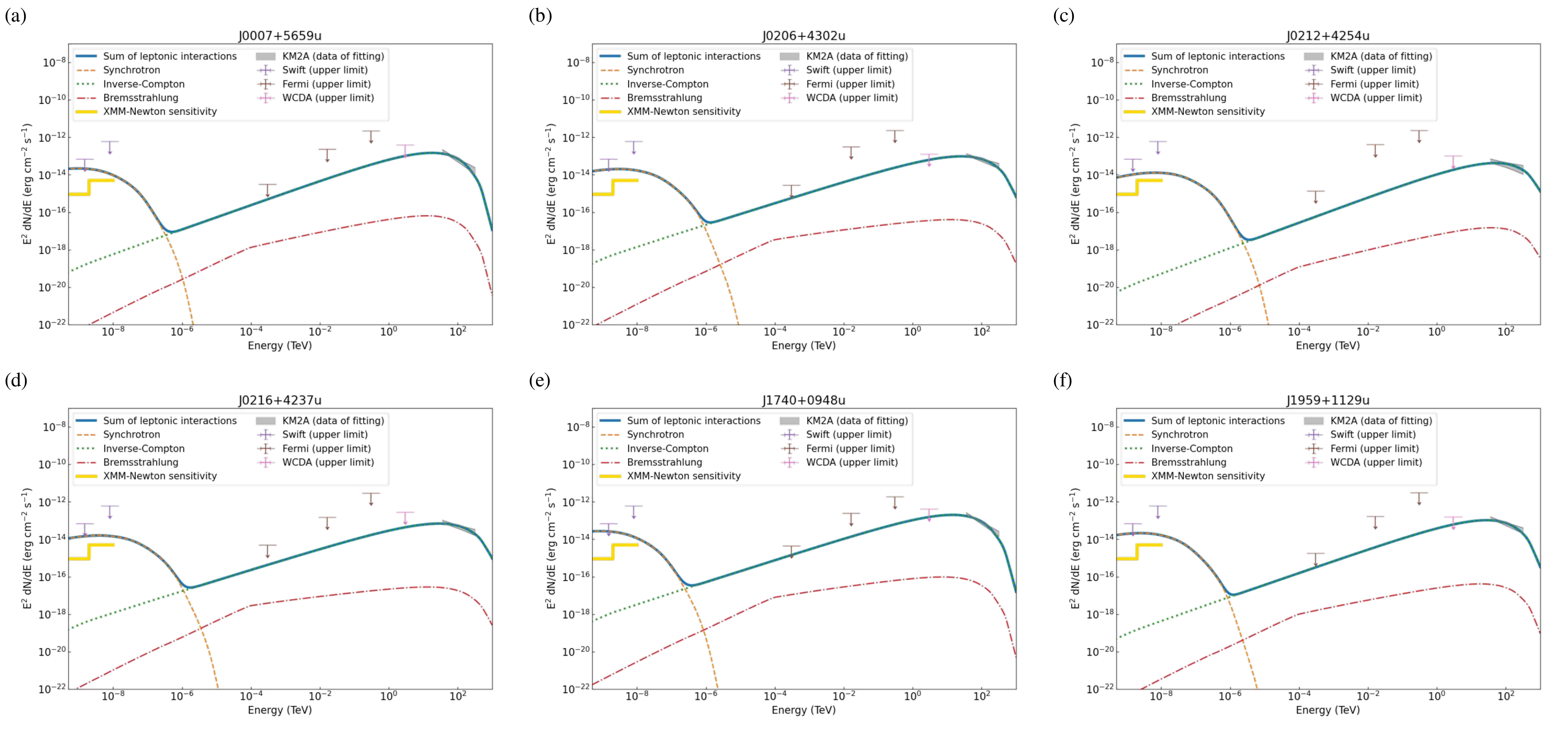}
    \caption{\label{fig:4}Results of leptonic modeling research. Panel (a) shows the result of J0007$+$5659u. The solid blue line represents the total SED of $\gamma$-ray generated by leptonic interactions. The dashed and dotted lines in different colors are the SEDs contributed from synchrotron radiation, IC scattering, and bremsstrahlung. The pointing down arrows stand for the upper limits obtained from the observations of \textit{Swift}-XRT, \textit{Fermi}-LAT, and WCDA. The expected sensitivity of 3 $\times$ 10$^{4}$ seconds' observation with EPIC-pn of \textit{XMM-Newton} has been shown by the thick golden line. The fitting SED of KM2A data with error bar is displayed by a butterfly plot filled in grey. Panels (b)-(f) are the results of J0206$+$4302u, J0212$+$4254u, J0216$+$4237u, J1740$+$0948u, and J1959$+$1129u, respectively.}
\end{figure*}

Figure \ref{fig:4} shows the results of our leptonic modeling research. The corner plots showing the posterior probability distribution function (PDF) of each fitting parameter in the model are given in Fig. \ref{fig:B1} of Appendix \ref{sec:cor}. According to these corner plots, the MCMC fittings have reached convergence. The best-fit $\alpha_{e}$, $E_{c}$, and $d$ are presented in Table \ref{tab:2}. For all of the six sources, the best-fit spectral indexes are within the range of 1.6-1.9, while the best-fit cutoff energies are above 100 TeV. These best-fit values are reasonable for electron injection of a PWN \citep{gaensler2006evolution, de2022potential}, implying a scenario where the six LHAASO sources may originate from energetic PWNe like Geminga. The relatively large uncertainties in the best-fit values of $E_{c}$ and $d$ listed in Table \ref{tab:2} stem from the scarcity of observational data, as the multiwavelength and multimessenger study has only provided upper limits. This represents the primary limitation of our modeling research.

\renewcommand\arraystretch{1.5}
\begin{table}
  \centering
  \caption{\label{tab:2}Best-fit parameters in leptonic modeling research}
  \begin{tabular}{cccc}
  \hline
    Source & $\alpha_{e}$ & $E_{c}$ (TeV) & $d$ (kpc) \\ \hline
    J0007$+$5659u & $1.70^{+0.22}_{-0.41}$ & $119.90^{+35.05}_{-29.92}$ & $2.49^{+1.06}_{-0.84}$ \\
    J0206$+$4302u & $1.85^{+0.19}_{-0.33}$ & $238.89^{+103.16}_{-67.75}$ & $2.67^{+1.33}_{-1.05}$ \\
    J0212$+$4254u & $1.64^{+0.31}_{-0.58}$ & $300.96^{+267.43}_{-139.33}$ & $5.57^{+1.76}_{-2.31}$ \\
    J0216$+$4237u & $1.87^{+0.24}_{-0.45}$ & $281.65^{+158.81}_{-106.30}$ & $3.10^{+2.03}_{-1.64}$ \\
    J1740$+$0948u & $1.83^{+0.14}_{-0.30}$ & $125.44^{+27.47}_{-23.49}$ & $1.74^{+0.91}_{-0.51}$ \\
    J1959$+$1129u & $1.73^{+0.19}_{-0.39}$ & $196.45^{+74.77}_{-58.63}$ & $3.05^{+1.22}_{-0.94}$ \\
    \hline
  \end{tabular}
\end{table}
\renewcommand\arraystretch{1}

In Fig. \ref{fig:4}, the expected sensitivity of 3 $\times$ 10$^{4}$ seconds' observation with the EUROPEAN PHOTON IMAGING CAMERA (EPIC) integrated on \textit{XMM-Newton} in the energy range of 0.5-10 keV is also presented.\footnote{The sensitivity is based on the \textit{XMM-Newton} Users Handbook, which can be downloaded from \url{http://xmm-tools.cosmos.esa.int/external/xmm_user_support/documentation/uhb/XMM_UHB.pdf}.} It can be seen that, even with a weak magnetic field strength of 1 $\mu$G, the sensitivity limit of \textit{XMM-Newton} can reach the predicted X-ray SED in our leptonic model, while a proper exposure time is required. That means, it is possible for some sensitive X-ray telescopes like \textit{XMM-Newton} to detect the potential X-ray fluxes from the six LHAASO sources, which is essential for judging whether the six sources belonging to leptonic origins such as the IC scattering between UHE electrons from PWN and CMB. Thus, further research in the X-ray band with state-of-the-art X-ray telescopes is anticipated.

\subsection{Hadronic modeling}\label{subsec:had}

In addition to the three cooling mechanisms of relativistic electrons appeared in the last subsection, p-p interactions also play a significant role in the production of UHE $\gamma$-rays. In the category of astronomy, p-p interactions conventionally happen between accelerated CR protons and cold MC protons. Inelastic collisions among these protons will produce light mesons, which can decay into $\gamma$-photons via reactions like $\pi^{0}\rightarrow\gamma\gamma$ and $\eta\rightarrow\gamma\gamma$. This process is always accompanied by the production of neutrinos. Thus, if an UHE $\gamma$-ray source is confirmed as hadronic origin, the study combined with neutrino detection will be very helpful in revealing the exact physical mechanisms behind it.

As previously mentioned, for the p-p interaction, the MC which provides cold protons is always essential. According to Sects. \ref{subsec:planck} and \ref{subsec:co}, we can only find a thin MC in the vicinity of J0007$+$5659u. Although the position offset of approximately 0\fdg5 may decrease the possibility of J0007$+$5659u being associated with the MC, we still explored a p-p interaction model to evaluate the possibility of this source belonging to the hadronic scenario. In the model, we assumed J0007$+$5659u originating from the p-p interaction between injection CR protons and cold protons from the MC indicated in Fig. \ref{fig:3} (a). The modeling research was also performed with the GAMERA toolkit \citep{hahn2015gamera}. The injection CR protons were assumed to have a simple power-law spectrum in the form of 
\begin{equation}\label{eq-plh}
	\frac{dN}{dE_{p}}=N_{p}\left(\frac{E_{p}}{1\,\rm{TeV}}\right)^{-\alpha_{p}},
\end{equation}
where $E_{p}$ denotes the energy of injection CR proton. $N_{p}$ represents the normalization factor depending on the total energy $W_{p}$ of the injection CR protons. Since there was no information about the source responsible for accelerating CR protons, $W_{p}$ was set as a free parameter in our model. $\alpha_{p}$ is the spectral index. It was also left free in the model. The maximum energy $E^{p}_{\rm{max}}$ of CR protons was chosen to be the CR knee energy of $\sim$ 3.16 PeV, while the minimum energy $E^{p}_{\rm{min}}$ was another free parameter. 

Except for the energy spectrum of CR protons, the total inelastic cross section is also of great importance in p-p interaction. In the GAMERA toolkit we adopted, it has the expression of \citep{kafexhiu2014parametrization} 
\begin{equation}\label{eq-cs}
    \sigma_{\rm{pp}}(E_{p})=(30.7-0.96{\rm{log}}L+0.18{\rm{log}^{2}}L)\times(1-L^{-1.9})^{3}.
\end{equation}
In Eq. (\ref{eq-cs}), the cross section is presented in the unit mb (mbarn). $L=E_{p}/(0.2797\,\rm{GeV})$ is a energy dependent dimensionless variable. 

As previously discussed, there were three free parameters to fit in our p-p interaction modeling research\textemdash $W_{p}$, $\alpha_{p}$, and $E^{p}_{\rm{min}}$. To simplify the calculation, we assumed that $W_{p}$ satisfies the formula $W_{p}=1\times10^{\Gamma_{1}}$ erg and $E^{p}_{\rm{min}}$ satisfies the formula $E^{p}_{\rm{min}}=1\times10^{\Gamma_{2}}$ TeV. Thus, the fitting parameters were transformed into three dimensionless indexes $\Gamma_{1}$, $\Gamma_{2}$, and $\alpha_{p}$. As in Sect. \ref{subsec:lep}, the input and output in our p-p interaction model were the energy of $\gamma$-photons and the corresponding SED, respectively. An MCMC method with each chain containing 2000 steps (the first 200 steps were burn-in steps) was adopted, while the fitting KM2A SED of J0007$+$5659u was regarded as experimental data. The upper limits of SEDs in different energy bands based on \textit{Fermi}-LAT (see Sect. \ref{subsec:fermi}) and WCDA \citep{cao2024first} datasets were used to constrain the output of model. The upper limits of neutrino flux obtained from the IceCube dataset (see Sect. \ref{subsec:icecube}) were much higher than those of \textit{Fermi}-LAT and WCDA. Thus, they were not involved in the fitting. During the processing, the distance and mean particle number density of the MC were the same as presented in Sect. \ref{subsec:co}. 

\begin{figure}
    \centering
    \includegraphics[width=0.5\textwidth]{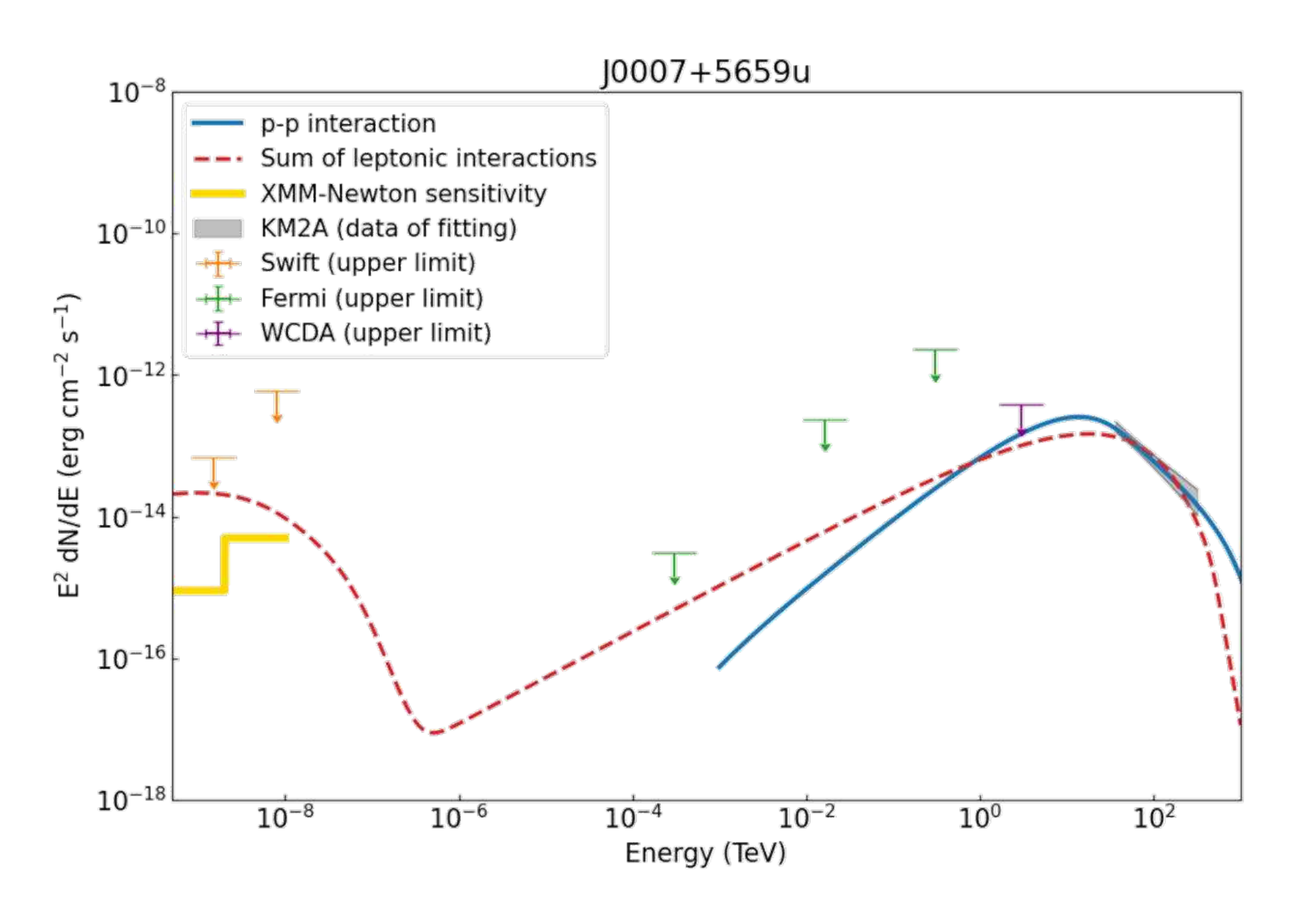}
    \caption{\label{fig:5}The best-fit result of the hadronic model for J0007$+$5659u, fitted to the multiwavelength upper limits, is compared with the best-fit result of the leptonic model. The solid blue line represents the SED of $\gamma$-ray generated by p-p interaction. The dashed line in red stands for the total SED of $\gamma$-ray generated by leptonic interactions. The pointing down arrows denote the upper limits of SED according to the observations of \textit{Swift}-XRT, \textit{Fermi}-LAT, and WCDA. Also, the expected sensitivity of 3 $\times$ 10$^{4}$ seconds' observation with EPIC-pn of \textit{XMM-Newton} has been shown by the thick golden line. The fitting SED of KM2A data with error bar is displayed by a butterfly plot filled in grey.}
\end{figure}

Figure \ref{fig:5} shows the result of our hadronic modeling research on J0007$+$5659u compared with that in the leptonic scenario. Notably, the $\gamma$-ray SED generated by p-p interaction seems more consistent with the fitting SED of KM2A. Also, the corner plot showing the posterior PDF of each fitting parameter in the model is presented in Fig. \ref{fig:B2} of Appendix \ref{sec:cor}. The MCMC fitting in hadronic model is converged, as shown by the corner plot. The best-fit $\Gamma_{1}$ is $46.19^{+0.21}_{-0.19}$, which will result in $W_{p}$ of $1.56^{+0.74}_{-0.67}\times10^{46}$ erg. Adopting the assumption of symmetric spatial distribution of the MC, which is previously discussed in Sect. \ref{subsec:co}, the required energy density $U_{p}$ of the injection CR protons is $0.82^{+0.41}_{-0.38}$ eV cm$^{-3}$. This value is approximately 1.7 times lower than the average value in the Milky Way, while higher than that of the Large Magellanic Cloud (LMC) \citep{yoast2016equipartition}. The best-fit $\Gamma_{2}$ is $1.84^{+0.21}_{-0.20}$, corresponding to $E^{p}_{\rm{min}}$ of $68.84^{+34.06}_{-31.99}$ TeV. The best-fit $\alpha_{p}$ is $3.12^{+0.22}_{-0.18}$, which is reasonable for CR protons \citep{de2022exploring}. The sensitivity limit of \textit{XMM-Newton} with an exposure time of 3 $\times$ 10$^{4}$ seconds, which can reach the X-ray flux predicted by our leptonic model, is also presented in Fig. \ref{fig:5}. Observations of the exact location of J0007$+$5659u with current X-ray telescopes, such as \textit{XMM-Newton}, are expected to be highly useful in determining whether this source has a leptonic or hadronic origin, as the hadronic scenario does not predict X-ray emission.

The minimum energy of the injection protons expected in our hadronic model is relatively high. This might imply a scenario where a supernova remnant (SNR) possessing a long Sedov time \citep{de2022exploring} is associated with the MC in proximity to J0007$+$5659u. The SNR traps low-energy CR protons in a small region and leads to the lonely $>$ 25 TeV emission from J0007$+$5659u. However, our examination within the online high-energy SNR catalog \citep{ferrand2012census} appeared not to support this explanation, as no SNR spatially close to J0007$+$5659u was found.\footnote{\url{http://snrcat.physics.umanitoba.ca}} Searching for some other CR proton accelerator candidates like blazars \citep {murase2012blazars} and active galactic nuclei (AGN) \citep{george2008active} in this sky region, along with a CO survey possessing higher angular resolution, are required for further research.

As previously mentioned, the generation of $\gamma$-photons in p-p interaction is always accompanied by the production of neutrinos via the decay of pions and muons. Thus, neutrino detection is also essential for identifying the origin of a $\gamma$-ray source and can provide helpful supplementary information about the source. According to our hadronic modeling research, the expected neutrino SED from J0007$+$5659u is approximately an order of magnitude below the sensitivity limit of IceCube-Gen2 after 10 years of observation at 10 TeV, and about two orders of magnitude lower at 100 TeV. It seems unlikely to observe the neutrino fluxes emitted from this source at present, even the hadronic origin has been confirmed. However, with improved technologies, there are still prospects of detecting associated neutrinos with some other proposed next generation neutrino observatories, such as TRIDENT \citep{ye2022proposal} and NEON \citep{zhang2024proposed}. The details of neutrino SED calculation are provided in Appendix \ref{sec:neu}.

\section{Traveling-PWN modeling attempt}\label{sec:tra}

One of the most attractive discoveries presented by the LHAASO catalog is the dumbbell-like structure comprising J0206$+$4302u, J0212$+$4254u, and J0216$+$4237u.\footnote{The significance map on the dumbbell-like structure is presented at the left bottom of Fig. 10 in \citet{cao2024first}.} The similarities in position and spectral shape of the three sources suggest a physical association among them. The low UHE $\gamma$-ray source density at such high galactic latitude makes this assumption more plausible. Some previous studies have attempted to explain the physical association proposing mechanisms such as the asymmetric propagation of UHE electrons caused by turbulent magnetic fields surrounding pulsars to account for this structure and identifying PSR J0218$+$4232 as a potential counterpart \citep{bao2024mirages, bao2024mirage}. Furthermore, the relativistic jet from a microquasar, which can form a double-peaked $\gamma$-ray morphology \citep{hess2024acceleration}, is also a potential explanation for the generation of the dumbbell-like structure. However, the current multiwavelength and multimessenger observation on this structure have not shown enough evidence. In this section, we attempt to use the traveling-PWN model with one traveling-PWN, two traveling-PWNe, and three PWNe under the isotropic and homogeneous diffusion condition to explain the origin of the dumbbell-like structure.

\subsection{Single traveling-PWN modeling attempt}\label{subsec:sin}

Firstly, we discuss the hypothesis that the dumbbell-like structure was formed by a single traveling-PWN. For this scenario, the model only requires one corresponding pulsar, and its complexity can be significantly reduced, since there is no need to account for the probability of spatial coincidences among multiple pulsars. Similar to the propagation of CR particles, the relativistic electrons accelerated by the central pulsar associated with a PWN will undergo spatial diffusion. This process is accompanied by the proper motion of the central pulsar and leaves $\gamma$-ray and X-ray filaments. According to the statistical study, a pulsar can travel with a velocity of up to $\sim$ 1600 km s$^{-1}$ \citep{hobbs2005statistical}. If the distance $d$ between the PWN and Earth is short enough (e.g. $<$ 0.5 kpc), with such proper motion velocity, the significant $\gamma$-ray filament is probably to be detected by LHAASO-KM2A and misunderstood as three point-like sources.

The diffusion of electrons accelerated by a pulsar can be described by a diffusion equation \citep{delahaye2008positrons, delahaye2010galactic, fang2018two, zhang2021morphology, fang2022interpretation}
\begin{equation}\label{eq-tr}
	\frac{\partial\psi}{\partial t}-\nabla\cdot[D(E_{e})\nabla\psi]-\frac{\partial}{\partial E_{e}}[b(E_{e})\psi]=q(\vec{r},E_{e},t),
\end{equation}
where $\psi(\vec{r},E_{e},t)$ denotes the electron number density per unit energy. $D(E_{e})\equiv D_{0}\epsilon^{\delta}$ is the energy-dependent diffusion coefficient assumed to be isotropic and homogeneous \citep{zhang2021morphology, fang2022interpretation}. $D_{0}$ represents the normalization factor. $\epsilon\equiv E_{e}/E_{d}$ is the dimensionless energy of electron, where $E_{d}=1$ GeV \citep{delahaye2008positrons, delahaye2010galactic}. $b(E_{e})$ describes the energy-loss rate of electrons due to synchrotron radiation and IC scattering on cosmic radiation fields \citep{delahaye2008positrons}. The exact form of $b(E_{e})$, considering the Klein-Nishina effect \citep{nakar2009klein}, can be written as \citep{zhang2021morphology}
\begin{eqnarray}\label{eq-b}
    b(E_{e})&=&-\frac{dE_{e}}{dt}\nonumber\\&=&\frac{4}{3}\sigma_{T}c\left(\frac{E_{e}}{m_{e}c^{2}}\right)^{2}\left[U_{B}+\frac{U_{\rm{ph}}}{\left(1+\frac{4E\epsilon_{0}}{m_{e}^{2}c^{4}}\right)^{3/2}}\right]\nonumber\\&=&\frac{4}{3}\sigma_{T}c\epsilon^{2}\left(\frac{E_{d}}{m_{e}c^{2}}\right)^{2}\left[U_{B}+\frac{U_{\rm{ph}}}{\left(1+\frac{4E_{d}\epsilon\epsilon_{0}}{m_{e}^{2}c^{4}}\right)^{3/2}}\right].
\end{eqnarray}
Here, $\sigma_{T}$ is the Thomson cross section. $c$ denotes the velocity of light. $m_{e}$ stands for the mass of electron. $U_{B}=B^{2}/(2\mu_{0})$ is the energy density of magnetic field, where $B$ represents the strength of magnetic field and $\mu_{0}$ stands for the permeability of vacuum. $U_{\rm{ph}}$ is the energy density of radiation field. $\epsilon_{0} = 2.82k_{b}\rm{T}$ is the typical photon energy of the blackbody/graybody target radiation field, as $k_{b}$ and $\rm{T}$ represent the Boltzmann constant and temperature, respectively. 

Assuming the PWN to be a constant injection source, the source term $q(\vec{r},E_{e},t)$ is expected to follow a power-law spectrum with an exponential cutoff
\begin{eqnarray}\label{eq-q}
	q(\vec{r},E_{e},t)&=&q_{0}\left(\frac{E_{e}}{1\,\rm{GeV}}\right)^{-\Gamma}{\rm{exp}}\left(-\frac{E_{e}}{E_{c}}\right)\delta(\vec{r}-\vec{r}_{0}-\vec{v}t)\nonumber\\
    &=&q_{0}\epsilon^{-\Gamma}{\rm{exp}}\left(-\frac{E_{d}}{E_{c}}\epsilon\right)\delta(\vec{r}-\vec{r}_{0}-\vec{v}t),
\end{eqnarray}
where $\Gamma$ is the spectral index of injection spectrum. $\vec{r}_{0}$ denotes the position of the traveling-PWN at $t=0$ and $\vec{v}$ stands for the proper motion velocity of the central pulsar. $q_{0}$ is the normalization factor depending on the total energy of PWN injection, which was denoted by $W_{e}$ in the previous text. The relationship between $q_{0}$ and $W_{e}$ is similar to that between $N_{e}$ and $W_{e}$, as presented in Eq. (\ref{eq-re}). The $W_{e}$ is a fraction $\eta_{0}$ of the total spin-down energy $W_{0}$ of the corresponding pulsar, while $W_{0}$ satisfies the equation \citep{gaensler2006evolution, fang2022interpretation}
\begin{equation}\label{eq-w0}
    W_{0}=\dot{\xi}t_{a}\left(1+\frac{t_{a}}{\tau_{0}}\right)\left(\frac{t_{a}+\tau_{0}}{t+\tau_{0}}\right)^{\frac{n+1}{n-1}}.
\end{equation}
Here, $\dot{\xi}$ denotes the rotational kinetic energy dissipation rate or called "spin-down luminosity" of the pulsar at $t=t_{a}$, and $t_{a}$ represents the evolutionary time of the pulsar. In our model, the spin-down luminosity decays with time following \citet{ding2021implications}, meanwhile, $\Gamma$ and $E_{c}$ of the accelerated electrons remain constant through the traveling history. $\tau_{0}\sim10$ kyr is the typical spin-down luminosity decay time \citep{ding2021implications}. 

Based on Eqs. (\ref{eq-re}), (\ref{eq-b}), (\ref{eq-q}), and (\ref{eq-w0}), the full expansion of Eq. (\ref{eq-tr}) can be expressed as
\begin{eqnarray}\label{eq-fu}
    &&\frac{\partial\psi}{\partial t}-D_{0}\epsilon^{\delta}\Delta\psi-\frac{\partial}{\partial\epsilon}\left(\frac{\epsilon^{2}}{\tau_{l}}\psi\right)\nonumber\\&=&q(\vec{r},\epsilon,t)\nonumber\\
    &=&\frac{\eta_{0}\dot{\xi}}{E_{d}^{2}I_{\epsilon}}\left(1+\frac{t_{a}}{\tau_{0}}\right)\left(\frac{t_{a}+\tau_{0}}{t+\tau_{0}}\right)^{\frac{n+1}{n-1}}\nonumber\\&&\times\,\epsilon^{-\Gamma}{\rm{exp}}\left(-\frac{E_{d}}{E_{c}}\epsilon\right)\delta(\vec{r}-\vec{r}_{0}-\vec{v}t)\gamma_{n}(t),
\end{eqnarray}
while
\begin{eqnarray}\label{eq-tau}
    \frac{1}{\tau_{l}}&=&\frac{4}{3}\sigma_{T}c\frac{E_{d}}{m_{e}^{2}c^{4}}\left[U_{B}+\frac{U_{\rm{ph}}}{\left(1+\frac{4E_{d}\epsilon\epsilon_{0}}{m_{e}^{2}c^{4}}\right)^{3/2}}\right],\\
    \label{eq-i}I_{\epsilon}&=&\int_{\epsilon_{\rm{min}}}^{\epsilon_{\rm{max}}}\epsilon_{1}^{1-\Gamma}{\rm{exp}}\left(-\frac{E_{d}}{E_{c}}\epsilon_{1}\right)d\epsilon_{1}.
\end{eqnarray}
Here, $\epsilon_{\rm{min}}$ and $\epsilon_{\rm{max}}$ represent the constraints on the energy of electrons emitted by PWN. $\gamma_{n}(t)$ is a step function, which returns 1 when $t\geq0$.

\begin{figure*}[!h]
    \centering
    \includegraphics[width=0.95\textwidth]{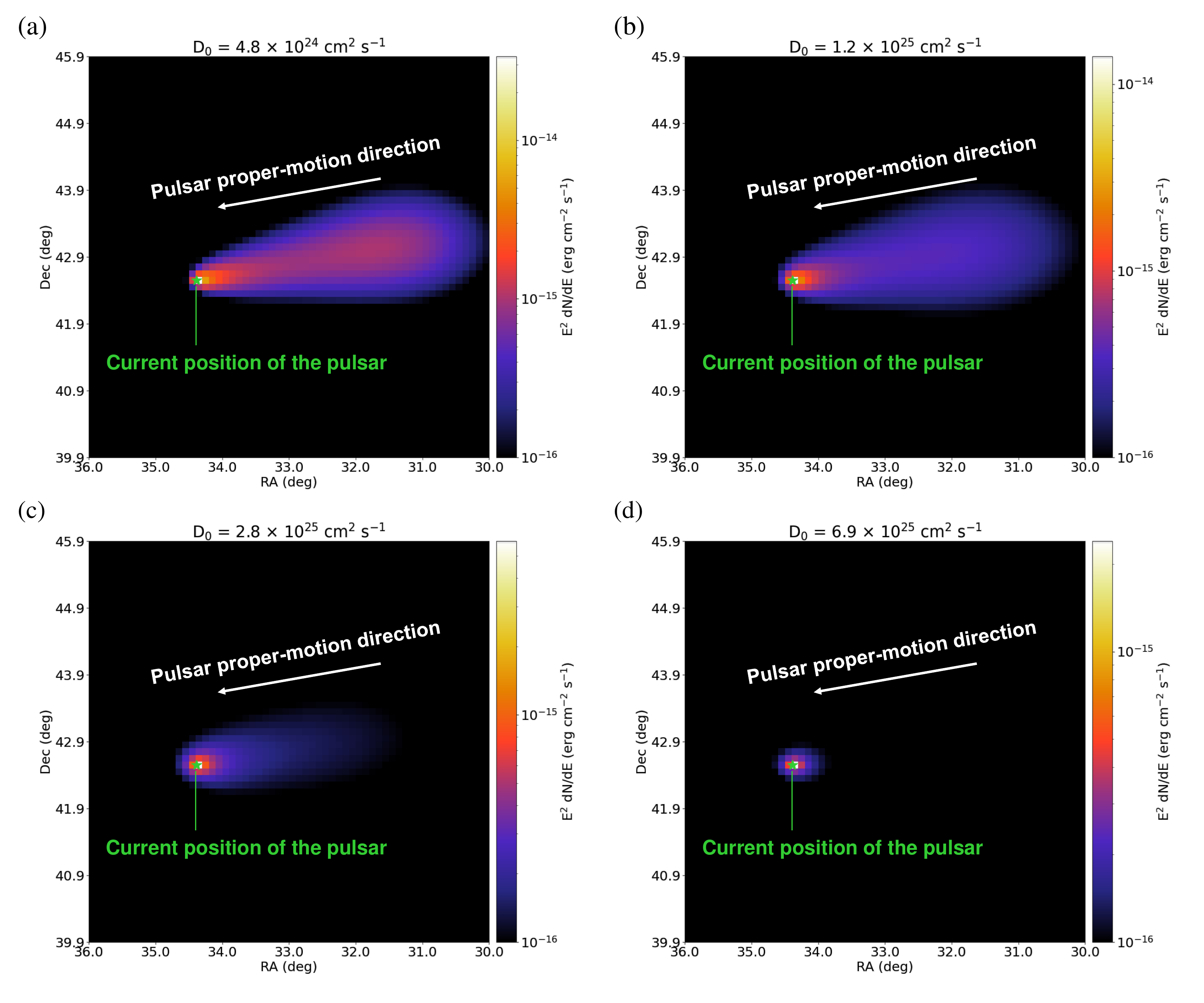}
    \caption{\label{fig:6} The $\gamma$-ray flux intensity maps at 50 TeV generated for different values of $D_{0}$. Panels (a)-(d) give the results for $D_{0}=4.8$ $\times$ 10$^{24}$ cm$^{2}$ s$^{-1}$, $1.2\times$ 10$^{25}$ cm$^{2}$ s$^{-1}$, $2.8$ $\times$ 10$^{25}$ cm$^{2}$ s$^{-1}$, and $6.9$ $\times$ 10$^{25}$ cm$^{2}$ s$^{-1}$, respectively. For comparing with the physical image shown in the LHAASO catalog, the coordinate system used here is the J2000 equatorial frame. Other parameters adopted in the calculation are $d=0.4$ kpc and $v=1600$ km s$^{-1}$. The white arrow indicates the proper motion direction of the pulsar, while the green star marks the current position of the pulsar.}
\end{figure*}

As the diffusion coefficient is isotropic and homogeneous, we can use the Green's function method to solve Eq. (\ref{eq-fu}) analytically. For our model, the Green's function can be written as \citep{fang2022interpretation}  \begin{equation}\label{eq-gr}
    G(\vec{r},t\leftarrow\vec{r}_{s},t_{s})=\frac{b(\epsilon^{\ast})}{b(\epsilon)}\frac{1}{(\pi\lambda^{2})^{3/2}}{\rm{exp}}\left[-\frac{(\vec{r}-\vec{r}_{s})^{2}}{\lambda^{2}}\right],
\end{equation}
where $\vec{r}_{s}$ and $t_{s}$ represent the position and time of an arbitrary electron injection in the Green's function method. The diffusion length $\lambda$ and the modified dimensionless energy $\epsilon^{\ast}$ are expressed as follows:
\begin{eqnarray}\label{eq-exp}
    \lambda^{2}&=&\frac{4D_{0}\tau_{l}}{\delta-2}({\epsilon^{\ast}}^{\delta-1}-\epsilon^{\delta-1}),\\
    \label{eq-mod}\epsilon^{\ast}&=&\frac{\epsilon}{1-\epsilon/\tau_{l}(t-t_{s})}.
\end{eqnarray}
For a free boundary condition, the analytical solution of Eq. (\ref{eq-fu}) based on Green's function method can be expressed as 
\begin{equation}\label{eq-so}
    \psi(\vec{r},\epsilon,t)=\int_{\rm{R}^{3}}d^{3}\vec{r}_s\int_{t_{i}}^{t}dt_{s}G(\vec{r},t\leftarrow\vec{r}_{s},t_{s})q(\vec{r}_{s},\epsilon^{\ast},t_{s}).
\end{equation}
Here, $t_{i}={\rm{max}}\{t-\tau_{l}/\epsilon,0\}$ is the injection time, while $\tau_{l}/\epsilon=E_{e}/b(E_{e})$ describes the cooling timescale of the injection electrons \citep{zhang2021morphology}.

The morphology of PWN electrons diffusion, combined with proper motion, depends on several parameters, with the most significant being the diffusion coefficient $D(E_{e})$, the distance $d$, and the proper motion velocity $v$. Thus, we firstly explore the influence of varying $D_{0}$, $d$, and $v$. From Eq. (\ref{eq-exp}), it can be seen that the diffusion length of electron is in direct proportion to $\sqrt{D_{0}}$. With a higher $D_{0}$, the electrons diffuse more rapidly, resulting in a decrease in the local electron number density and a lower $\gamma$-ray flux intensity. The parameters $d$ and $v$ mainly influence the evolutionary time $t_{a}$ of the pulsar. To form the $>$ 25 TeV dumbbell-like structure, $t_{a}$ needs to be shorter than the cooling timescale of $>$ 100 TeV electrons which are essential to the generation of $>$ 25 TeV $\gamma$-photons. Moreover, $d$ also have impact on the extension size of the source and the $\gamma$-ray flux intensity experimentally observed.

\begin{figure*}[!h]
    \centering
    \includegraphics[width=0.95\textwidth]{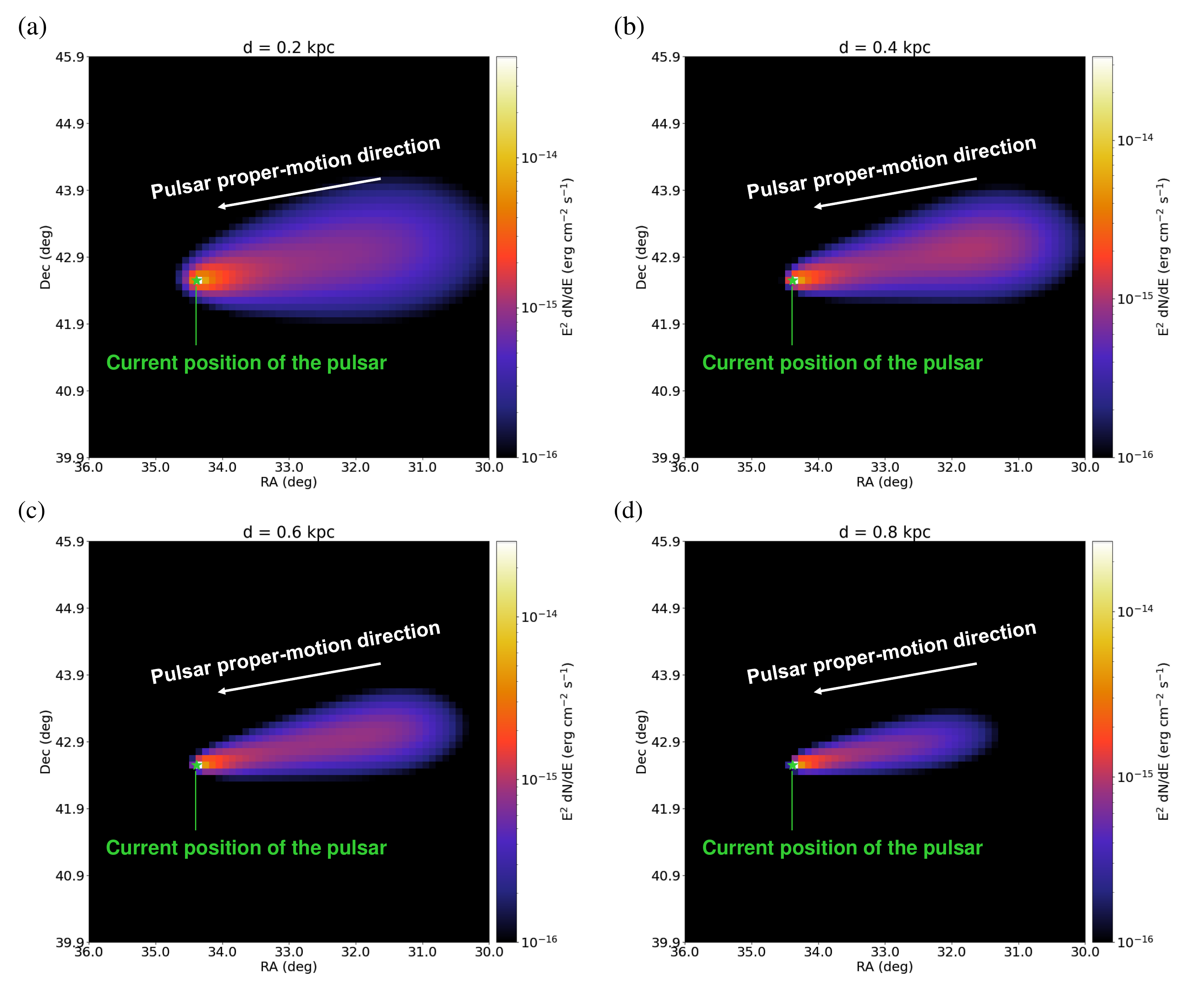}
    \caption{\label{fig:7} The $\gamma$-ray flux intensity maps at 50 TeV generated for different values of $d$. Panels (a)-(d) give the results for $d=0.2$ kpc, 0.4 kpc, 0.6 kpc, and 0.8 kpc, respectively. Other parameters adopted in the calculation are $D_{0}=4.8$ $\times$ 10$^{24}$ cm$^{2}$ s$^{-1}$ and $v=1600$ km s$^{-1}$.}
\end{figure*}

The spatial distribution of electron number density per unit energy can be directly calculated by Eq. (\ref{eq-so}) and transformed into $\gamma$-ray flux intensity map with the magnetic field and radiation field model adopted in Sect. \ref{subsec:lep}. Figure \ref{fig:6} presents the $\gamma$-ray flux intensity maps at 50 TeV corresponding to different values of $D_{0}$. Here, a pulsar proper motion perpendicular to the line of sight (LOS) is assumed \citep{zhang2021morphology}, and $D_{0}$ varies from an extremely low observed value of 4.8 $\times$ 10$^{24}$ cm$^{2}$ s$^{-1}$ around HESS J1026$-$582 \citep{di2020evidences} to the typical value of 6.9 $\times$ 10$^{25}$ cm$^{2}$ s$^{-1}$ around Geminga \citep{fang2022interpretation}. The $D_{0}$ of 4.8 $\times$ 10$^{24}$ cm$^{2}$ s$^{-1}$ is much lower than the Bohm limit \citep{abeysekara2017extended, mukhopadhyay2022self}. Such value can only be reached under a strong magnetic turbulence condition \citep{shalchi2009diffusive, hussein2014detailed}, which can result in anisotropic diffusion. In Fig. \ref{fig:6}, the lower bound of $\gamma$-ray SED is set at 1 $\times$ 10$^{-16}$ erg cm$^{-2}$ s$^{-1}$. As shown in Fig. \ref{fig:6}, with the increase of $D_{0}$, the rapid diffusion of electrons will decrease the local electron number density and consequently diminish the $\gamma$-ray flux intensity. This leads to the TeV $\gamma$-ray blob becoming smaller (due to the SED cut) and dimmer as $D_{0}$ increases. Thus, for the single traveling-PWN scenario, only if the diffusion coefficient is significantly lower than the Bohm limit, the TeV $\gamma$-photons emitted from the PWN can cover the entire sky region of the dumbbell-like structure as presented in Fig. \ref{fig:6} (a).

The $\gamma$-ray flux intensity maps at 50 TeV with different values of $d$ are presented in Fig. \ref{fig:7}. Also, the direction of pulsar proper motion is assumed to be perpendicular to the LOS. $d$ varies from 0.2 kpc to 0.8 kpc, which are reasonable values. As in Fig. \ref{fig:6}, the lower bound of the $\gamma$-ray SED in Fig. \ref{fig:7} is set at 1 $\times$ 10$^{-16}$ erg cm$^{-2}$ s$^{-1}$. As shown in the figure, when $d$ is too small, leading to an excessively short $t_{a}$, the recently injected energetic electrons form a large and bright $\gamma$-ray blob. In this case, a double-peaked structure, like the one observed in the LHAASO significance map, cannot be reproduced. With the increase of $d$, $t_{a}$ can be longer than the cooling timescale of UHE electrons. Due to the cooling of previously injected electrons, the TeV $\gamma$-ray emission ceases to cover the entire sky region of the dumbbell-like structure. Therefore, the dumbbell-like structure can only be produced by a single pulsar located at an appropriate distance, as demonstrated in Fig. \ref{fig:7} (b).

\begin{figure*}[!h]
    \centering
    \includegraphics[width=0.95\textwidth]{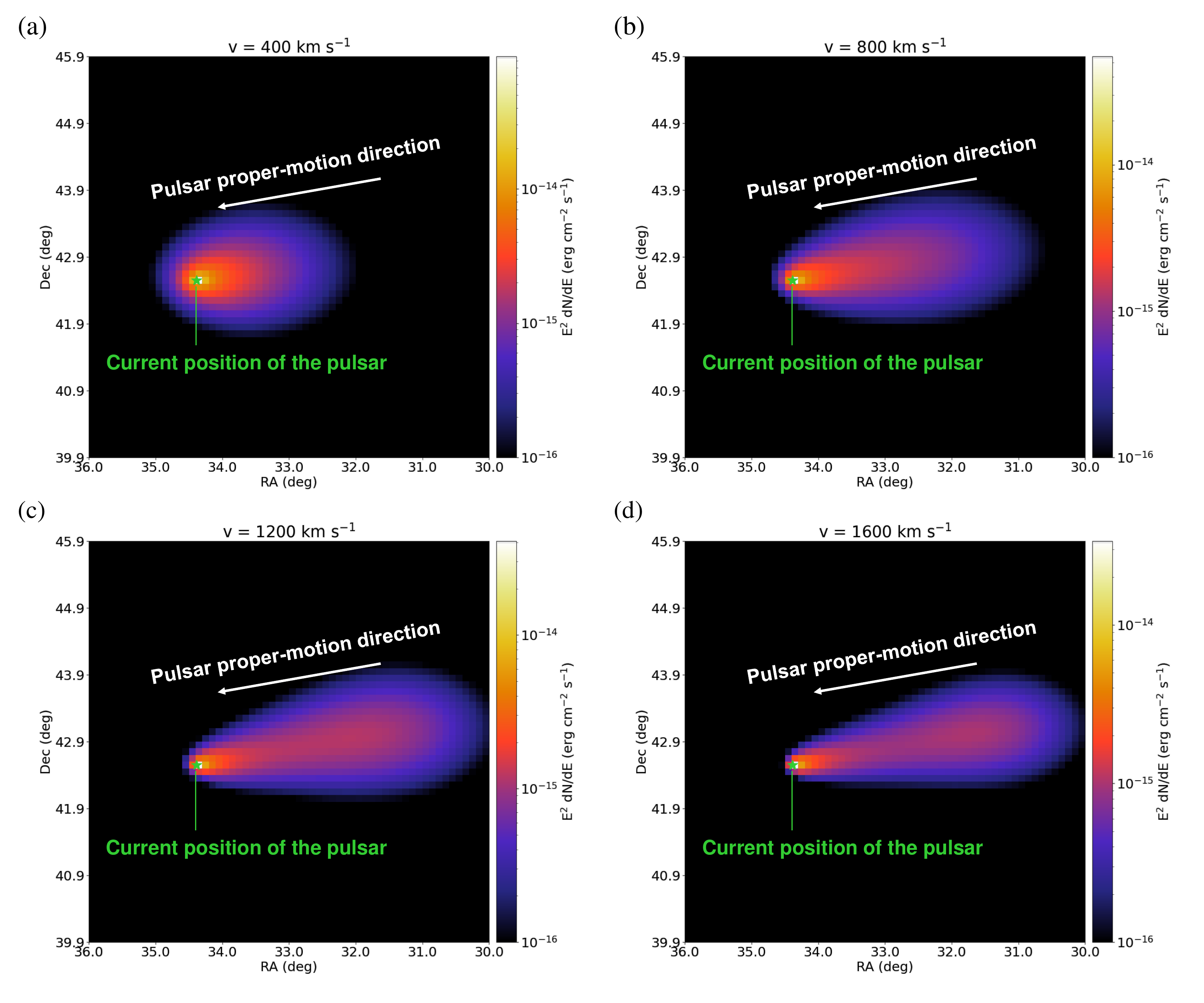}
    \caption{\label{fig:8} The $\gamma$-ray flux intensity maps at 50 TeV generated for different values of $v$. Panels (a)-(d) give the results for $v=400$ km s$^{-1}$, $800$ km s$^{-1}$, $1200$ km s$^{-1}$, and $1600$ km s$^{-1}$, respectively. Other parameters adopted in the calculation are $D_{0}=4.8$ $\times$ 10$^{24}$ cm$^{2}$ s$^{-1}$ and $d=0.4$ kpc.}
\end{figure*}

Finally, the $\gamma$-ray flux intensity maps at 50 TeV, generated for different values of $v$, are shown in Fig. \ref{fig:8}. Here, $v$ varies from a common value of 400 km s$^{-1}$ to the maximum observed value of 1600 km s$^{-1}$ \citep{hobbs2005statistical}. As presented in Fig. \ref{fig:8}, even with a sufficiently short distance, the PWN would still need to travel at an extraordinary velocity exceeding 1200 km s$^{-1}$ to prevent the cooling of UHE electrons, which is essential for generating the observed dumbbell-like structure. As the proper motion velocity increased to 1600 km s$^{-1}$, the morphology gradually becomes double-peaked.

\begin{figure*}[!h]
    \centering
    \includegraphics[width=0.95\textwidth]{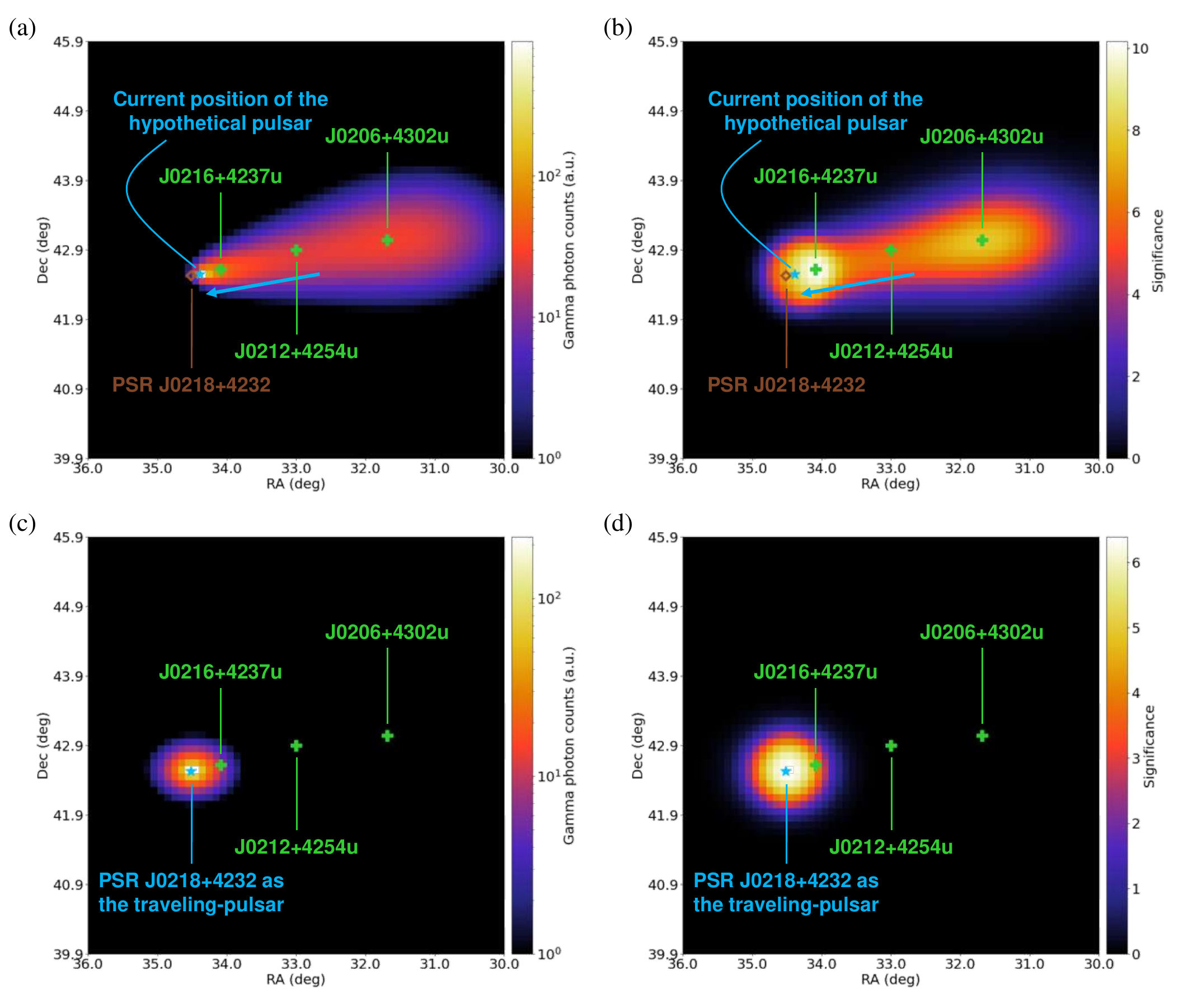}
    \caption{\label{fig:9} The predicted $\gamma$-photon counts maps and smoothed significance maps for energies exceeding 25 TeV in single traveling-PWN modeling attempt. Panel (a) shows the $\gamma$-photon counts map calculated with $D_{0}=4.8$ $\times$ 10$^{24}$ cm$^{2}$ s$^{-1}$, $d=0.4$ kpc, and $v=1600$ km s$^{-1}$. The blue arrow indicates the proper motion direction of the hypothetical pulsar proposed in our model. Panel (b) shows the predicted significance map seen from LHAASO. The result has been smoothed with the LHAASO PSF at energies above 25 TeV (1$\sigma=0\fdg2$). As a comparison, the LHAASO significance map on the dumbbell-like structure can be found at the left bottom of Fig. 10 in the LHAASO catalog \citep{cao2024first}. Panels (c)-(d) give the results of $\gamma$-photon counts map and smoothed significance map, where PSR J0218$+$4232 is considered as the traveling-pulsar.}
\end{figure*}

Based on our exploration about the impact of $D_{0}$, $d$, and $v$ on $\gamma$-ray morphology, we set $D_{0}=4.8$ $\times$ 10$^{24}$ cm$^{2}$ s$^{-1}$, $d=0.4$ kpc, and $v=1600$ km s$^{-1}$ in the single traveling-PWN scenario. The above parameters can be regarded as some typical values required for a dumbbell-like structure to be created by a single pulsar. Using these parameters, along with the previously adopted magnetic field and radiation field model, the $\gamma$-photon counts map for energies exceeding 25 TeV is presented in Fig. \ref{fig:9} (a). Here, we assumed that the PWN travel from ($\rm{RA}=30\fdg8000$, $\rm{Dec}=43\fdg1850$) to ($\rm{RA}=34\fdg4000$, $\rm{Dec}=42\fdg5575$) during the evolution. Some other major parameters used are presented in Table \ref{tab:3}. The parameter $\eta_{0}\dot{\xi}$ reflecting the total energy of injection electrons has been adjusted to a proper value that leads to the conformity between the calculated and observed $\gamma$-ray flux intensities from the three LHAASO sources. The parameter $\delta=0.33$ is a typical value predicted by the Kolmogorov's theory \citep{fang2022interpretation}. The evolutionary time $t_{a}$ is calculated from the selected values of $v$ and $d$. 

\begin{table}
  \centering
  \caption{\label{tab:3}Major parameters adopted in the single traveling-PWN modeling attempt}
  \begin{tabular}{cccc}
  \hline
    Parameter & Value & Parameter & Value \\ \hline
    $\eta_{0}\dot{\xi}$ (erg s$^{-1}$) & 1.2 $\times$ 10$^{28}$ & $\delta$ & 0.33 \\
    $D_{0}$ (cm$^{2}$ s$^{-1}$) & 4.8 $\times$ 10$^{24}$$^{\rm{a}}$ & $n$ & 2 \\
    $E_{c}$ (TeV) & 800 & $\Gamma$ & 1.8 \\
    $v$ (km s$^{-1}$) & 1600 & $\epsilon_{\rm{min}}$ & 0.1 \\
    $d$ (kpc) & 0.4 & $\epsilon_{\rm{max}}$ & 1.0 $\times$ 10$^{7}$ \\
    $t_{a}$ (kyr) & 15.6 & & \\
    \hline
  \end{tabular}
  \tablefoot{${\rm{a}}$. The value is approximately 7\% of that around Geminga.} 
\end{table}

By considering the diffuse Galactic $\gamma$-ray emission (GDE) background \citep{cao2024first} and applying a Gaussian smooth with $1\sigma=0\fdg2$, it is not difficult to predict the significance map seen from LHAASO based on the $\gamma$-photon counts map. The $1\sigma$ value used here is similar to that of the LHAASO point-spread function (PSF) at energies above 25 TeV \citep{cao2025data}. Figure \ref{fig:9} (b) presents the calculated significance map for energies exceeding 25 TeV. The results demonstrate that, under the assumption of an extremely low diffusion coefficient and an ultra-fast proper motion velocity of the central pulsar, the filament of a single traveling-PWN\textemdash produced through isotropic and homogeneous diffusion of electrons\textemdash can exhibit a double-peaked morphology resembling the dumbbell-like structure observed by LHAASO-KM2A. Comparing Figs. \ref{fig:9} (a) and (b), it is evident that the western $>$ 25 TeV blob (J0206$+$4302u) is formed by the diffused electrons from the central pulsar with higher spin-down luminosity at the early time, meanwhile, the eastern $>$ 25 TeV blob (J0216$+$4237u) is generated by the fresh PWN electrons recently injected and extended with the LHAASO PSF. However, the necessity of a diffusion coefficient below the Bohm limit under the assumption of isotropic diffusion renders this explanation seemingly implausible. The proper motion of a single pulsar may provide a more valid explanation for certain double-peaked extended structures with smaller scales (angular sizes of approximately 1$^{\circ}$) observed in lower energy bands ($<$ 10 TeV).

In Figs. \ref{fig:9} (a) and (b), except for the three LHAASO sources, the position of the only known energetic pulsar nearby\textemdash PSR J0218$+$4232 is also marked. Notably, PSR J0218$+$4232 lies in close proximity to the current position of the hypothetical traveling-pulsar proposed in our model. However, based on our calculation, it seems impossible that PSR J0218$+$4232 is the traveling-pulsar required. In the calculation, we adopted a diffusion coefficient of 6.9 $\times$ 10$^{25}$ cm$^{2}$ s$^{-1}$, consistent with the value observed around Geminga. This value falls within the one standard deviation range of the statistical results for the diffusion coefficients derived from 27 pulsars \citep{di2020evidences}. As reported in the Australia Telescope National Facility (ATNF) pulsar catalog \citep{manchester2005australia}, the spin-down luminosity, distance, and proper motion velocity of PSR J0218$+$4232 used in the calculation were 2.4 $\times$ 10$^{35}$ erg s$^{-1}$, 3.15 kpc, and 97.48 km s$^{-1}$, respectively. Some other common values such as the spectral index of 1.4 and the cutoff energy of 800 TeV for injection electrons were also adopted. Figure \ref{fig:9} (c)-(d) present the predicted $\gamma$-photon counts map and the smoothed significance map, respectively. Unsurprisingly, there is no filament produced by this pulsar and only the electrons recently injected can contribute to $\gamma$-ray emission, due to its long distance to Earth and slow proper motion velocity. Therefore, under the isotropic and homogeneous diffusion condition, the dumbbell-like structure observed by LHAASO-KM2A appears unlikely to be generated solely by PSR J0218+4232.

\subsection{Multiple traveling-PWNe modeling attempt}\label{subsec:mul}

In the previous subsection, we explored the scenario of a single traveling-PWN. By adopting extreme parameter values, such as a diffusion coefficient $D_{0}$ of 4.8 $\times$ 10$^{24}$ cm$^{2}$ s$^{-1}$ and a proper motion velocity $v$ of 1600 km s$^{-1}$, the predicted $\gamma$-ray morphology aligns well with the observations from LHAASO-KM2A. However, since this model requires an isotropic diffusion coefficient significantly lower than the Bohm limit, it is nearly unattainable from a physical perspective. If the diffusion coefficient becomes larger, the rapid diffusion of electrons will disrupt the double-peaked morphology, causing the dumbbell-like structure to vanish. Thus, we favor a traveling-PWN model involving multiple pulsars to account for the origin of the dumbbell-like structure.

\begin{figure*}[!h]
    \centering
    \includegraphics[width=0.95\textwidth]{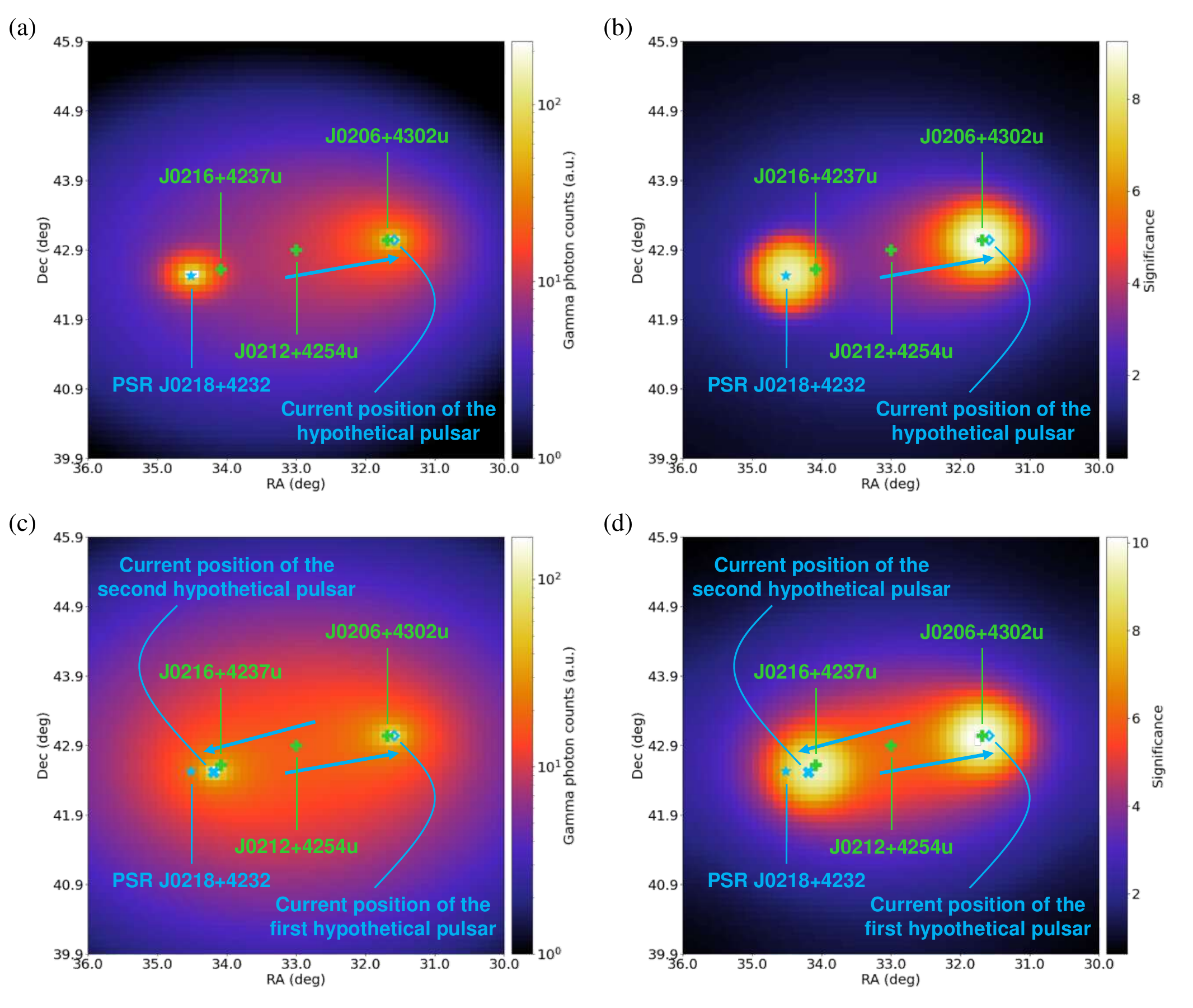}
    \caption{\label{fig:10}The predicted $\gamma$-photon counts maps and smoothed significance maps for energies exceeding 25 TeV in multiple traveling-PWNe modeling attempt. Panels (a)-(b) show the $\gamma$-photon counts map and significance map calculated with the model including PSR J0218$+$4232 and a hypothetical pulsar. The blue arrow indicates the proper motion direction of the hypothetical pulsar. Panels (c)-(d) show the $\gamma$-photon counts map and significance map calculated with the model including two hypothetical pulsars. Here, the $\gamma$-ray emission from PSR J0218$+$4232 has been excluded.}
\end{figure*}

In the analysis, we firstly assumed PSR J0218$+$4232 as one of the traveling-PWNe in our model, and expected its TeV $\gamma$-ray emission combined with the contribution from a hypothetical traveling-PWN possessing conventional characteristics can form a morphology resembling the dumbbell-like structure. The result is shown in Fig. \ref{fig:10} (a)-(b). The calculation was accomplished by adding an extra source term to Eq. (\ref{eq-fu}) and analytically solving the equation with the Green's function method. The diffusion coefficient was also set as the typical value of 6.9 $\times$ 10$^{25}$ cm$^{2}$ s$^{-1}$. All parameters associated with PSR J0218$+$4232 had the same values as adopted in Sect. \ref{subsec:sin}. The hypothetical PWN was assumed to travel from ($\rm{RA}=34\fdg3500$, $\rm{Dec}=42\fdg5763$) to ($\rm{RA}=31\fdg6000$, $\rm{Dec}=43\fdg0500$) during the evolution. The distance $d$ and the proper motion velocity $v$ for its central pulsar were set to 0.4 kpc and 800 km s$^{-1}$, respectively. All the parameters applied here no longer exhibit extreme values; instead, they fall within reasonable ranges consistent with those of a conventional pulsar. The proper motion velocity of the hypothetical pulsar is relatively fast but still acceptable. As presented in Fig. \ref{fig:10} (a)-(b), the TeV $\gamma$-photons emitted from a conventional pulsar appear unlikely to cover a sky region with an angular size of approximately 3$^{\circ}$, due to the cooling of UHE electrons \citep{zhang2021morphology}. Consequently, the traveling-PWN model including PSR J0218$+$4232 and a conventional pulsar is still far from explaining the origin of the dumbbell-like structure.

According to the current results, the only known pulsar PSR J0218$+$4232 can be ruled out under the isotropic and homogeneous diffusion environment. However, this pulsar is still an important potential counterpart of the dumbbell-like structure in some other theories, such as the asymmetric propagation of UHE electrons due to the turbulent magnetic fields around pulsars \citep{bao2024mirages, bao2024mirage}. As PSR J0218$+$4232 had been excluded, we explored the possibility of using two relatively fast-traveling PWNe located at short distances to account for the dumbbell-like structure. One of the PWNe was assumed to travel from ($\rm{RA}=34\fdg3500$, $\rm{Dec}=42\fdg5763$) to ($\rm{RA}=31\fdg6000$, $\rm{Dec}=43\fdg0500$) during the evolution, while the other one was assumed to travel from ($\rm{RA}=31\fdg4500$, $\rm{Dec}=43\fdg2364$) to ($\rm{RA}=34\fdg2000$, $\rm{Dec}=42\fdg5300$). The major parameters adopted for the two corresponding central pulsars are presented in Table \ref{tab:4}. The contribution of PSR J0218$+$4232 was not considered in this scenario. 
\begin{table}
  \centering
  \caption{\label{tab:4}Major parameters adopted in the double traveling-PWNe modeling attempt}
  \begin{tabular}{ccc}
  \hline
    Parameter & The first pulsar & The second pulsar \\ \hline
    $\eta_{0}\dot{\xi}$ (erg s$^{-1}$) & 3.7 $\times$ 10$^{28}$ & 2.0 $\times$ 10$^{28}$ \\
    $D_{0}$ (cm$^{2}$ s$^{-1}$) & 6.9 $\times$ 10$^{25}$ & 6.9 $\times$ 10$^{25}$ \\
    $E_{c}$ (TeV) & 500 & 400 \\
    $v$ (km s$^{-1}$) & 800 & 600 \\
    $d$ (kpc) & 0.4 & 0.3 \\
    $t_{a}$ (kyr) & 23.8 & 24.2 \\
    $\delta$ & 0.33 & 0.33 \\
    $n$ & 2 & 2 \\
    $\Gamma$ & 1.65 & 1.50 \\
    $\epsilon_{\rm{min}}$ & 0.1 & 0.1 \\
    $\epsilon_{\rm{max}}$ & 1.0 $\times$ 10$^{7}$ & 1.0 $\times$ 10$^{7}$ \\
    \hline
  \end{tabular}
\end{table}
Figure \ref{fig:10} (c)-(d) show the expected $\gamma$-photon counts map and the smoothed significance map, respectively. Compared with Fig. \ref{fig:10} (a)-(b), the $\gamma$-ray morphology in the scenario involving two hypothetical pulsars aligns more closely with the observational result of LHAASO-KM2A. For this scenario, both of the two $>$ 25 TeV blobs are generated by the fresh PWN electrons recently injected and extended with the LHAASO PSF while the entire extended structure is a combination of the diffused filaments left by the two PWNe. As all the parameters applied on the two pulsars possess common values, the double traveling-PWNe explanation is physically plausible. However, some strong constraints on the characteristics of the two pulsars are still necessary. As indicated in Table \ref{tab:4}, the conditions of short distance and fast traveling of the two pulsars are supposed to be satisfied simultaneously. Moreover, to form the dumbbell-like structure, the proper motions of the two pulsars would need to be opposite in direction, which significantly reduces the likelihood of this scenario. The probability of this explanation, as well as the more conventional interpretation involving the three LHAASO UHE sources corresponding to three PWNe\textemdash whose leptonic modeling was explored in Sect. \ref{subsec:lep}\textemdash will be further discussed in Sect. \ref{subsec:occ}.

\subsection{Probabilities of occurrence for different scenarios}\label{subsec:occ}

As previously discussed, to form a dumbbell-like structure as observed by LHAASO-KM2A with isotropic and homogeneous diffusion of PWN electrons, some constraints need to be set on the corresponding central pulsars. Furthermore, since more than one pulsar is required, the spatial coincidences among multiple pulsars are also essential. Thus, the probability of occurrence $P_{k}$ for each scenario can be calculated by
\begin{equation}\label{eq-pk}
    P_{k}=(P_{\rm{diff}}P_{d}P_{v})^{k}P_{c},
\end{equation}
where $k$ is the number of PWN. $P_{\rm{diff}}$, $P_{d}$, and $P_{v}$ describe the probability of the diffusion coefficient, distance, and proper motion velocity fall within their constraint ranges, respectively. $P_{c}$ denotes the probability of pulsars spatial coincidence by chance. 

\begin{table*}
  \centering
  \caption{\label{tab:5}Probabilities of occurrence for different scenarios}
  \begin{tabular}{ccccccc}
  \hline
    Scenario & $k$ & $P_{\rm{diff}}$ & $P_{d}$ & $P_{v}$ & $P_{c}$ & $P_{k}$ \\ \hline
    Single traveling-PWN & 1 & $\ll$ 0.32 & 0.0327 & 0.00013 & 1 & $\ll$ 0.00014\% \\
    Double traveling-PWNe & 2 & 1 & 0.0327 & 0.1628 & 0.0273 & $\sim$ 0.00008\% \\
    Triple PWNe & 3 & 1 & 0.4862 & 1 & 0.0319 & $\sim$ 0.37\% \\
    \hline
  \end{tabular}
\end{table*}

For the single traveling-PWN model, although the requirement for an isotropic diffusion coefficient below the Bohm limit renders the explanation seemingly improbable, the applied value finds observational support\textemdash potentially due to anisotropic diffusion\textemdash around HESS J1026$-$582 \citep{di2020evidences}. Consequently, we proceed to estimate the likelihood of this scenario. Since the diffusion coefficient around HESS J1026$-$582 is far out of the one standard deviation interval, the $P_{\rm{diff}}$ in this scenario is much lower than 0.32. The exact result cannot be provided here due to the scarcity of statistical samples. For the multiple traveling-PWNe model with two and three pulsars, a conventional diffusion coefficient like the typical value around Geminga can satisfy the requirement of generating the dumbbell-like structure. Thus, no specific constraints on diffusion coefficient are required, which will result in $P_{\rm{diff}}=1$. 

The radial density profile of Galactic pulsars can be described by a gamma function \citep{lorimer2006parkes}
\begin{equation}\label{eq-rhor}
    \rho(R)=A\left(\frac{R}{R_{\odot}}\right)^{F}{\rm{exp}}\left[-C\left(\frac{R-R_{\odot}}{R_{\odot}}\right)\right],
\end{equation}
where $R$ represents the galactocentric radius. $R_{\odot}=8.3$ kpc is the distance between the Sun and Galactic center. $A$, $F$, and $C$ are all fitting parameters depending on model. Along the direction perpendicular to the Galactic plane, the distribution of Galactic pulsars is assumed to satisfy a simple exponential function \citep{lorimer2006parkes}
\begin{equation}\label{eq-nz}
    N(z)=D{\rm{exp}}\left(-\frac{|z|}{H}\right).
\end{equation}
Here, $z$ denotes the height above the Galactic plane. $D$ and $H$ are also fitting parameters. Based on the fundamental geometrical relationship, it is not difficult to extract the Galactic pulsar distribution depending on $d$ via Eqs. (\ref{eq-rhor}) and (\ref{eq-nz}). For the scenarios involving one pulsar and two pulsars, we assumed $d<0.5$ kpc, while for the triple pulsars scenario, based on the considerations outlined in Sect. \ref{subsec:lep}, we implemented a more relaxed constraint of $d<6$ kpc. Thus, $P_{d}$ of the three scenarios with one pulsar, two pulsars, and three pulsars calculated by integrating the $d$ profile are 0.0327, 0.0327, and 0.4862, respectively. The fitting parameters used in the calculation follows the Model C fit in \citet{lorimer2006parkes}.

Based on the statistical result on proper motion velocities of 233 pulsars, the distribution of $v$ is well described by a Maxwellian distribution \citep{hobbs2005statistical} which can be written as
\begin{equation}\label{eq-mv}
    N(v)=\sqrt{\frac{2}{\pi}}\frac{v^{2}}{\sigma_{v}^{3}}{\rm{exp}}\left(-\frac{v^{2}}{2\sigma_{v}^{2}}\right),
\end{equation}
where the best-fit 1D root-mean-square $\sigma_{v}$ is 265 km s$^{-1}$. For the single traveling-PWN model, we assumed $v$ $>$ 1200 km s$^{-1}$. Thus, based on Eq. (\ref{eq-mv}), $P_{v}$ of this scenario is 0.00013. For the double traveling-PWNe model, we assumed $v$ $>$ 600 km s$^{-1}$ and obtained $P_{v}=0.1628$. In the conventional triple PWNe model, no constraint on $v$ is required. Thus, $P_{v}$ of this scenario is 1. 

At last, we considered the calculation of $P_{c}$ in multiple traveling-PWNe scenarios. The probability of two pulsars accidentally appear in the same sky region with position offset $r_{1}$ can be expressed as \citep{mattox1997identification}
\begin{equation}\label{eq-pc}
    P(r_{1})=1-{\rm{exp}}\left(-\frac{r_{1}^{2}}{r_{0}^{2}}\right),
\end{equation}
while
\begin{equation}\label{eq-dr0}
    r_{0}=[\pi\rho(\dot{\xi})]^{-1/2}.
\end{equation}
Here, $\rho(\dot{\xi})$ represents the number density of local pulsars. Consulting the method used in \citet{cao2024first}, we examined the ATNF catalog to check the number of pulsars within a 20$^{\circ}$ $\times$ 5$^{\circ}$ sky region centered on the exact location of J0212$+$4254u. Considering the pulsar needs to be energetic enough to be detected by LHAASO-KM2A, a cut at $\dot{\xi}/d^{2}$ $>$ 10$^{34}$ erg s$^{-1}$ kpc$^{-2}$ was applied during the examination, following \citet{cao2024first}. As a result, before the cut on $\dot{\xi}/d^{2}$, there were 4 pulsars found within the 20$^{\circ}$ $\times$ 5$^{\circ}$ sky region, while only 2 pulsars were left after the cut. The value extracted from the ATNF catalog is merely the number density of observed radio pulsars, which should be corrected to the actual number density of pulsars based on the product of beaming fraction and detection fraction \citep{johnston2020galactic}. Following \citet{johnston2020galactic}, we assumed a ratio of 0.09 between the number density of observed pulsars and the actual value. Thus, the $\rho(\dot{\xi})$ around the dumbbell-like structure is $\sim$ 765.36 rad$^{-2}$.

For the double traveling-PWNe scenario, based on the separation between the two hypothetical pulsars and $\rho(\dot{\xi})$ of 765.36 rad$^{-2}$, $P_{c}$ can be directly calculated by Eqs. (\ref{eq-pc}) and (\ref{eq-dr0}). The result is 0.9814. However, since the two pulsars need to travel in almost opposite directions, this value should be multiplied by the probability of proper motion directions coincidence by chance to acquire the actual $P_{c}$. According to our simulation, the appropriate angle between the proper motion directions of the two pulsars to form the dumbbell-like structure is $180\pm5^{\circ}$. Thus, the actual $P_{c}$ of this scenario is $0.9814/36=0.0273$. 

The problem will be more complicated in the triple PWNe scenario. We firstly calculated the probability of accidentally having two pulsars with position offset equals to that between J0206$+$4302u and J0216$+$4237u via Eqs. (\ref{eq-pc}) and (\ref{eq-dr0}). Afterward, the result was multiplied by the probability of having another pulsar coincidentally appeared in a 0\fdg2 radius circle region (the radius was assumed to be the 1$\sigma$ value of LHAASO PSF at energies exceeding 25 TeV) centered on the exact location of J0212$+$4254u, which can be calculated by
\begin{equation}\label{eq-op}
    P(S)=\rho(\dot{\xi})S{\rm{exp}}[-\rho(\dot{\xi})S],
\end{equation}
where $S$ is the area of the circle region in the unit rad$^{2}$. Consequently, we could obtain the $P_{c}$ of triple PWNe scenario as 0.0319.

As all the values of different probabilities have been estimated, based on Eq. (\ref{eq-pk}), $P_{k}$ of the three scenarios with one pulsar, two pulsars, and three pulsars are $\ll$ 0.00014\%, $\sim$ 0.00008\%, and $\sim$ 0.37\%, respectively. As a brief summary, Table \ref{tab:5} presents all the results obtained from our calculation. It is worth noting that, even the conventional triple PWNe model appears to be more complicated, the probability of occurrence for this scenario is still much higher than that of the other two scenarios. The underlying rationale is straightforward. To explain the dumbbell-like structure comprising three UHE sources through either the proper motion of one pulsar or two pulsars in the isotropic and homogeneous diffusion environment would necessitate imposing numerous stringent constraints on the characteristics of pulsars, which can significantly increase the complexity of model and reduce the probability of occurrence. Thus, it might not be a good choice to use the proper motion of a single pulsar, or even several pulsars, to explain such a giant extended structure in UHE band.

\section{Conclusion and Discussion}\label{sec:dis}

The first LHAASO catalog presents six enigmatic UHE sources solely detected by LHAASO-KM2A \citep{cao2024first}. All of them are located far from the Galactic plane and have no confirmed counterparts. Only two energetic pulsars PSR J0218$+$4232 and PSR J1740$+$1000 were found in proximity to J0216$+$4237u and J1740$+$0948u, respectively. A more intriguing thing is that J0206$+$4302u, J0212$+$4254u, and J0216$+$4237u are spatially linked and share a similar spectral shape. On the significance map, these three sources constitute a dumbbell-like structure. 

To reveal the physical mechanisms hiding behind the six LHAASO UHE sources, especially the intriguing dumbbell-like structure, we conducted a multiwavelength and multimessenger study based on the \textit{Fermi}-LAT, \textit{Swift}-XRT, \textit{Planck}, CfA $^{12}$CO survey, and IceCube neutrino datasets. As a result, only two $\gamma$-ray sources were observed by \textit{Fermi}-LAT, which could be identified as the two known pulsars PSR J0218$+$4232 and PSR J1740$+$1000. For the X-ray band study based on the \textit{Swift}-XRT dataset, no abundant efficient photons were detected. Within the \textit{Planck} and CfA $^{12}$CO survey datasets, only the source J0007$+$5659u appeared to have a thin MC nearby and the potential hadronic origin of this source has been considered.  CO surveys with enhanced angular resolution and LSR velocity coverage are anticipated. Based on the HESE data release of 12 years and 10 years of track-like events within the IceCube neutrino dataset, no neutrino event was found to be associated with the six sources. Moreover, with the RATAN-600 radio telescope, the three sources constituting the dumbbell-like structure were observed at 3 to 5 epochs in February and April, 2024. No positive detection was found, and some preliminary results on upper limits were given: J0206$+$4302u ($<$ 4 mJy), J0218$+$4232 ($<$ 6 mJy), and J0212$+$4254u ($<$ 6 mJy). Since the current multiwavelength and multimessenger study has not identified any confirmed counterparts, the origins of these six LHAASO sources remain uncertain.

Based on the results of the multiwavelength and multimessenger study, we performed leptonic and hadronic modeling research on the six LHAASO sources. In the leptonic scenario, we assumed the six sources originated from Geminga-like PWNe whose injection electron spectra satisfy the simple power-law function with exponential cutoff. Our leptonic model indicated very high exponential cutoff energies $>$ 100 TeV for all six sources, which was common for pulsar injection \citep{de2022potential}. In the hadronic scenario, we focus on the p-p interaction origin of J0007$+$5659u, where the CRs were assumed to interact with the MC spatially close to J0007$+$5659u. The results of modeling research were physically reasonable. Our model suggests a notable observational implication that, X-ray band observations with sensitive telescopes such as \textit{XMM-Newton} could provide helpful diagnostic information to determine whether J0007$+$5659u belonging to leptonic or hadronic origins. Based on the best-fit parameters in hadronic model, we also calculated the expected neutrino flux from J0007$+$5659u. The result was significantly below the sensitivity limit of IceCube-Gen2 after 10 years of observation. However, there still remains the possibility of detecting associated neutrino fluxes with some other proposed next generation neutrino observatories like TRIDENT \citep{ye2022proposal} and NEON \citep{zhang2024proposed}.

For the enigmatic dumbbell-like structure we are most interested in, which is also recognized as three point-like sources in the LHAASO catalog \citep{cao2024first}, a traveling-PWN model under the isotropic and homogeneous diffusion condition was studied. We firstly explored the scenario including only one traveling-PWN. The impact of diffusion coefficient, distance, and proper motion velocity on the final morphology of $\gamma$-ray emission was discussed. As a result, we found that to generate a dumbbell-like structure with a single pulsar, the diffusion coefficient had to be much lower than the Bohm limit, while the proper motion velocity needed to be approximately 1600 km s$^{-1}$. Furthermore, the distance within an appropriate range was also essential. Such extreme precondition denied the possibility of the only known pulsar PSR J0218$+$4232, which possesses long distance to Earth and slow proper motion velocity, to be the assumed traveling-PWN in our model. With all the required parameters being applied and adopting a typical PWN spin-down evolution history following \citet{ding2021implications}, we demonstrated that the filament left by the proper motion of a single traveling-PWN could exhibit a double-peaked morphology and generate a significance map aligning with the result of LHAASO-KM2A. However, based on the present theories, it is very difficult for electrons to diffuse much slower than Bohm diffusion under the premise of isotropic diffusion. Thus, the single traveling-PWN model seems invalid in explaining the origin of such a giant extended structure, which agrees with the results presented in \citet{zhang2021morphology}. As the isotropic and homogeneous diffusion condition is not a constraint on our model, we plan to explore the single traveling-PWN scenario under the anisotropic diffusion environment in the future. 

We also studied the model including two traveling-PWNe. PSR J0218$+$4232 was involved in the study at first, and was then ruled out, since the expected TeV emission of this pulsar was far from explaining even the eastern part of the dumbbell-like structure. However, if this pulsar is surrounded by a turbulent magnetic field environment, the "mirage source" formed by the asymmetric propagation of electrons \citep{bao2024mirages, bao2024mirage} can lead to a $\gamma$-ray morphology resembling the KM2A observation. Thus, this pulsar is still an important potential counterpart of the dumbbell-like structure. Assuming two hypothetical pulsars with relatively fast proper motion velocities and short distances, the double traveling-PWNe model could also generate an extended structure aligning well with the dumbbell-like structure observed by LHAASO-KM2A. Since no extreme parameters were applied, we considered this explanation was acceptable. 

The probabilities of occurrence for the two scenarios previously mentioned, along with that of the most conventional explanation with the three LHAASO UHE sources corresponding to three PWNe, as discussed in Sect. \ref{subsec:lep}, were also estimated. As a result, the probability of occurrence for the triple PWNe scenario is more than three orders of magnitude higher than those of the scenarios requiring less PWNe. It is straightforward that the stringent constraints on the characteristics of the corresponding central pulsars can significantly increase the complexity of model and reduce the probability of occurrence, even for the double traveling-PWNe scenario. Thus, it appears unreasonable to use the proper motion of a single pulsar or several pulsars to account for a giant extended structure in UHE band. As the conventional triple PWNe explanation has the highest probability of occurrence, we expect the radio band observation with the Five-hundred-meter Aperture Spherical radio Telescope (FAST) and X-ray band observation with \textit{XMM-Newton} on the exact locations of J0206$+$4302u, J0212$+$4254u, and J0216$+$4237u can be helpful in revealing the origin of the dumbbell-like structure. Furthermore, the up-to-date experimental data provided by the LHAASO Collaboration in the future is also expected.

\section*{Data availability}\label{sec:data}

The LHAASO spectral data (WCDA upper limits and KM2A fitting SEDs with error bars) presented in Figs. \ref{fig:4} and \ref{fig:5} are available in electronic form at the CDS via \url{http://cdsweb.u-strasbg.fr/cgi-bin/qcat?J/A+A/}.

\begin{acknowledgements}

We thank the anonymous referee for the constructive comments and suggestions. We thank Ruoyu Liu, Shaoqiang Xi, Sujie Lin, Renfeng Xu, Wenjun Huang, and Sheng Tang for the helpful discussion. We thank Timur Mufakharov and Yulia Sotnikova for the RATAN-600 observation. This work was supported by the National Natural Science Foundation of China (NSFC) Grants No. 12261141691 and No. 12005313. We thank the support from Fundamental Research Funds for the Central Universities, Sun Yat-sen University, No. 24qnpy123.

\end{acknowledgements}

\begin{appendix} 

\onecolumn
\section{Multiwavelength and multimessenger data analysis}\label{sec:ana}

\subsection{\textit{Fermi}-LAT data analysis}\label{subsec:fermi ana}

The \textit{Fermi}-LAT data is publicly available on the \textit{Fermi}-LAT website and can be analyzed with the Fermitools developed by the \textit{Fermi}-LAT Collaboration.\footnote{\url{https://fermi.gsfc.nasa.gov/cgi-bin/ssc/LAT/LATDataQuery.cgi}} After acquiring the \textit{Fermi}-LAT data from 10 years of observation, we used the gtselect function to filter the $\gamma$-photons outside the ROI centered on the source to be studied. The ROI radius was chosen as 1\fdg 5 for the $\gamma$-ray flux calculations and 5$^{\circ}$ for the TS map calculations. The $\gamma$-photons were also selected by their energies. Only the $\gamma$-photons possessing energies that fall within 300 MeV to 300 GeV were involved in the analysis. The background contribution was subtracted by including the Galactic diffuse model (gll\_iem\_v07) and the isotropic background (iso\_P8R3\_SOURCE\_V3\_v1), along with the 4FGL catalog.

The $\gamma$-ray fluxes of the six LHAASO UHE sources were obtained by binned maximum likelihood analysis. The recommended power-law spectral models \citep{tam2020multiwavelength, cao2024first} for each source were added to the input model files generated by LATSourceModel Python package, as shown below
\begin{equation}\label{eq-pl}
	\frac{dN}{dE_{\gamma}}=N_{0}\left(\frac{E_{\gamma}}{E_{0}}\right)^{-\alpha_{0}}.
\end{equation}
Here, $E_{\gamma}$ denotes the energy of $\gamma$-photon. The normalization $N_{0}$ and spectral index $\alpha_{0}$ were allowed to change, while the reference energy $E_{0}$ was fixed. After running the maximum likelihood analysis function gtlike, the $\gamma$-ray fluxes in the MeV to GeV band could be acquired. Due to the scarcity of observational data, the results had error bars larger than the truth values. Thus, we also calculated the upper limits of SED with UpperLimits, a Python package of Fermitools, to set constraints on our leptonic and hadronic models.

After the binned maximum likelihood analysis, the TS maps of the six LHAASO sources were calculated by running the gttsmap function. The input model files adopted in this procedure were the fitted model files as the outputs of the maximum likelihood analysis with some modifications. All free parameters in these model files were fixed, and for calculating TS maps containing 4FGL sources in the vicinity of LHAASO sources with a position offset less than 0\fdg5, the corresponding 4FGL source models were removed.

\subsection{\textit{Swift}-XRT data analysis}\label{subsec:swift ana}

We checked the \textit{Swift}-XRT dataset and only found that the source J0206$+$4302u had X-ray data within the 0\fdg3 ROI centered on it. No accurate energy spectrum could be extracted from these data due to the scarcity of efficient photons. However, the photon flux upper limit could be calculated via the X-ray image analysis package XIMAGE integrated into the HEASOFT software, whose value was 1.55 $\times$ 10$^{-2}$ s$^{-1}$.\footnote{The software can be downloaded from \url{https://heasarc.gsfc.nasa.gov/lheasoft/download.html}.} Based on the photon flux upper limit and the photo-electric absorption model describing the photon flux attenuation during the transmitting process, the upper limit of SED can be obtained. Here, the effect of photo-electric absorption was considered to satisfy an exponential damping function as shown below
\begin{equation}\label{eq-ab}
	M(E_{\rm{X}})={\rm{exp}}[-N_{\rm{H}}\sigma_{\rm{pe}}(E_{\rm{X}})],
\end{equation}
where $E_{\rm{X}}$ denotes the energy of X-ray photon. $N_{\rm{H}}$ is the column density of hydrogen. It was set to the calculated Galactic value in the sky region around J0206$+$4302u of 9.84 $\times$ 10$^{20}$ cm$^{-2}$.\footnote{The Galactic value of hydrogen column density can be calculated online at \url{https://www.swift.ac.uk/analysis/nhtot/}.} $\sigma_{\rm{pe}}(E_{\rm{X}})$ represents the photo-electric cross section.

Similar to the analysis of \textit{Fermi}-LAT data, a power-law model was used to describe the original energy spectrum of X-ray without considering photo-electric absorption. Due to the limited number of efficient photons, an accurate spectral shape could not be determined. Therefore, the spectral index was assumed to be 1.8 \citep{giommi2021x}. By integrating the product of the power-law spectrum and Eq. (\ref{eq-ab}), and comparing it with the previously calculated photon flux upper limit, the upper limit of the normalization was derived. Afterward, the upper limit of SED could be obtained with this result. Since there were no observational data, we applied the SED upper limit of J0206$+$4302u to all six LHAASO sources, with the assumption that the other five sources had X-ray photon fluxes similar to or lower than that of J0206$+$4302u.

\subsection{CfA $^{12}$CO survey data analysis}\label{subsec:co ana}

The recent 115 GHz $^{12}$CO line data acquired from the CfA 1.2 m telescope and its twin instrument in Chile, combined with the data previously presented in 2001 \citep{dame2001milky}, were adopted in the analysis. The moment masking technique was applied to these data to suppress noise \citep{dame2011optimization}. This whole dataset contained 677 $\times$ 1441 $\times$ 146 pixels, while some of them were blank pixels without detected emission. In the analysis procedure, these blanks were filled with the zero value. Along the LSR velocity axis, the dataset was divided into 146 $^{12}$CO brightness temperature maps with each map comprising 677 $\times$ 1441 pixels, covering the entire northern sky. By checking these maps, we can explore whether there are MCs spatially linked to the LHAASO sources. For J0007$+$5659u, a thin MC with a position offset of $\sim$ 0\fdg5 was found. The LSR velocity coverage of this MC is approximately -4.9 km s$^{-1}$ to 1.6 km s$^{-1}$. For the other five LHAASO sources, within the LSR velocity limitation of $\pm47.1$ km s$^{-1}$, no significant MC nearby was observed. The LSR velocity integration was conducted in a discrete way, where pixels possessing the same spatial coordinates but different LSR velocities were multiplied by the velocity resolution of 0.65 km s$^{-1}$ and added up.

\begin{figure}[h!]
    \centering
    \includegraphics[width=0.5\textwidth]{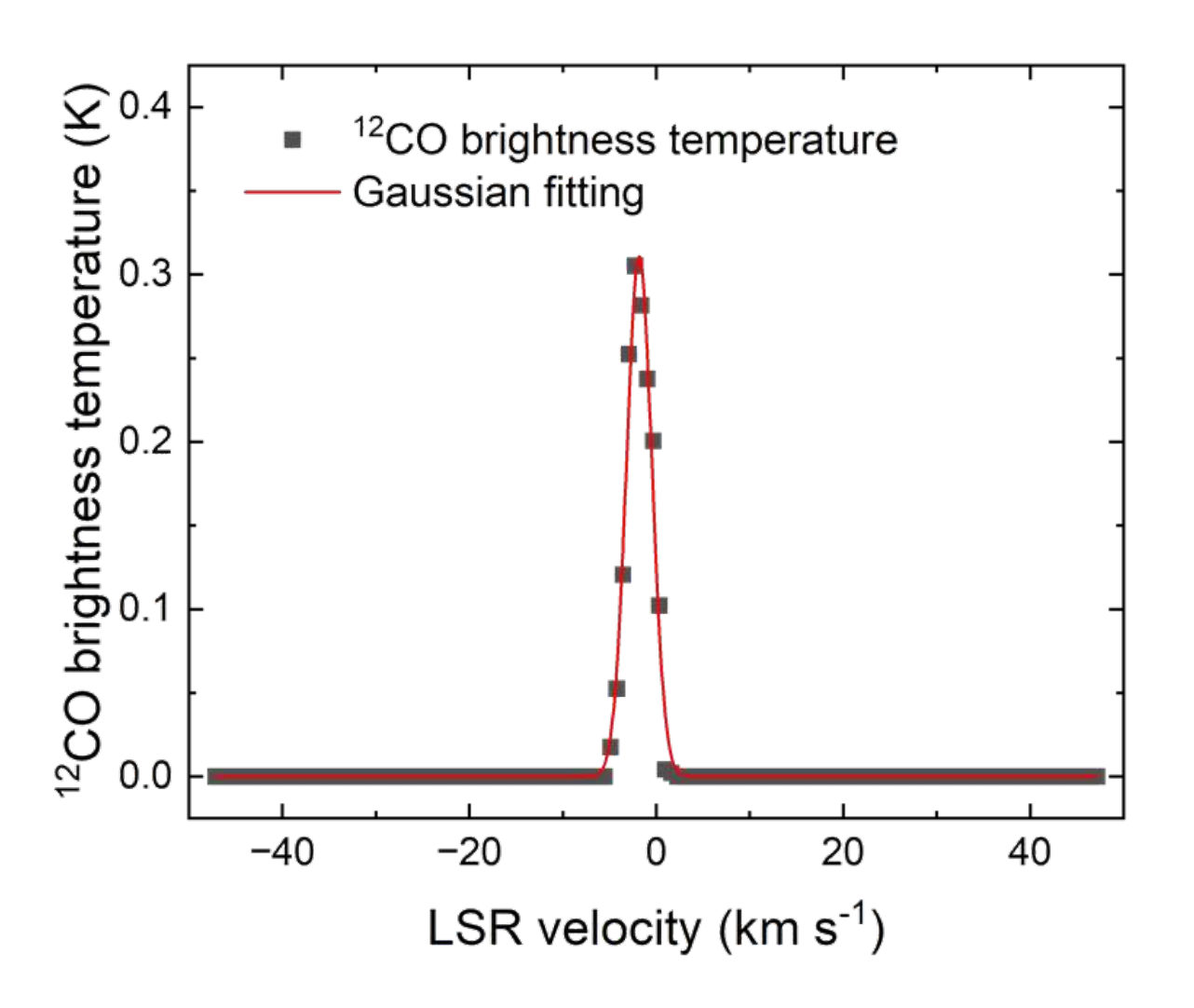}
    \caption{\label{fig:A1}The $^{12}$CO spectrum of the MC proposed to be associated with J0007$+$5659u. The spectral data appears very smooth due to the application of moment masking. The red line shows the result of Gaussian fitting.}
\end{figure}

Figure \ref{fig:A1} shows the $^{12}$CO spectrum of the MC proposed to be associated with J0007$+$5659u. The x-axis represents the LSR velocity, and the y-axis stands for the mean $^{12}$CO brightness temperature over the sky region covered by the MC. The spectrum does not show significant line broadening, which can be seen in the case of MC interacting with SNR \citep{slane2015supernova, sano2018molecular, enokiya2023discovery}.

\FloatBarrier
\twocolumn

\onecolumn
\section{Corner plots associated with MCMC fittings}\label{sec:cor}

For a fitting based on the MCMC method, it is conventional to use the corner plot showing the posterior PDF of each fitting parameter to reflect the convergence of the fitting and the correlation within parameters. If the MCMC fitting has reached convergence and the joint distribution function between parameters satisfies the normal distribution, the 2-D scatter diagram in the corner plot is supposed to exhibit a stable shape, approximately a circle or ellipse. 

\begin{figure*}[h!]
    \centering
    \includegraphics[width=1.0\textwidth]{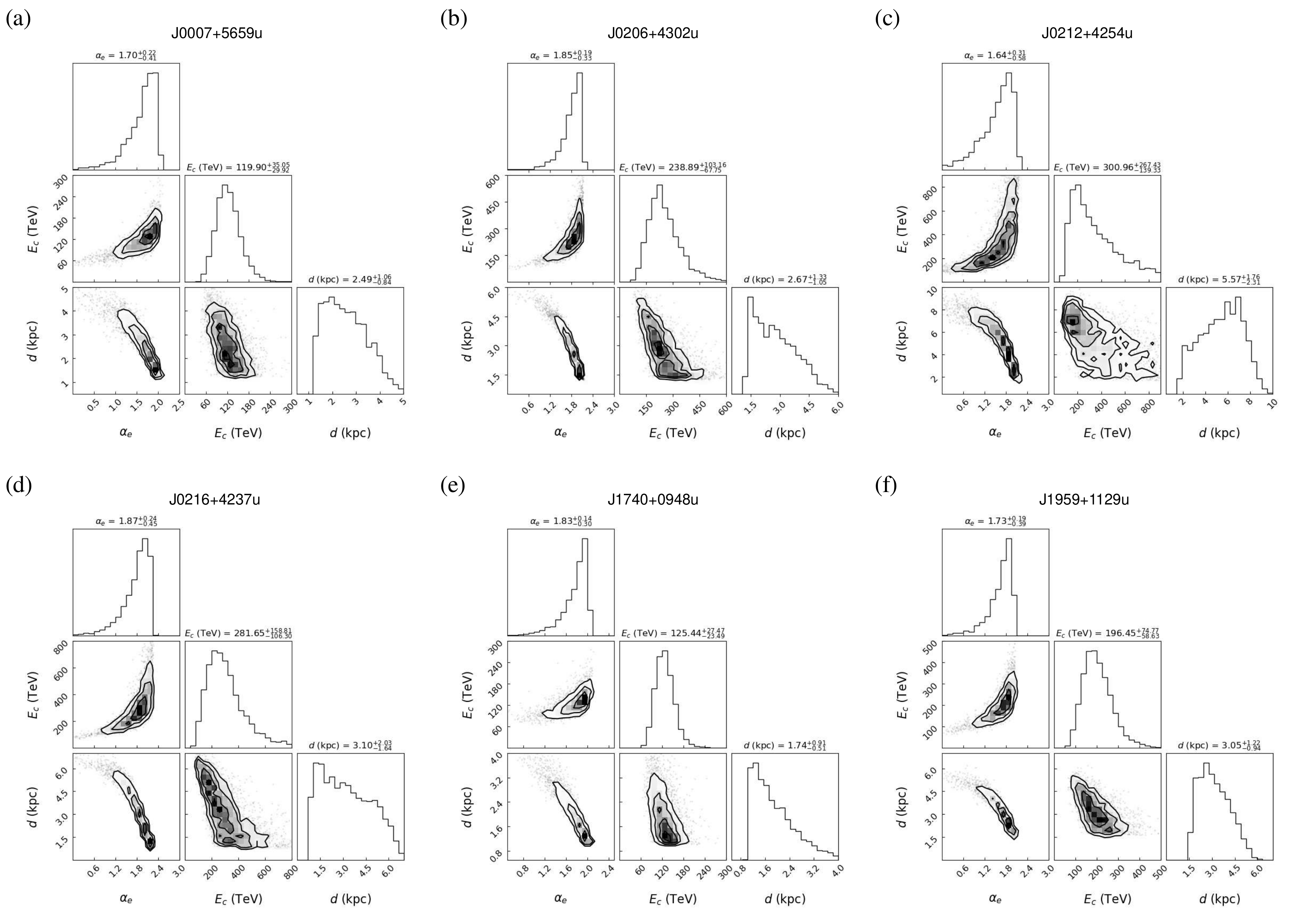}
    \caption{\label{fig:B1}Corner plots showing the posterior PDF of $\alpha_{e}$, $E_{c}$, and $d$ for the leptonic modeling research. Panels (a)-(f) present the results of J0007$+$5659u, J0206$+$4302u, J0212$+$4254u, J0216$+$4237u, J1740$+$0948u, and J1959$+$1129u, respectively.}
\end{figure*}

\begin{figure}[h!]
    \centering
    \includegraphics[width=0.33\textwidth]{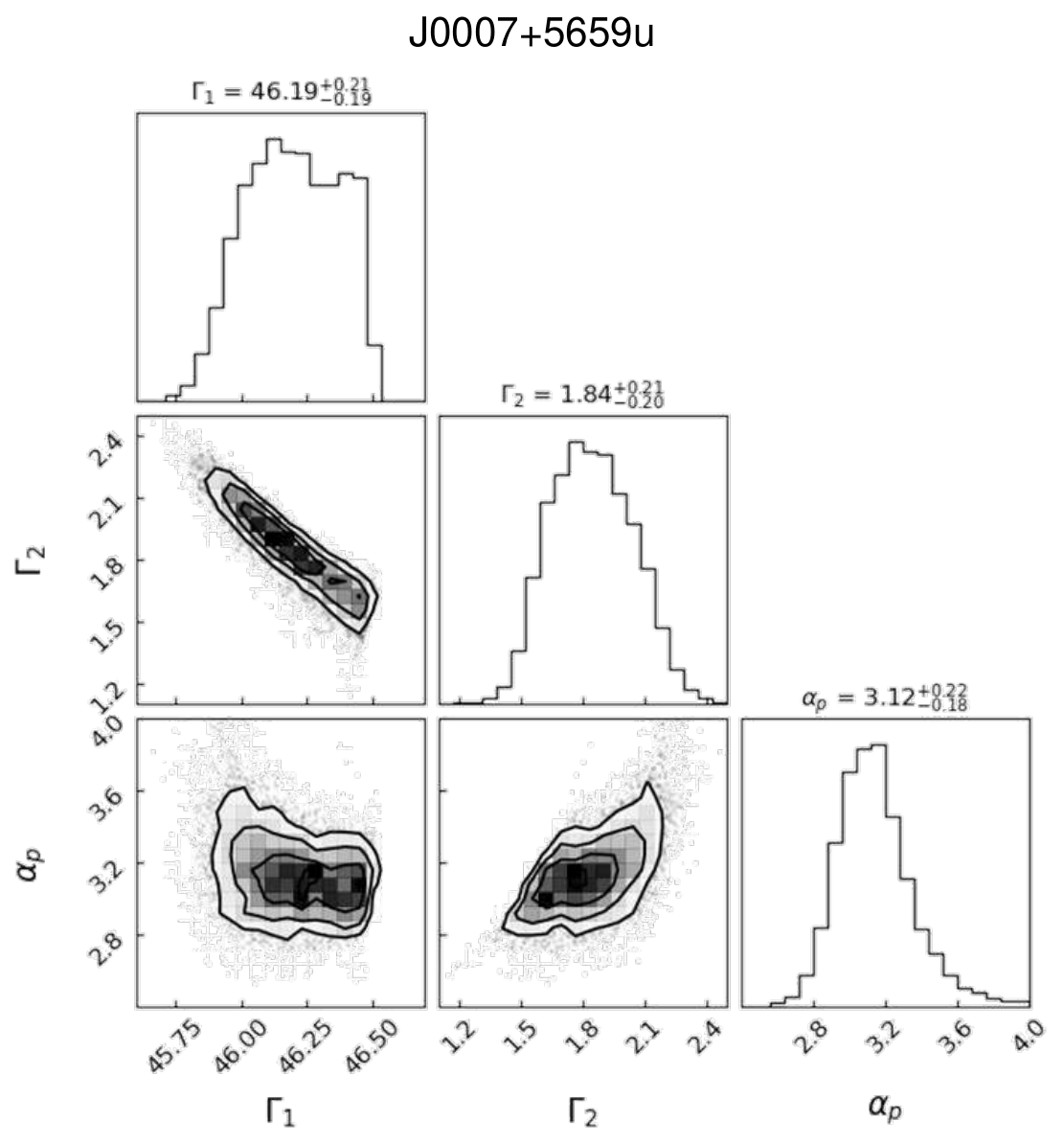}
    \caption{\label{fig:B2}Corner plot showing the posterior PDF of $\Gamma_{1}$, $\Gamma_{2}$, and $\alpha_{p}$ for the hadronic modeling research on J0007$+$5659u.}
\end{figure}

In Figs. \ref{fig:B1} and \ref{fig:B2}, we present the corner plots associated with the MCMC fittings in our leptonic and hadronic modeling research. As shown by these figures, all the fitting are converged, while the distributions of some parameters seem to be more complicated than the expectations, due to the scarcity of constraints from experimental data. For our modeling research, there is still some room to improve by exploring better-suited fitting parameters in the models and obtaining more observational data in other energy bands.

\FloatBarrier
\twocolumn

\onecolumn
\section{Expected neutrino flux calculation}\label{sec:neu}

The production rate and spectrum of neutrinos from p-p interaction depend on the type of meson decay and the flavor of generated neutrinos. For muonic neutrinos, the neutrino flux per unit energy can be calculated via the equation below \citep{kelner2006energy, de2022exploring}
\begin{equation}\label{eq-nef}
	\Phi_{\nu}(E_{\nu})=\frac{c\overline{n}({\rm{H_{2}}})}{4\pi d^{2}}\int\frac{1}{x}\sigma_{\rm{pp}}(E_{\nu}/x)J_{p}(E_{\nu}/x)F_{\nu}(x, E_{\nu}/x)dx,
\end{equation}
where $x=E_{\nu}/E_{p}$ is the variable of integration. $\overline{n}({\rm{H_{2}}})$ stands for the mean particle number density of $10.50\pm3.08$ cm$^{-3}$ as presented in Sect. \ref{subsec:co}. $E_{\nu}$ denotes the energy of produced neutrino. $\sigma_{\rm{pp}}(E_{\nu}/x)$ is the total inelastic cross section in p-p interaction, as presented by Eq. (\ref{eq-cs}). $J_{p}(E_{\nu}/x)$ represents the energy distribution of protons as shown in Eq. (\ref{eq-plh}). $F_{\nu}(x, E_{\nu}/x)$ describes the energy spectra of neutrinos produced through meson decay. For muonic neutrinos generated from the decays of charged pions and secondary muons respectively, $F_{\nu}(x, E_{\nu}/x)$ have different forms \citep{kelner2006energy}. The lower and upper limits of integration are also determined by the type of meson decay. In the pion decay scenario, since the $F_{\nu}(x, E_{\nu}/x)$ has a sharp cutoff at $x=0.427$ \citep{kelner2006energy}, the upper limit of integration can be simply set at 0.427. Meanwhile, in the muon decay scenario, the integration should be performed with $x$ varying from 0 to 1.

\begin{figure*}[h!]
    \centering
    \includegraphics[width=0.5\textwidth]{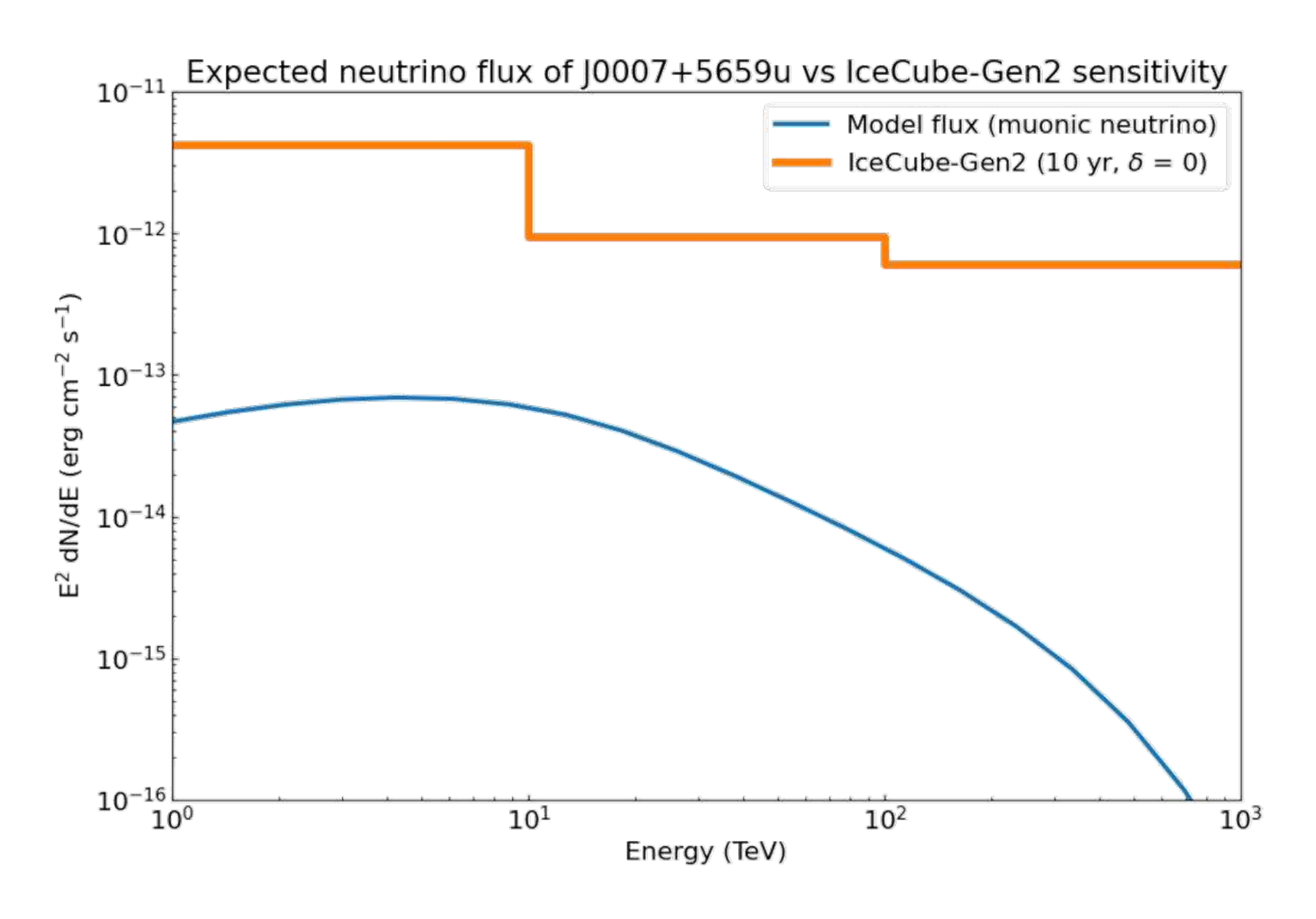}
    \caption{\label{fig:C1}The expected muonic neutrino flux of J0007$+$5659u generated by p-p interaction, comparing with the 10 years of observation sensitivity of IceCube-Gen2 \citep{aartsen2019neutrino}.}
\end{figure*}

Based on Eqs. (\ref{eq-plh}), (\ref{eq-cs}), and (\ref{eq-nef}), we calculated the expected muonic neutrino flux from J0007$+$5659u with the best-fit parameters in our hadronic model as presented in Sect. \ref{subsec:had} of the main text and compared with the 10 years of observation sensitivity of IceCube-Gen2 \citep{aartsen2019neutrino}. Figure \ref{fig:C1} shows the result. The fluxes of other flavor neutrinos are expected to be the same as that of muonic neutrinos, as $\nu_{e}:\nu_{\mu}:\nu_{tau} = 1:1:1$ at Earth. As shown in Fig. \ref{fig:C1}, the calculated neutrino SED of J0007$+$5659u is significantly lower than the sensitivity limit of IceCube-Gen2 with energy ranging from 1 TeV to 1 PeV.

Moreover, the neutrino fluxes from $\gamma$-p interaction \citep{hummer2010simplified} for the six LHAASO UHE sources were also considered. We studied the interactions between the $\gamma$-photons emitted from the six sources and Galactic CR protons. Due to the weak $\gamma$-ray flux intensities of the six sources, the expected neutrino fluxes in $\gamma$-p interaction calculated with the Galactic CR proton spectrum \citep{dampe2019measurement} are much lower than those in p-p interaction.

\end{appendix}


\begin{thebibliography}{80}
\expandafter\ifx\csname natexlab\endcsname\relax\def\natexlab#1{#1}\fi

\bibitem[{{Aartsen} {et~al.}(2013){Aartsen}, {Abbasi}, {Abdou}, {Ackermann}, {Adams}, {Aguilar}, {Ahlers}, {Altmann}, {Auffenberg}, {Bai}, {Baker}, {Barwick}, {Baum}, {Bay}, {Beatty}, {Bechet}, {Becker Tjus}, {Becker}, {Bell}, {Benabderrahmane}, {BenZvi}, {Berdermann}, {Berghaus}, {Berley}, {Bernardini}, {Bernhard}, {Bertrand}, {Besson}, {Binder}, {Bindig}, {Bissok}, {Blaufuss}, {Blumenthal}, {Boersma}, {Bohaichuk}, {Bohm}, {Bose}, {B{\"o}ser}, {Botner}, {Brayeur}, {Bretz}, {Brown}, {Bruijn}, {Brunner}, {Carson}, {Casey}, {Casier}, {Chirkin}, {Christov}, {Christy}, {Clark}, {Clevermann}, {Coenders}, {Cohen}, {Cowen}, {Cruz Silva}, {Danninger}, {Daughhetee}, {Davis}, {De Clercq}, {De Ridder}, {Desiati}, {de With}, {DeYoung}, {D{\'\i}az-V{\'e}lez}, {Dunkman}, {Eagan}, {Eberhardt}, {Eisch}, {Ellsworth}, {Euler}, {Evenson}, {Fadiran}, {Fazely}, {Fedynitch}, {Feintzeig}, {Feusels}, {Filimonov}, {Finley}, {Fischer-Wasels}, {Flis}, {Franckowiak}, {Franke}, {Frantzen}, {Fuchs}, {Gaisser}, {Gallagher}, {Gerhardt},
  {Gladstone}, {Gl{\"u}senkamp}, {Goldschmidt}, {Golup}, {Gonzalez}, {Goodman}, {G{\'o}ra}, {Grant}, {Gro{\ss}}, {Gurtner}, {Ha}, {Haj Ismail}, {Hallen}, {Hallgren}, {Halzen}, {Hanson}, {Heereman}, {Heinen}, {Helbing}, {Hellauer}, {Hickford}, {Hill}, {Hoffman}, {Hoffmann}, {Homeier}, {Hoshina}, {Huelsnitz}, {Hulth}, {Hultqvist}, {Hussain}, {Ishihara}, {Jacobi}, {Jacobsen}, {Jagielski}, {Japaridze}, {Jero}, {Jlelati}, {Kaminsky}, {Kappes}, {Karg}, {Karle}, {Kelley}, {Kiryluk}, {Kislat}, {Kl{\"a}s}, {Klein}, {K{\"o}hne}, {Kohnen}, {Kolanoski}, {K{\"o}pke}, {Kopper}, {Kopper}, {Koskinen}, {Kowalski}, {Krasberg}, {Krings}, {Kroll}, {Kunnen}, {Kurahashi}, {Kuwabara}, {Labare}, {Landsman}, {Larson}, {Lesiak-Bzdak}, {Leuermann}, {Leute}, {L{\"u}nemann}, {Madsen}, {Maruyama}, {Mase}, {Matis}, {McNally}, {Meagher}, {Merck}, {M{\'e}sz{\'a}ros}, {Meures}, {Miarecki}, {Middell}, {Milke}, {Miller}, {Mohrmann}, {Montaruli}, {Morse}, {Nahnhauer}, {Naumann}, {Niederhausen}, {Nowicki}, {Nygren}, {Obertacke}, {Odrowski},
  {Olivas}, {Olivo}, {O'Murchadha}, {Paul}, {Pepper}, {P{\'e}rez de los Heros}, {Pfendner}, {Pieloth}, {Pinat}, {Pirk}, {Posselt}, {Price}, {Przybylski}, {R{\"a}del}, {Rameez}, {Rawlins}, {Redl}, {Reimann}, {Resconi}, {Rhode}, {Ribordy}, \& {Richman}}]{aartsen2013first}
{Aartsen}, M.~G., {Abbasi}, R., {Abdou}, Y., {et~al.} 2013, \prl, 111, 021103

\bibitem[{Aartsen {et~al.}(2017)Aartsen, Ackermann, Adams, Aguilar, Ahlers, Ahrens, Altmann, Andeen, Anderson, Ansseau, {et~al.}}]{aartsen2017icecube}
Aartsen, M.~G., Ackermann, M., Adams, J., {et~al.} 2017, JInst, 12, P03012

\bibitem[{{Aartsen} {et~al.}(2020){Aartsen}, {Ackermann}, {Adams}, {Aguilar}, {Ahlers}, {Ahrens}, {Alispach}, {Andeen}, {Anderson}, {Ansseau}, {Anton}, {Arg{\"u}elles}, {Auffenberg}, {Axani}, {Backes}, {Bagherpour}, {Bai}, {Balagopal}, {Barbano}, {Barwick}, {Bastian}, {Baum}, {Baur}, {Bay}, {Beatty}, {Becker}, {Becker Tjus}, {BenZvi}, {Berley}, {Bernardini}, {Besson}, {Binder}, {Bindig}, {Blaufuss}, {Blot}, {Bohm}, {B{\"o}rner}, {B{\"o}ser}, {Botner}, {B{\"o}ttcher}, {Bourbeau}, {Bourbeau}, {Bradascio}, {Braun}, {Bron}, {Brostean-Kaiser}, {Burgman}, {Buscher}, {Busse}, {Carver}, {Chen}, {Cheung}, {Chirkin}, {Choi}, {Clark}, {Classen}, {Coleman}, {Collin}, {Conrad}, {Coppin}, {Correa}, {Cowen}, {Cross}, {Dave}, {De Clercq}, {DeLaunay}, {Dembinski}, {Deoskar}, {De Ridder}, {Desiati}, {de Vries}, {de Wasseige}, {de With}, {DeYoung}, {Diaz}, {D{\'\i}az-V{\'e}lez}, {Dujmovic}, {Dunkman}, {Dvorak}, {Eberhardt}, {Ehrhardt}, {Eller}, {Engel}, {Evenson}, {Fahey}, {Fazely}, {Felde}, {Filimonov}, {Finley}, {Fox},
  {Franckowiak}, {Friedman}, {Fritz}, {Gaisser}, {Gallagher}, {Ganster}, {Garrappa}, {Gerhardt}, {Ghorbani}, {Glauch}, {Gl{\"u}senkamp}, {Goldschmidt}, {Gonzalez}, {Grant}, {Griffith}, {Griswold}, {G{\"u}nder}, {G{\"u}nd{\"u}z}, {Haack}, {Hallgren}, {Halliday}, {Halve}, {Halzen}, {Hanson}, {Haungs}, {Hebecker}, {Heereman}, {Heix}, {Helbing}, {Hellauer}, {Henningsen}, {Hickford}, {Hignight}, {Hill}, {Hoffman}, {Hoffmann}, {Hoinka}, {Hokanson-Fasig}, {Hoshina}, {Huang}, {Huber}, {Huber}, {Hultqvist}, {H{\"u}nnefeld}, {Hussain}, {In}, {Iovine}, {Ishihara}, {Japaridze}, {Jeong}, {Jero}, {Jones}, {Jonske}, {Joppe}, {Kang}, {Kang}, {Kappes}, {Kappesser}, {Karg}, {Karl}, {Karle}, {Katz}, {Kauer}, {Kelley}, {Kheirandish}, {Kim}, {Kintscher}, {Kiryluk}, {Kittler}, {Klein}, {Koirala}, {Kolanoski}, {K{\"o}pke}, {Kopper}, {Kopper}, {Koskinen}, {Kowalski}, {Krings}, {Kr{\"u}ckl}, {Kulacz}, {Kurahashi}, {Kyriacou}, {Labare}, {Lanfranchi}, {Larson}, {Lauber}, {Lazar}, {Leonard}, {Leszczy{\'n}ska}, {Leuermann}, {Liu},
  {Lohfink}, {Lozano Mariscal}, {Lu}, {Lucarelli}, {L{\"u}nemann}, {Luszczak}, {Lyu}, {Ma}, {Madsen}, {Maggi}, {Mahn}, {Makino}, {Mallik}, {Mallot}, {Mancina}, {Mari{\c{s}}}, {Maruyama}, {Mase}, \& {Matis}}]{IceCube:10yeardata}
{Aartsen}, M.~G., {Ackermann}, M., {Adams}, J., {et~al.} 2020, \prl, 124, 051103

\bibitem[{{Abbasi} {et~al.}(2023){Abbasi}, {Ackermann}, {Adams}, {Aggarwal}, {Aguilar}, {Ahlers}, {Alameddine}, {Alves}, {Amin}, {Andeen}, {Anderson}, {Anton}, {Arg{\"u}elles}, {Ashida}, {Athanasiadou}, {Axani}, {Bai}, {Balagopal}, {Baricevic}, {Barwick}, {Basu}, {Bay}, {Beatty}, {Becker}, {Becker Tjus}, {Beise}, {Bellenghi}, {Benda}, {BenZvi}, {Berley}, {Bernardini}, {Besson}, {Binder}, {Bindig}, {Blaufuss}, {Blot}, {Bontempo}, {Book}, {Borowka}, {Meneguolo}, {B{\"o}ser}, {Botner}, {B{\"o}ttcher}, {Bourbeau}, {Braun}, {Brinson}, {Brostean-Kaiser}, {Burley}, {Busse}, {Campana}, {Carnie-Bronca}, {Chang}, {Chen}, {Chen}, {Chirkin}, {Choi}, {Clark}, {Classen}, {Coleman}, {Collin}, {Connolly}, {Conrad}, {Coppin}, {Correa}, {Countryman}, {Cowen}, {Dappen}, {Dave}, {De Clercq}, {DeLaunay}, {L{\'o}pez}, {Dembinski}, {Deoskar}, {Desai}, {Desiati}, {de Vries}, {de Wasseige}, {DeYoung}, {Diaz}, {D{\'\i}az-V{\'e}lez}, {Dittmer}, {Dujmovic}, {DuVernois}, {Ehrhardt}, {Eller}, {Engel}, {Erpenbeck}, {Evans}, {Evenson},
  {Fan}, {Fazely}, {Fedynitch}, {Feigl}, {Fiedlschuster}, {Fienberg}, {Finley}, {Fischer}, {Fox}, {Franckowiak}, {Friedman}, {Fritz}, {F{\"u}rst}, {Gaisser}, {Gallagher}, {Ganster}, {Garcia}, {Garrappa}, {Gerhardt}, {Ghadimi}, {Glaser}, {Glauch}, {Gl{\"u}senkamp}, {Goehlke}, {Gonzalez}, {Goswami}, {Grant}, {Gray}, {Gr{\'e}goire}, {Griswold}, {G{\"u}nther}, {Gutjahr}, {Haack}, {Hallgren}, {Halliday}, {Halve}, {Halzen}, {Hamdaoui}, {Minh}, {Hanson}, {Hardin}, {Harnisch}, {Hatch}, {Haungs}, {Helbing}, {Hellrung}, {Henningsen}, {Heuermann}, {Hickford}, {Hidvegi}, {Hill}, {Hill}, {Hoffman}, {Hoshina}, {Hou}, {Huber}, {Hultqvist}, {H{\"u}nnefeld}, {Hussain}, {Hymon}, {In}, {Iovine}, {Ishihara}, {Jansson}, {Japaridze}, {Jeong}, {Jin}, {Jones}, {Kang}, {Kang}, {Kang}, {Kappes}, {Kappesser}, {Kardum}, {Karg}, {Karl}, {Karle}, {Katz}, {Kauer}, {Kelley}, {Kheirandish}, {Kin}, {Kiryluk}, {Klein}, {Kochocki}, {Koirala}, {Kolanoski}, {Kontrimas}, {K{\"o}pke}, {Kopper}, {Koskinen}, {Koundal}, {Kovacevich}, {Kowalski},
  {Kozynets}, {Kruiswijk}, {Krupczak}, {Kun}, {Kurahashi}, {Lad}, {Lagunas Gualda}, {Lamoureux}, {Larson}, {Lauber}, {Lazar}, {Lee}, {Leonard DeHolton}, {Leszczy{\'n}ska}, {Lincetto}, {Liu}, \& {Liubarska}}]{IceCube:2022heu}
{Abbasi}, R., {Ackermann}, M., {Adams}, J., {et~al.} 2023, \apjl, 945, L8

\bibitem[{{Abdo} {et~al.}(2009){Abdo}, {Ackermann}, {Ajello}, {Atwood}, {Axelsson}, {Baldini}, {Ballet}, {Barbiellini}, {Baring}, {Bastieri}, {Bechtol}, {Bellazzini}, {Berenji}, {Bloom}, {Bonamente}, {Borgland}, {Bregeon}, {Brez}, {Brigida}, {Bruel}, {Burnett}, {Caliandro}, {Cameron}, {Caraveo}, {Casandjian}, {Cavazzuti}, {Cecchi}, {{\c{C}}elik}, {Chekhtman}, {Cheung}, {Chiang}, {Ciprini}, {Claus}, {Cohen-Tanugi}, {Cominsky}, {Conrad}, {Cutini}, {de Angelis}, {de Palma}, {Di Bernardo}, {Silva}, {Drell}, {Drlica-Wagner}, {Dubois}, {Dumora}, {Farnier}, {Favuzzi}, {Fegan}, {Finke}, {Focke}, {Fortin}, {Foschini}, {Frailis}, {Fukazawa}, {Funk}, {Fusco}, {Gargano}, {Gasparrini}, {Gehrels}, {Germani}, {Giavitto}, {Giebels}, {Giglietto}, {Giommi}, {Giordano}, {Glanzman}, {Godfrey}, {Grenier}, {Grondin}, {Grove}, {Guillemot}, {Guiriec}, {Hanabata}, {Hayashida}, {Hays}, {Horan}, {Hughes}, {Jackson}, {J{\'o}hannesson}, {Johnson}, {Johnson}, {Johnson}, {Kamae}, {Katagiri}, {Kataoka}, {Kawai}, {Kerr}, {Kn{\"o}dlseder},
  {Kocian}, {Kuss}, {Lande}, {Latronico}, {Lemoine-Goumard}, {Longo}, {Loparco}, {Lott}, {Lovellette}, {Lubrano}, {Madejski}, {Makeev}, {Mazziotta}, {McConville}, {McEnery}, {Meurer}, {Michelson}, {Mitthumsiri}, {Mizuno}, {Moiseev}, {Monte}, {Monzani}, {Morselli}, {Moskalenko}, {Murgia}, {Nolan}, {Norris}, {Nuss}, {Ohsugi}, {Omodei}, {Orlando}, {Ormes}, {Ozaki}, {Paneque}, {Panetta}, {Parent}, {Pelassa}, {Pepe}, {Pesce-Rollins}, {Piron}, {Porter}, {Rain{\`o}}, {Rando}, {Razzano}, {Reimer}, {Reimer}, {Reposeur}, {Reyes}, {Ritz}, {Rochester}, {Rodriguez}, {Roth}, {Ryde}, {Sadrozinski}, {Sanchez}, {Sander}, {Saz Parkinson}, {Scargle}, {Schalk}, {Sellerholm}, {Sgr{\`o}}, {Shaw}, {Siskind}, {Smith}, {Smith}, {Spandre}, {Spinelli}, {Strickman}, {Suson}, {Tajima}, {Takahashi}, {Takahashi}, {Tanaka}, {Tanaka}, {Thayer}, {Thayer}, {Thompson}, {Tibaldo}, {Torres}, {Tosti}, {Tramacere}, {Uchiyama}, {Usher}, {Vasileiou}, {Vilchez}, {Vitale}, {Waite}, {Wang}, {Winer}, {Wood}, {Ylinen}, \& {Ziegler}}]{abdo2009fermi}
{Abdo}, A.~A., {Ackermann}, M., {Ajello}, M., {et~al.} 2009, \apj, 707, 1310

\bibitem[{{Abdollahi} {et~al.}(2020){Abdollahi}, {Acero}, {Ackermann}, {Ajello}, {Atwood}, {Axelsson}, {Baldini}, {Ballet}, {Barbiellini}, {Bastieri}, {Becerra Gonzalez}, {Bellazzini}, {Berretta}, {Bissaldi}, {Blandford}, {Bloom}, {Bonino}, {Bottacini}, {Brandt}, {Bregeon}, {Bruel}, {Buehler}, {Burnett}, {Buson}, {Cameron}, {Caputo}, {Caraveo}, {Casandjian}, {Castro}, {Cavazzuti}, {Charles}, {Chaty}, {Chen}, {Cheung}, {Chiaro}, {Ciprini}, {Cohen-Tanugi}, {Cominsky}, {Coronado-Bl{\'a}zquez}, {Costantin}, {Cuoco}, {Cutini}, {D'Ammando}, {DeKlotz}, {de la Torre Luque}, {de Palma}, {Desai}, {Digel}, {Di Lalla}, {Di Mauro}, {Di Venere}, {Dom{\'\i}nguez}, {Dumora}, {Fana Dirirsa}, {Fegan}, {Ferrara}, {Franckowiak}, {Fukazawa}, {Funk}, {Fusco}, {Gargano}, {Gasparrini}, {Giglietto}, {Giommi}, {Giordano}, {Giroletti}, {Glanzman}, {Green}, {Grenier}, {Griffin}, {Grondin}, {Grove}, {Guiriec}, {Harding}, {Hayashi}, {Hays}, {Hewitt}, {Horan}, {J{\'o}hannesson}, {Johnson}, {Kamae}, {Kerr}, {Kocevski}, {Kovac'evic'},
  {Kuss}, {Landriu}, {Larsson}, {Latronico}, {Lemoine-Goumard}, {Li}, {Liodakis}, {Longo}, {Loparco}, {Lott}, {Lovellette}, {Lubrano}, {Madejski}, {Maldera}, {Malyshev}, {Manfreda}, {Marchesini}, {Marcotulli}, {Mart{\'\i}-Devesa}, {Martin}, {Massaro}, {Mazziotta}, {McEnery}, {Mereu}, {Meyer}, {Michelson}, {Mirabal}, {Mizuno}, {Monzani}, {Morselli}, {Moskalenko}, {Negro}, {Nuss}, {Ojha}, {Omodei}, {Orienti}, {Orlando}, {Ormes}, {Palatiello}, {Paliya}, {Paneque}, {Pei}, {Pe{\~n}a-Herazo}, {Perkins}, {Persic}, {Pesce-Rollins}, {Petrosian}, {Petrov}, {Piron}, {Poon}, {Porter}, {Principe}, {Rain{\`o}}, {Rando}, {Razzano}, {Razzaque}, {Reimer}, {Reimer}, {Remy}, {Reposeur}, {Romani}, {Saz Parkinson}, {Schinzel}, {Serini}, {Sgr{\`o}}, {Siskind}, {Smith}, {Spandre}, {Spinelli}, {Strong}, {Suson}, {Tajima}, {Takahashi}, {Tak}, {Thayer}, {Thompson}, {Tibaldo}, {Torres}, {Torresi}, {Valverde}, {Van Klaveren}, {van Zyl}, {Wood}, {Yassine}, \& {Zaharijas}}]{abdollahi2020fermi}
{Abdollahi}, S., {Acero}, F., {Ackermann}, M., {et~al.} 2020, \apjs, 247, 33

\bibitem[{{Abeysekara} {et~al.}(2017){Abeysekara}, {Albert}, {Alfaro}, {Alvarez}, {{\'A}lvarez}, {Arceo}, {Arteaga-Vel{\'a}zquez}, {Avila Rojas}, {Ayala Solares}, {Barber}, {Bautista-Elivar}, {Becerril}, {Belmont-Moreno}, {BenZvi}, {Berley}, {Bernal}, {Braun}, {Brisbois}, {Caballero-Mora}, {Capistr{\'a}n}, {Carrami{\~n}ana}, {Casanova}, {Castillo}, {Cotti}, {Cotzomi}, {Couti{\~n}o de Le{\'o}n}, {De Le{\'o}n}, {De la Fuente}, {Dingus}, {DuVernois}, {D{\'\i}az-V{\'e}lez}, {Ellsworth}, {Engel}, {Enr{\'\i}quez-Rivera}, {Fiorino}, {Fraija}, {Garc{\'\i}a-Gonz{\'a}lez}, {Garfias}, {Gerhardt}, {Gonz{\'a}lez Mu{\~n}oz}, {Gonz{\'a}lez}, {Goodman}, {Hampel-Arias}, {Harding}, {Hern{\'a}ndez}, {Hern{\'a}ndez-Almada}, {Hinton}, {Hona}, {Hui}, {H{\"u}ntemeyer}, {Iriarte}, {Jardin-Blicq}, {Joshi}, {Kaufmann}, {Kieda}, {Lara}, {Lauer}, {Lee}, {Lennarz}, {Vargas}, {Linnemann}, {Longinotti}, {Luis Raya}, {Luna-Garc{\'\i}a}, {L{\'o}pez-Coto}, {Malone}, {Marinelli}, {Martinez}, {Martinez-Castellanos}, {Mart{\'\i}nez-Castro},
  {Mart{\'\i}nez-Huerta}, {Matthews}, {Miranda-Romagnoli}, {Moreno}, {Mostaf{\'a}}, {Nellen}, {Newbold}, {Nisa}, {Noriega-Papaqui}, {Pelayo}, {Pretz}, {P{\'e}rez-P{\'e}rez}, {Ren}, {Rho}, {Rivi{\`e}re}, {Rosa-Gonz{\'a}lez}, {Rosenberg}, {Ruiz-Velasco}, {Salazar}, {Salesa Greus}, {Sandoval}, {Schneider}, {Schoorlemmer}, {Sinnis}, {Smith}, {Springer}, {Surajbali}, {Taboada}, {Tibolla}, {Tollefson}, {Torres}, {Ukwatta}, {Vianello}, {Weisgarber}, {Westerhoff}, {Wisher}, {Wood}, {Yapici}, {Yodh}, {Younk}, {Zepeda}, {Zhou}, {Guo}, {Hahn}, {Li}, \& {Zhang}}]{abeysekara2017extended}
{Abeysekara}, A.~U., {Albert}, A., {Alfaro}, R., {et~al.} 2017, Science, 358, 911

\bibitem[{{Acciari} {et~al.}(2021){Acciari}, {Ansoldi}, {Antonelli}, {Arbet Engels}, {Artero}, {Asano}, {Baack}, {Babi{\'c}}, {Baquero}, {Barres de Almeida}, {Barrio}, {Batkovi{\'c}}, {Becerra Gonz{\'a}lez}, {Bednarek}, {Bellizzi}, {Bernardini}, {Bernardos}, {Berti}, {Besenrieder}, {Bhattacharyya}, {Bigongiari}, {Biland}, {Blanch}, {Bonnoli}, {Bo{\v{s}}njak}, {Busetto}, {Carosi}, {Ceribella}, {Cerruti}, {Chai}, {Chilingarian}, {Cikota}, {Colak}, {Colombo}, {Contreras}, {Cortina}, {Covino}, {D'Amico}, {D'Elia}, {Da Vela}, {Dazzi}, {De Angelis}, {De Lotto}, {Delfino}, {Delgado}, {Delgado Mendez}, {Depaoli}, {Di Pierro}, {Di Venere}, {Do Souto Espi{\~n}eira}, {Dominis Prester}, {Donini}, {Dorner}, {Doro}, {Elsaesser}, {Fallah Ramazani}, {Fattorini}, {Ferrara}, {Fonseca}, {Font}, {Fruck}, {Fukami}, {Garc{\'\i}a L{\'o}pez}, {Garczarczyk}, {Gasparyan}, {Gaug}, {Giglietto}, {Giordano}, {Gliwny}, {Godinovi{\'c}}, {Green}, {Green}, {Hadasch}, {Hahn}, {Heckmann}, {Herrera}, {Hoang}, {Hrupec}, {H{\"u}tten}, {Inada},
  {Inoue}, {Ishio}, {Iwamura}, {Jim{\'e}nez}, {Jormanainen}, {Jouvin}, {Kajiwara}, {Karjalainen}, {Kerszberg}, {Kobayashi}, {Kubo}, {Kushida}, {Lamastra}, {Lelas}, {Leone}, {Lindfors}, {Lombardi}, {Longo}, {L{\'o}pez-Coto}, {L{\'o}pez-Moya}, {L{\'o}pez-Oramas}, {Loporchio}, {Machado de Oliveira Fraga}, {Maggio}, {Majumdar}, {Makariev}, {Mallamaci}, {Maneva}, {Manganaro}, {Mannheim}, {Maraschi}, {Mariotti}, {Mart{\'\i}nez}, {Mazin}, {Menchiari}, {Mender}, {Mi{\'c}anovi{\'c}}, {Miceli}, {Miener}, {Minev}, {Miranda}, {Mirzoyan}, {Molina}, {Moralejo}, {Morcuende}, {Moreno}, {Moretti}, {Neustroev}, {Nigro}, {Nilsson}, {Nishijima}, {Noda}, {Nozaki}, {Ohtani}, {Oka}, {Otero-Santos}, {Paiano}, {Palatiello}, {Paneque}, {Paoletti}, {Paredes}, {Pavleti{\'c}}, {Pe{\~n}il}, {Perennes}, {Persic}, {Prada Moroni}, {Prandini}, {Priyadarshi}, {Puljak}, {Rhode}, {Rib{\'o}}, {Rico}, {Righi}, {Rugliancich}, {Saha}, {Sahakyan}, {Saito}, {Sakurai}, {Satalecka}, {Saturni}, {Schleicher}, {Schmidt}, {Schweizer}, {Sitarek},
  {{\v{S}}nidari{\'c}}, {Sobczynska}, {Spolon}, {Stamerra}, {Strom}, {Strzys}, {Suda}, {Suri{\'c}}, {Takahashi}, {Tavecchio}, {Temnikov}, {Terzi{\'c}}, {Teshima}, {Tosti}, {Truzzi}, {Tutone}, {Ubach}, {van Scherpenberg}, {Vanzo}, {Vazquez Acosta}, {Ventura}, {Verguilov}, {Vigorito}, {Vitale}, {Vovk}, {Will}, {Wunderlich}, {Zari{\'c}}, {Zari{\'c}}, {Caraveo}, {Cognard}, {Guillemot}, {Harding}, {Li}, {Limyansky}, \& {Ng}}]{acciari2021search}
{Acciari}, V.~A., {Ansoldi}, S., {Antonelli}, L.~A., {et~al.} 2021, \apj, 922, 251

\bibitem[{{An} {et~al.}(2019){An}, {Asfandiyarov}, {Azzarello}, {Bernardini}, {Bi}, {Cai}, {Chang}, {Chen}, {Chen}, {Chen}, {Chen}, {Cui}, {Cui}, {Dai}, {D'Amone}, {De Benedittis}, {De Mitri}, {Di Santo}, {Ding}, {Dong}, {Dong}, {Dong}, {Donvito}, {Droz}, {Duan}, {Duan}, {D'Urso}, {Fan}, {Fan}, {Fang}, {Feng}, {Feng}, {Fusco}, {Gallo}, {Gan}, {Gao}, {Gargano}, {Gong}, {Gong}, {Guo}, {Guo}, {Guo}, {Han}, {Hu}, {Huang}, {Huang}, {Huang}, {Ionica}, {Jiang}, {Jin}, {Kong}, {Lei}, {Li}, {Li}, {Li}, {Li}, {Li}, {Liang}, {Liang}, {Liao}, {Liu}, {Liu}, {Liu}, {Liu}, {Liu}, {Liu}, {Loparco}, {Luo}, {Ma}, {Ma}, {Ma}, {Ma}, {Ma}, {Marsella}, {Mazziotta}, {Mo}, {Niu}, {Pan}, {Peng}, {Peng}, {Qiao}, {Rao}, {Salinas}, {Shang}, {Shen}, {Shen}, {Shen}, {Song}, {Su}, {Su}, {Sun}, {Surdo}, {Teng}, {Tykhonov}, {Vitillo}, {Wang}, {Wang}, {Wang}, {Wang}, {Wang}, {Wang}, {Wang}, {Wang}, {Wang}, {Wang}, {Wang}, {Wang}, {Wang}, {Wei}, {Wei}, {Wei}, {Wen}, {Wu}, {Wu}, {Wu}, {Wu}, {Wu}, {Xi}, {Xia}, {Xu}, {Xu}, {Xu}, {Xu}, {Xue},
  {Yang}, {Yang}, {Yang}, {Yang}, {Yao}, {Yu}, {Yuan}, {Yue}, {Zang}, {Zhang}, {Zhang}, {Zhang}, {Zhang}, {Zhang}, {Zhang}, {Zhang}, {Zhang}, {Zhang}, {Zhang}, {Zhang}, {Zhang}, {Zhang}, {Zhao}, {Zhao}, {Zhao}, {Zhou}, {Zhou}, {Zhu}, {Zhu}, \& {Zimmer}}]{dampe2019measurement}
{An}, Q., {Asfandiyarov}, R., {Azzarello}, P., {et~al.} 2019, Sci. Adv., 5, eaax3793

\bibitem[{{Bao} {et~al.}(2024{\natexlab{a}}){Bao}, {Giacinti}, {Liu}, {Zhang}, \& {Chen}}]{bao2024mirages}
{Bao}, Y., {Giacinti}, G., {Liu}, R.-Y., {Zhang}, H.-M., \& {Chen}, Y. 2024{\natexlab{a}}, ArXiv e-prints [\eprint[arXiv]{2407.02478}]

\bibitem[{{Bao} {et~al.}(2024{\natexlab{b}}){Bao}, {Liu}, {Giacinti}, {Zhang}, \& {Chen}}]{bao2024mirage}
{Bao}, Y., {Liu}, R.-Y., {Giacinti}, G., {Zhang}, H.-M., \& {Chen}, Y. 2024{\natexlab{b}}, ArXiv e-prints [\eprint[arXiv]{2407.02829}]

\bibitem[{{Baring} {et~al.}(1999){Baring}, {Ellison}, {Reynolds}, {Grenier}, \& {Goret}}]{baring1999radio}
{Baring}, M.~G., {Ellison}, D.~C., {Reynolds}, S.~P., {Grenier}, I.~A., \& {Goret}, P. 1999, \apj, 513, 311

\bibitem[{{Bednarek} \& {Bartosik}(2003)}]{bednarek2003gamma}
{Bednarek}, W. \& {Bartosik}, M. 2003, \aap, 405, 689

\bibitem[{{Blumenthal} \& {Gould}(1970)}]{blumenthal1970bremsstrahlung}
{Blumenthal}, G.~R. \& {Gould}, R.~J. 1970, Rev. Mod. Phys., 42, 237

\bibitem[{{Bolatto} {et~al.}(2013){Bolatto}, {Wolfire}, \& {Leroy}}]{bolatto2013co}
{Bolatto}, A.~D., {Wolfire}, M., \& {Leroy}, A.~K. 2013, \araa, 51, 207

\bibitem[{{Burrows} {et~al.}(2005){Burrows}, {Hill}, {Nousek}, {Kennea}, {Wells}, {Osborne}, {Abbey}, {Beardmore}, {Mukerjee}, {Short}, {Chincarini}, {Campana}, {Citterio}, {Moretti}, {Pagani}, {Tagliaferri}, {Giommi}, {Capalbi}, {Tamburelli}, {Angelini}, {Cusumano}, {Br{\"a}uninger}, {Burkert}, \& {Hartner}}]{burrows2005swift}
{Burrows}, D.~N., {Hill}, J.~E., {Nousek}, J.~A., {et~al.} 2005, \ssr, 120, 165

\bibitem[{{Cao} {et~al.}(2024){Cao}, {Aharonian}, {An}, {Axikegu}, {Bai}, {Bao}, {Bastieri}, {Bi}, {Bi}, {Cai}, {Cao}, {Cao}, {Cao}, {Chang}, {Chang}, {Chen}, {Chen}, {Chen}, {Chen}, {Chen}, {Chen}, {Chen}, {Chen}, {Chen}, {Chen}, {Chen}, {Chen}, {Cheng}, {Cheng}, {Cui}, {Cui}, {Cui}, {Cui}, {Dai}, {Dai}, {Dai}, {Danzengluobu}, {Della Volpe}, {Dong}, {Duan}, {Fan}, {Fan}, {Fang}, {Fang}, {Feng}, {Feng}, {Feng}, {Feng}, {Feng}, {Gabici}, {Gao}, {Gao}, {Gao}, {Gao}, {Gao}, {Gao}, {Ge}, {Geng}, {Giacinti}, {Gong}, {Gou}, {Gu}, {Guo}, {Guo}, {Guo}, {Guo}, {Han}, {He}, {He}, {He}, {He}, {He}, {Heller}, {Hor}, {Hou}, {Hou}, {Hou}, {Hu}, {Hu}, {Hu}, {Huang}, {Huang}, {Huang}, {Huang}, {Huang}, {Huang}, {Huang}, {Ji}, {Jia}, {Jia}, {Jiang}, {Jiang}, {Jiang}, {Jin}, {Kang}, {Ke}, {Kuleshov}, {Kurinov}, {Li}, {Li}, {Li}, {Li}, {Li}, {Li}, {Li}, {Li}, {Li}, {Li}, {Li}, {Li}, {Li}, {Li}, {Li}, {Li}, {Li}, {Li}, {Li}, {Liang}, {Liang}, {Lin}, {Liu}, {Liu}, {Liu}, {Liu}, {Liu}, {Liu}, {Liu}, {Liu}, {Liu}, {Liu}, {Liu},
  {Liu}, {Liu}, {Liu}, {Lu}, {Luo}, {Lv}, {Ma}, {Ma}, {Ma}, {Mao}, {Min}, {Mitthumsiri}, {Mu}, {Nan}, {Neronov}, {Ou}, {Pang}, {Pattarakijwanich}, {Pei}, {Qi}, {Qi}, {Qiao}, {Qin}, {Ruffolo}, {S{\'a}iz}, {Semikoz}, {Shao}, {Shao}, {Shchegolev}, {Sheng}, {Shu}, {Song}, {Stenkin}, {Stepanov}, {Su}, {Sun}, {Sun}, {Sun}, {Tam}, {Tang}, {Tang}, {Tian}, {Wang}, {Wang}, {Wang}, {Wang}, {Wang}, {Wang}, {Wang}, {Wang}, {Wang}, {Wang}, {Wang}, {Wang}, {Wang}, {Wang}, {Wang}, {Wang}, {Wang}, {Wang}, {Wang}, {Wang}, {Wang}, {Wei}, {Wei}, {Wei}, {Wen}, {Wu}, \& {Wu}}]{cao2024first}
{Cao}, Z., {Aharonian}, F., {An}, Q., {et~al.} 2024, \apjs, 271, 25

\bibitem[{{Cao} {et~al.}(2025){Cao}, {Aharonian}, {Axikegu}, {Bai}, {Bao}, {Bastieri}, {Bi}, {Bi}, {Bian}, {Bukevich}, {Cao}, {Cao}, {Cao}, {Chang}, {Chang}, {Chen}, {Chen}, {Chen}, {Chen}, {Chen}, {Chen}, {Chen}, {Chen}, {Chen}, {Chen}, {Chen}, {Chen}, {Chen}, {Chen}, {Cheng}, {Cheng}, {Cui}, {Cui}, {Cui}, {Cui}, {Dai}, {Dai}, {Dai}, {Danzengluobu}, {Dong}, {Duan}, {Fan}, {Fan}, {Fang}, {Fang}, {Fang}, {Feng}, {Feng}, {Feng}, {Feng}, {Feng}, {Feng}, {Feng}, {Gabici}, {Gao}, {Gao}, {Gao}, {Gao}, {Gao}, {Ge}, {Geng}, {Giacinti}, {Gong}, {Gou}, {Gu}, {Guo}, {Guo}, {Guo}, {Guo}, {Han}, {Hasan}, {He}, {He}, {He}, {He}, {Hor}, {Hou}, {Hou}, {Hou}, {Hu}, {Hu}, {Hu}, {Huang}, {Huang}, {Huang}, {Huang}, {Huang}, {Huang}, {Ji}, {Jia}, {Jia}, {Jiang}, {Jiang}, {Jiang}, {Jin}, {Kang}, {Karpikov}, {Kuleshov}, {Kurinov}, {Li}, {Li}, {Li}, {Li}, {Li}, {Li}, {Li}, {Li}, {Li}, {Li}, {Li}, {Li}, {Li}, {Li}, {Li}, {Li}, {Li}, {Li}, {Li}, {Liang}, {Liang}, {Lin}, {Liu}, {Liu}, {Liu}, {Liu}, {Liu}, {Liu}, {Liu}, {Liu}, {Liu},
  {Liu}, {Liu}, {Liu}, {Liu}, {Liu}, {Luo}, {Luo}, {Lv}, {Ma}, {Ma}, {Ma}, {Mao}, {Min}, {Mitthumsiri}, {Mu}, {Nan}, {Neronov}, {Ou}, {Pattarakijwanich}, {Pei}, {Qi}, {Qi}, {Qiao}, {Qin}, {Raza}, {Ruffolo}, {S{\'a}iz}, {Saeed}, {Semikoz}, {Shao}, {Shchegolev}, {Sheng}, {Shu}, {Song}, {Stenkin}, {Stepanov}, {Su}, {Sun}, {Sun}, {Sun}, {Sun}, {Takata}, {Tam}, {Tang}, {Tang}, {Tang}, {Tian}, {Wang}, {Wang}, {Wang}, {Wang}, {Wang}, {Wang}, {Wang}, {Wang}, {Wang}, {Wang}, {Wang}, {Wang}, {Wang}, {Wang}, {Wang}, {Wang}, {Wang}, {Wang}, {Wang}, {Wang}, {Wang}, {Wang}, \& {Wei}}]{cao2025data}
{Cao}, Z., {Aharonian}, F., {Axikegu}, {et~al.} 2025, Astropart. Phys., 164, 103029

\bibitem[{{Dame}(2011)}]{dame2011optimization}
{Dame}, T.~M. 2011, ArXiv e-prints [\eprint[arXiv]{1101.1499}]

\bibitem[{{Dame} {et~al.}(2001){Dame}, {Hartmann}, \& {Thaddeus}}]{dame2001milky}
{Dame}, T.~M., {Hartmann}, D., \& {Thaddeus}, P. 2001, \apj, 547, 792

\bibitem[{{Dame} \& {Thaddeus}(2022)}]{dame2022co}
{Dame}, T.~M. \& {Thaddeus}, P. 2022, \apjs, 262, 5

\bibitem[{{de O{\~n}a Wilhelmi} {et~al.}(2022){de O{\~n}a Wilhelmi}, {L{\'o}pez-Coto}, {Amato}, \& {Aharonian}}]{de2022potential}
{de O{\~n}a Wilhelmi}, E., {L{\'o}pez-Coto}, R., {Amato}, E., \& {Aharonian}, F. 2022, \apjl, 930, L2

\bibitem[{{De Sarkar} \& {Gupta}(2022)}]{de2022exploring}
{De Sarkar}, A. \& {Gupta}, N. 2022, \apj, 934, 118

\bibitem[{{Delahaye} {et~al.}(2010){Delahaye}, {Lavalle}, {Lineros}, {Donato}, \& {Fornengo}}]{delahaye2010galactic}
{Delahaye}, T., {Lavalle}, J., {Lineros}, R., {Donato}, F., \& {Fornengo}, N. 2010, \aap, 524, A51

\bibitem[{{Delahaye} {et~al.}(2008){Delahaye}, {Lineros}, {Donato}, {Fornengo}, \& {Salati}}]{delahaye2008positrons}
{Delahaye}, T., {Lineros}, R., {Donato}, F., {Fornengo}, N., \& {Salati}, P. 2008, \prd, 77, 063527

\bibitem[{{Di Mauro} {et~al.}(2020){Di Mauro}, {Manconi}, \& {Donato}}]{di2020evidences}
{Di Mauro}, M., {Manconi}, S., \& {Donato}, F. 2020, \prd, 101, 103035

\bibitem[{{di Sciascio} \& {LHAASO Collaboration}(2016)}]{di2016lhaaso}
{di Sciascio}, G. \& {LHAASO Collaboration}. 2016, Nucl. Part. Phys. Proc., 279-281, 166

\bibitem[{{Ding} {et~al.}(2021){Ding}, {Li}, {Wei}, {Wu}, \& {Zhou}}]{ding2021implications}
{Ding}, Y.-C., {Li}, N., {Wei}, C.-C., {Wu}, Y.-L., \& {Zhou}, Y.-F. 2021, \prd, 103, 115010

\bibitem[{{Enokiya} {et~al.}(2023){Enokiya}, {Sano}, {Filipovi{\'c}}, {Alsaberi}, {Inoue}, \& {Oka}}]{enokiya2023discovery}
{Enokiya}, R., {Sano}, H., {Filipovi{\'c}}, M.~D., {et~al.} 2023, \pasj, 75, 970

\bibitem[{{Fang} \& {Bi}(2022)}]{fang2022interpretation}
{Fang}, K. \& {Bi}, X.-J. 2022, \prd, 105, 103007

\bibitem[{{Fang} {et~al.}(2018){Fang}, {Bi}, {Yin}, \& {Yuan}}]{fang2018two}
{Fang}, K., {Bi}, X.-J., {Yin}, P.-F., \& {Yuan}, Q. 2018, \apj, 863, 30

\bibitem[{{Fang} \& {Halzen}(2024)}]{Fang:2024LHAASOnu}
{Fang}, K. \& {Halzen}, F. 2024, JHEAp, 43, 140

\bibitem[{{Ferrand} \& {Safi-Harb}(2012)}]{ferrand2012census}
{Ferrand}, G. \& {Safi-Harb}, S. 2012, Adv. Space Res., 49, 1313

\bibitem[{{Funk} {et~al.}(2013){Funk}, {Hinton}, \& {CTA Consortium}}]{funk2013comparison}
{Funk}, S., {Hinton}, J.~A., \& {CTA Consortium}. 2013, Astropart. Phys., 43, 348

\bibitem[{{Gaensler} \& {Slane}(2006)}]{gaensler2006evolution}
{Gaensler}, B.~M. \& {Slane}, P.~O. 2006, \araa, 44, 17

\bibitem[{{George} {et~al.}(2008){George}, {Fabian}, {Baumgartner}, {Mushotzky}, \& {Tueller}}]{george2008active}
{George}, M.~R., {Fabian}, A.~C., {Baumgartner}, W.~H., {Mushotzky}, R.~F., \& {Tueller}, J. 2008, \mnras, 388, L59

\bibitem[{Geyer \& Charles(1992)}]{geyer1992practical}
Geyer \& Charles, J. 1992, Stat. Sci., 7, 473

\bibitem[{{Ghisellini} {et~al.}(1988){Ghisellini}, {Guilbert}, \& {Svensson}}]{ghisellini1988synchrotron}
{Ghisellini}, G., {Guilbert}, P.~W., \& {Svensson}, R. 1988, \apjl, 334, L5

\bibitem[{{Giommi} {et~al.}(2021){Giommi}, {Perri}, {Capalbi}, {D'Elia}, {Barres de Almeida}, {Brandt}, {Pollock}, {Arneodo}, {Di Giovanni}, {Chang}, {Civitarese}, {De Angelis}, {Leto}, {Verrecchia}, {Ricard}, {Di Pippo}, {Middei}, {Penacchioni}, {Ruffini}, {Sahakyan}, {Israyelyan}, \& {Turriziani}}]{giommi2021x}
{Giommi}, P., {Perri}, M., {Capalbi}, M., {et~al.} 2021, \mnras, 507, 5690

\bibitem[{Grant {et~al.}(2019)Grant, Ackermann, Karle, \& Kowalski}]{aartsen2019neutrino}
Grant, D., Ackermann, M., Karle, A., \& Kowalski, M. 2019, BAAS, 51, 288

\bibitem[{{H.~E.~S.~S. Collaboration} {et~al.}(2024){H.~E.~S.~S. Collaboration}, {Aharonian}, {Ait Benkhali}, {Aschersleben}, {Ashkar}, {Backes}, {Barbosa Martins}, {Batzofin}, {Becherini}, {Berge}, {Bernl{\"o}hr}, {Bi}, {B{\"o}ttcher}, {Boisson}, {Bolmont}, {de Lavergne}, {Borowska}, {Bouyahiaoui}, {Breuhaus}, {Brose}, {Brown}, {Brun}, {Bruno}, {Bulik}, {Burger-Scheidlin}, {Caroff}, {Casanova}, {Cecil}, {Celic}, {Cerruti}, {Chand}, {Chandra}, {Chen}, {Chibueze}, {Chibueze}, {Cotter}, {Dai}, {Mbarubucyeye}, {Djannati-Ata{\"\i}}, {Dmytriiev}, {Doroshenko}, {Egberts}, {Einecke}, {Ernenwein}, {Filipovic}, {Fontaine}, {F{\"u}{\ss}ling}, {Funk}, {Gabici}, {Ghafourizadeh}, {Giavitto}, {Glawion}, {Glicenstein}, {Grolleron}, {Haerer}, {Hinton}, {Hofmann}, {Holch}, {Holler}, {Horns}, {Jamrozy}, {Jankowsky}, {Jardin-Blicq}, {Joshi}, {Jung-Richardt}, {Kasai}, {Katarzy{\'n}ski}, {Khatoon}, {Kh{\'e}lifi}, {Klepser}, {Klu{\'z}niak}, {Komin}, {Kosack}, {Kostunin}, {Kundu}, {Lang}, {Le Stum}, {Leitl}, {Lemi{\`e}re},
  {Lenain}, {Leuschner}, {Lohse}, {Luashvili}, {Lypova}, {Mackey}, {Malyshev}, {Malyshev}, {Marandon}, {Marchegiani}, {Marcowith}, {Mart{\'\i}-Devesa}, {Marx}, {Mehta}, {Mitchell}, {Moderski}, {Mohrmann}, {Montanari}, {Moulin}, {Murach}, {Nakashima}, {de Naurois}, {Niemiec}, {Noel}, {Ohm}, {Olivera-Nieto}, {de Ona Wilhelmi}, {Ostrowski}, {Panny}, {Panter}, {Parsons}, {Peron}, {Prokhorov}, {P{\"u}hlhofer}, {Punch}, {Quirrenbach}, {Reichherzer}, {Reimer}, {Reimer}, {Ren}, {Renaud}, {Reville}, {Rieger}, {Rowell}, {Rudak}, {Ricarte}, {Ruiz-Velasco}, {Sahakian}, {Salzmann}, {Santangelo}, {Sasaki}, {Sch{\"a}fer}, {Sch{\"u}ssler}, {Schwanke}, {Shapopi}, {Sol}, {Specovius}, {Spencer}, {Stawarz}, {Steenkamp}, {Steinmassl}, {Steppa}, {Streil}, {Sushch}, {Suzuki}, {Takahashi}, {Tanaka}, {Taylor}, {Terrier}, {Tsirou}, {Tsuji}, {Unbehaun}, {van Eldik}, {Vecchi}, {Veh}, {Venter}, {Vink}, {Wach}, {Wagner}, {Werner}, {White}, {Wierzcholska}, {Wong}, {Zacharias}, {Zargaryan}, {Zdziarski}, {Zech}, {Zouari}, \&
  {{\.Z}ywucka}}]{hess2024acceleration}
{H.~E.~S.~S. Collaboration}, {Aharonian}, F., {Ait Benkhali}, F., {et~al.} 2024, Science, 383, 402

\bibitem[{{Hahn}(2015)}]{hahn2015gamera}
{Hahn}, J. 2015, in 34th International Cosmic Ray Conference (ICRC2015)

\bibitem[{{Hobbs} {et~al.}(2005){Hobbs}, {Lorimer}, {Lyne}, \& {Kramer}}]{hobbs2005statistical}
{Hobbs}, G., {Lorimer}, D.~R., {Lyne}, A.~G., \& {Kramer}, M. 2005, \mnras, 360, 974

\bibitem[{{H{\"u}mmer} {et~al.}(2010){H{\"u}mmer}, {R{\"u}ger}, {Spanier}, \& {Winter}}]{hummer2010simplified}
{H{\"u}mmer}, S., {R{\"u}ger}, M., {Spanier}, F., \& {Winter}, W. 2010, \apj, 721, 630

\bibitem[{{Hussein} \& {Shalchi}(2014)}]{hussein2014detailed}
{Hussein}, M. \& {Shalchi}, A. 2014, \apj, 785, 31

\bibitem[{{IceCube Collaboration}(2013)}]{icecube2013evidence}
{IceCube Collaboration}. 2013, Science, 342, 1242856

\bibitem[{{IceCube Collaboration} {et~al.}(2018{\natexlab{a}}){IceCube Collaboration}, {Aartsen}, {Ackermann}, {Adams}, {Aguilar}, {Ahlers}, {Ahrens}, {Al Samarai}, {Altmann}, {Andeen}, {Anderson}, {Ansseau}, {Anton}, {Arg{\"u}elles}, {Auffenberg}, {Axani}, {Bagherpour}, {Bai}, {Barron}, {Barwick}, {Baum}, {Bay}, {Beatty}, {Becker Tjus}, {Becker}, {BenZvi}, {Berley}, {Bernardini}, {Besson}, {Binder}, {Bindig}, {Blaufuss}, {Blot}, {Bohm}, {B{\"o}rner}, {Bos}, {B{\"o}ser}, {Botner}, {Bourbeau}, {Bourbeau}, {Bradascio}, {Braun}, {Brenzke}, {Bretz}, {Bron}, {Brostean-Kaiser}, {Burgman}, {Busse}, {Carver}, {Cheung}, {Chirkin}, {Christov}, {Clark}, {Classen}, {Coenders}, {Collin}, {Conrad}, {Coppin}, {Correa}, {Cowen}, {Cross}, {Dave}, {Day}, {de Andr{\'e}}, {De Clercq}, {DeLaunay}, {Dembinski}, {De Ridder}, {Desiati}, {de Vries}, {de Wasseige}, {de With}, {DeYoung}, {D{\'\i}az-V{\'e}lez}, {di Lorenzo}, {Dujmovic}, {Dumm}, {Dunkman}, {Dvorak}, {Eberhardt}, {Ehrhardt}, {Eichmann}, {Eller}, {Evenson}, {Fahey},
  {Fazely}, {Felde}, {Filimonov}, {Finley}, {Flis}, {Franckowiak}, {Friedman}, {Fritz}, {Gaisser}, {Gallagher}, {Gerhardt}, {Ghorbani}, {Glauch}, {Gl{\"u}senkamp}, {Goldschmidt}, {Gonzalez}, {Grant}, {Griffith}, {Haack}, {Hallgren}, {Halzen}, {Hanson}, {Hebecker}, {Heereman}, {Helbing}, {Hellauer}, {Hickford}, {Hignight}, {Hill}, {Hoffman}, {Hoffmann}, {Hoinka}, {Hokanson-Fasig}, {Hoshina}, {Huang}, {Huber}, {Hultqvist}, {H{\"u}nnefeld}, {Hussain}, {In}, {Iovine}, {Ishihara}, {Jacobi}, {Japaridze}, {Jeong}, {Jero}, {Jones}, {Kalaczynski}, {Kang}, {Kappes}, {Kappesser}, {Karg}, {Karle}, {Katz}, {Kauer}, {Keivani}, {Kelley}, {Kheirandish}, {Kim}, {Kim}, {Kintscher}, {Kiryluk}, {Kittler}, {Klein}, {Koirala}, {Kolanoski}, {K{\"o}pke}, {Kopper}, {Kopper}, {Koschinsky}, {Koskinen}, {Kowalski}, {Krings}, {Kroll}, {Kr{\"u}ckl}, {Kunwar}, {Kurahashi}, {Kuwabara}, {Kyriacou}, {Labare}, {Lanfranchi}, {Larson}, {Lauber}, {Leonard}, {Lesiak-Bzdak}, {Leuermann}, {Liu}, {Lozano Mariscal}, {Lu}, {L{\"u}nemann}, {Luszczak},
  {Madsen}, {Maggi}, {Mahn}, {Mancina}, {Maruyama}, {Mase}, {Maunu}, {Meagher}, {Medici}, {Meier}, {Menne}, {Merino}, {Meures}, {Miarecki}, {Micallef}, {Moment{\'e}}, {Montaruli}, {Moore}, {Morse}, {Moulai}, {Nahnhauer}, {Nakarmi}, {Naumann}, \& {Neer}}]{IceCube:2018dTXS}
{IceCube Collaboration}, {Aartsen}, M.~G., {Ackermann}, M., {et~al.} 2018{\natexlab{a}}, Science, 361, eaat1378

\bibitem[{{IceCube Collaboration} {et~al.}(2018{\natexlab{b}}){IceCube Collaboration}, {Aartsen}, {Ackermann}, {Adams}, {Aguilar}, {Ahlers}, {Ahrens}, {Samarai}, {Altmann}, {Andeen}, {Anderson}, {Ansseau}, {Anton}, {Arg{\"u}elles}, {Arsioli}, {Auffenberg}, {Axani}, {Bagherpour}, {Bai}, {Barron}, {Barwick}, {Baum}, {Bay}, {Beatty}, {Becker Tjus}, {Becker}, {BenZvi}, {Berley}, {Bernardini}, {Besson}, {Binder}, {Bindig}, {Blaufuss}, {Blot}, {Bohm}, {B{\"o}rner}, {Bos}, {B{\"o}ser}, {Botner}, {Bourbeau}, {Bourbeau}, {Bradascio}, {Braun}, {Brenzke}, {Bretz}, {Bron}, {Brostean-Kaiser}, {Burgman}, {Busse}, {Carver}, {Cheung}, {Chirkin}, {Christov}, {Clark}, {Classen}, {Coenders}, {Collin}, {Conrad}, {Coppin}, {Correa}, {Cowen}, {Cross}, {Dave}, {Day}, {de Andr{\'e}}, {De Clercq}, {DeLaunay}, {Dembinski}, {DeRidder}, {Desiati}, {de Vries}, {de Wasseige}, {de With}, {DeYoung}, {D{\'\i}az-V{\'e}lez}, {di Lorenzo}, {Dujmovic}, {Dumm}, {Dunkman}, {Dvorak}, {Eberhardt}, {Ehrhardt}, {Eichmann}, {Eller}, {Evenson}, {Fahey},
  {Fazely}, {Felde}, {Filimonov}, {Finley}, {Flis}, {Franckowiak}, {Friedman}, {Fritz}, {Gaisser}, {Gallagher}, {Gerhardt}, {Ghorbani}, {Giommi}, {Glauch}, {Gl{\"u}senkamp}, {Goldschmidt}, {Gonzalez}, {Grant}, {Griffith}, {Haack}, {Hallgren}, {Halzen}, {Hanson}, {Hebecker}, {Heereman}, {Helbing}, {Hellauer}, {Hickford}, {Hignight}, {Hill}, {Hoffman}, {Hoffmann}, {Hoinka}, {Hokanson-Fasig}, {Hoshina}, {Huang}, {Huber}, {Hultqvist}, {H{\"u}nnefeld}, {Hussain}, {In}, {Iovine}, {Ishihara}, {Jacobi}, {Japaridze}, {Jeong}, {Jero}, {Jones}, {Kalaczynski}, {Kang}, {Kappes}, {Kappesser}, {Karg}, {Karle}, {Katz}, {Kauer}, {Keivani}, {Kelley}, {Kheirandish}, {Kim}, {Kim}, {Kintscher}, {Kiryluk}, {Kittler}, {Klein}, {Koirala}, {Kolanoski}, {K{\"o}pke}, {Kopper}, {Kopper}, {Koschinsky}, {Koskinen}, {Kowalski}, {Krammer}, {Krings}, {Kroll}, {Kr{\"u}ckl}, {Kunwar}, {Kurahashi}, {Kuwabara}, {Kyriacou}, {Labare}, {Lanfranchi}, {Larson}, {Lauber}, {Leonard}, {Lesiak-Bzdak}, {Leuermann}, {Liu}, {Lozano Mariscal}, {Lu},
  {L{\"u}nemann}, {Luszczak}, {Madsen}, {Maggi}, {Mahn}, {Mancina}, {Maruyama}, {Mase}, {Maunu}, {Meagher}, {Medici}, {Meier}, {Menne}, {Merino}, {Meures}, {Miarecki}, {Micallef}, {Moment{\'e}}, {Montaruli}, {Moore}, {Morse}, {Moulai}, \& {Nahnhauer}}]{IceCube:2018TXSnu}
{IceCube Collaboration}, {Aartsen}, M.~G., {Ackermann}, M., {et~al.} 2018{\natexlab{b}}, Science, 361, 147

\bibitem[{{IceCube Collaboration} {et~al.}(2022){IceCube Collaboration}, {Abbasi}, {Ackermann}, {Adams}, {Aguilar}, {Ahlers}, {Ahrens}, {Alameddine}, {Alispach}, {Alves}, {Amin}, {Andeen}, {Anderson}, {Anton}, {Arg{\"u}elles}, {Ashida}, {Axani}, {Bai}, {Balagopal}, {Barbano}, {Barwick}, {Bastian}, {Basu}, {Baur}, {Bay}, {Beatty}, {Becker}, {Becker Tjus}, {Bellenghi}, {Benzvi}, {Berley}, {Bernardini}, {Besson}, {Binder}, {Bindig}, {Blaufuss}, {Blot}, {Boddenberg}, {Bontempo}, {Borowka}, {B{\"o}ser}, {Botner}, {B{\"o}ttcher}, {Bourbeau}, {Bradascio}, {Braun}, {Brinson}, {Bron}, {Brostean-Kaiser}, {Browne}, {Burgman}, {Burley}, {Busse}, {Campana}, {Carnie-Bronca}, {Chen}, {Chen}, {Chirkin}, {Choi}, {Clark}, {Clark}, {Classen}, {Coleman}, {Collin}, {Conrad}, {Coppin}, {Correa}, {Cowen}, {Cross}, {Dappen}, {Dave}, {de Clercq}, {Delaunay}, {Delgado L{\'o}pez}, {Dembinski}, {Deoskar}, {Desai}, {Desiati}, {de Vries}, {de Wasseige}, {de With}, {Deyoung}, {Diaz}, {D{\'\i}az-V{\'e}lez}, {Dittmer}, {Dujmovic}, {Dunkman},
  {Duvernois}, {Dvorak}, {Ehrhardt}, {Eller}, {Engel}, {Erpenbeck}, {Evans}, {Evenson}, {Fan}, {Fazely}, {Fedynitch}, {Feigl}, {Fiedlschuster}, {Fienberg}, {Filimonov}, {Finley}, {Fischer}, {Fox}, {Franckowiak}, {Friedman}, {Fritz}, {F{\"u}rst}, {Gaisser}, {Gallagher}, {Ganster}, {Garcia}, {Garrappa}, {Gerhardt}, {Ghadimi}, {Glaser}, {Glauch}, {Gl{\"u}senkamp}, {Goldschmidt}, {Gonzalez}, {Goswami}, {Grant}, {Gr{\'e}goire}, {Griswold}, {G{\"u}nther}, {Gutjahr}, {Haack}, {Hallgren}, {Halliday}, {Halve}, {Halzen}, {Hanson}, {Hardin}, {Harnisch}, {Haungs}, {Hebecker}, {Helbing}, {Henningsen}, {Hettinger}, {Hickford}, {Hignight}, {Hill}, {Hill}, {Hoffman}, {Hoffmann}, {Hokanson-Fasig}, {Hoshina}, {Huang}, {Huber}, {Huber}, {Hultqvist}, {H{\"u}nnefeld}, {Hussain}, {Hymon}, {in}, {Iovine}, {Ishihara}, {Jansson}, {Japaridze}, {Jeong}, {Jin}, {Jones}, {Kang}, {Kang}, {Kang}, {Kappes}, {Kappesser}, {Kardum}, {Karg}, {Karl}, {Karle}, {Katz}, {Kauer}, {Kellermann}, {Kelley}, {Kheirandish}, {Kin}, {Kintscher}, {Kiryluk},
  {Klein}, {Koirala}, {Kolanoski}, {Kontrimas}, {K{\"o}pke}, {Kopper}, {Kopper}, {Koskinen}, {Koundal}, {Kovacevich}, {Kowalski}, {Kozynets}, {Kun}, {Kurahashi}, {Lad}, {Lagunas Gualda}, {Lanfranchi}, {Larson}, {Lauber}, \& {Lazar}}]{IceCube:2022NGC}
{IceCube Collaboration}, {Abbasi}, R., {Ackermann}, M., {et~al.} 2022, Science, 378, 538

\bibitem[{{Icecube Collaboration} {et~al.}(2023){Icecube Collaboration}, {Abbasi}, {Ackermann}, {Adams}, {Aguilar}, {Ahlers}, {Ahrens}, {Alameddine}, {Alves}, {Amin}, {Andeen}, {Anderson}, {Anton}, {Arguelles}, {Ashida}, {Athanasiadou}, {Axani}, {Bai}, {Balagopal}, {Barwick}, {Basu}, {Baur}, {Bay}, {Beatty}, {Becker}, {Becker Tjus}, {Beise}, {Bellenghi}, {Benda}, {Benzvi}, {Berley}, {Bernardini}, {Besson}, {Binder}, {Bindig}, {Blaufuss}, {Blot}, {Boddenberg}, {Bontempo}, {Book}, {Borowka}, {Boser}, {Botner}, {Bottcher}, {Bourbeau}, {Bradascio}, {Braun}, {Brinson}, {Bron}, {Brostean-Kaiser}, {Burley}, {Busse}, {Campana}, {Carnie-Bronca}, {Chen}, {Chen}, {Chirkin}, {Choi}, {Clark}, {Clark}, {Classen}, {Coleman}, {Collin}, {Connolly}, {Conrad}, {Coppin}, {Correa}, {Cowen}, {Cross}, {Dappen}, {Dave}, {de Clercq}, {Delaunay}, {Delgado Lopez}, {Dembinski}, {Deoskar}, {Desai}, {Desiati}, {de Vries}, {de Wasseige}, {Deyoung}, {Diaz}, {Diaz-Velez}, {Dittmer}, {Dujmovic}, {Dunkman}, {Duvernois}, {Ehrhardt}, {Eller},
  {Engel}, {Erpenbeck}, {Evans}, {Evenson}, {Fan}, {Fazely}, {Fedynitch}, {Feigl}, {Fiedlschuster}, {Fienberg}, {Finley}, {Fischer}, {Fox}, {Franckowiak}, {Friedman}, {Fritz}, {Furst}, {Gaisser}, {Gallagher}, {Ganster}, {Garcia}, {Garrappa}, {Gerhardt}, {Ghadimi}, {Glaser}, {Glauch}, {Glusenkamp}, {Goehlke}, {Goldschmidt}, {Gonzalez}, {Goswami}, {Grant}, {Gregoire}, {Griswold}, {Gunther}, {Gutjahr}, {Haack}, {Hallgren}, {Halliday}, {Halve}, {Halzen}, {Ha}, {Hanson}, {Hardin}, {Harnisch}, {Haungs}, {Helbing}, {Henningsen}, {Hettinger}, {Hickford}, {Hignight}, {Hill}, {Hill}, {Hoffman}, {Hoshina}, {Hou}, {Huang}, {Huber}, {Huber}, {Hultqvist}, {Hunnefeld}, {Hussain}, {Hymon}, {in}, {Iovine}, {Ishihara}, {Jansson}, {Japaridze}, {Jeong}, {Jin}, {Jones}, {Kang}, {Kang}, {Kang}, {Kappes}, {Kappesser}, {Kardum}, {Karg}, {Karl}, {Karle}, {Katz}, {Kauer}, {Kellermann}, {Kelley}, {Kheirandish}, {Kin}, {Kiryluk}, {Klein}, {Kochocki}, {Koirala}, {Kolanoski}, {Kontrimas}, {Kopke}, {Kopper}, {Kopper}, {Koskinen},
  {Koundal}, {Kovacevich}, {Kowalski}, {Kozynets}, {Krupczak}, {Kun}, {Kurahashi}, {Lad}, {Lagunas Gualda}, {Lanfranchi}, {Larson}, {Lauber}, {Lazar}, {Lee}, \& {Leonard}}]{IceCube:2023Galactic}
{Icecube Collaboration}, {Abbasi}, R., {Ackermann}, M., {et~al.} 2023, Science, 380, 1338

\bibitem[{{IceCube Collaboration} {et~al.}(2021){IceCube Collaboration}, {Abbasi}, {Ackermann}, {Adams}, {Aguilar}, {Ahlers}, {Ahrens}, {Alispach}, {Amin}, {Andeen}, {Anderson}, {Ansseau}, {Anton}, {Arg{\"u}elles}, {Axani}, {Bai}, {Balagopal V.}, {Barbano}, {Barwick}, {Bastian}, {Basu}, {Baum}, {Baur}, {Bay}, {Beatty}, {Becker}, {Becker Tjus}, {Bellenghi}, {BenZvi}, {Berley}, {Bernardini}, {Besson}, {Binder}, {Bindig}, {Blaufuss}, {Blot}, {Bohm}, {B{\"o}ser}, {Botner}, {B{\"o}ttcher}, {Bourbeau}, {Bourbeau}, {Bradascio}, {Braun}, {Bron}, {Brostean-Kaiser}, {Burgman}, {Buscher}, {Busse}, {Campana}, {Carver}, {Chen}, {Cheung}, {Chirkin}, {Choi}, {Clark}, {Clark}, {Classen}, {Coleman}, {Collin}, {Conrad}, {Coppin}, {Correa}, {Cowen}, {Cross}, {Dave}, {De Clercq}, {DeLaunay}, {Dembinski}, {Deoskar}, {De Ridder}, {Desai}, {Desiati}, {de Vries}, {de Wasseige}, {de With}, {DeYoung}, {Dharani}, {Diaz}, {D{\'\i}az-V{\'e}lez}, {Dujmovic}, {Dunkman}, {DuVernois}, {Dvorak}, {Ehrhardt}, {Eller}, {Engel}, {Evenson},
  {Fahey}, {Fazely}, {Felde}, {Fienberg}, {Filimonov}, {Finley}, {Fischer}, {Fox}, {Franckowiak}, {Friedman}, {Fritz}, {Gaisser}, {Gallagher}, {Ganster}, {Garrappa}, {Gerhardt}, {Ghadimi}, {Glauch}, {Gl{\"u}senkamp}, {Goldschmidt}, {Gonzalez}, {Goswami}, {Grant}, {Gr{\'e}goire}, {Griffith}, {Griswold}, {G{\"u}nd{\"u}z}, {Haack}, {Hallgren}, {Halliday}, {Halve}, {Halzen}, {Minh}, {Hanson}, {Hardin}, {Haungs}, {Hauser}, {Hebecker}, {Heix}, {Helbing}, {Hellauer}, {Henningsen}, {Hickford}, {Hignight}, {Hill}, {Hill}, {Hoffman}, {Hoffmann}, {Hoinka}, {Hokanson-Fasig}, {Hoshina}, {Huang}, {Huber}, {Huber}, {Hultqvist}, {H{\"u}nnefeld}, {Hussain}, {In}, {Iovine}, {Ishihara}, {Jansson}, {Japaridze}, {Jeong}, {Jones}, {Jonske}, {Joppe}, {Kang}, {Kang}, {Kang}, {Kappes}, {Kappesser}, {Karg}, {Karl}, {Karle}, {Katz}, {Kauer}, {Kellermann}, {Kelley}, {Kheirandish}, {Kim}, {Kin}, {Kintscher}, {Kiryluk}, {Kittler}, {Klein}, {Koirala}, {Kolanoski}, {K{\"o}pke}, {Kopper}, {Kopper}, {Koskinen}, {Koundal}, {Kovacevich},
  {Kowalski}, {Krings}, {Kr{\"u}ckl}, {Kulacz}, {Kurahashi}, {Kyriacou}, {Lagunas Gualda}, {Lanfranchi}, {Larson}, {Lauber}, {Lazar}, {Leonard}, {Leszczy{\'n}ska}, {Li}, {Liu}, {Lohfink}, {Lozano Mariscal}, {Lu}, \& {Lucarelli}}]{IceCube:10year}
{IceCube Collaboration}, {Abbasi}, R., {Ackermann}, M., {et~al.} 2021, ArXiv e-prints [\eprint[arXiv]{2101.09836}]

\bibitem[{Jacob {et~al.}(2009)Jacob, Katz, Berriman, Good, Laity, Deelman, Kesselman, Singh, Su, Prince, {et~al.}}]{jacob2009montage}
Jacob, J.~C., Katz, D.~S., Berriman, G.~B., {et~al.} 2009, Int. J. Comput. Sci. Eng., 4, 73

\bibitem[{{Jansson} \& {Farrar}(2012)}]{jansson2012new}
{Jansson}, R. \& {Farrar}, G.~R. 2012, \apj, 757, 14

\bibitem[{{Johnston} {et~al.}(2020){Johnston}, {Smith}, {Karastergiou}, \& {Kramer}}]{johnston2020galactic}
{Johnston}, S., {Smith}, D.~A., {Karastergiou}, A., \& {Kramer}, M. 2020, \mnras, 497, 1957

\bibitem[{{Kafexhiu} {et~al.}(2014){Kafexhiu}, {Aharonian}, {Taylor}, \& {Vila}}]{kafexhiu2014parametrization}
{Kafexhiu}, E., {Aharonian}, F., {Taylor}, A.~M., \& {Vila}, G.~S. 2014, \prd, 90, 123014

\bibitem[{{Kelner} {et~al.}(2006){Kelner}, {Aharonian}, \& {Bugayov}}]{kelner2006energy}
{Kelner}, S.~R., {Aharonian}, F.~A., \& {Bugayov}, V.~V. 2006, \prd, 74, 034018

\bibitem[{{LHAASO Collaboration} {et~al.}(2025)}]{lhaaso2025ultra}
{LHAASO Collaboration} {et~al.} 2025, Innovation, 6, 100802

\bibitem[{{Li} {et~al.}(2024){Li}, {Huang}, {Xu}, \& {He}}]{Li:2024gnb}
{Li}, W., {Huang}, T.-Q., {Xu}, D., \& {He}, H. 2024, \apj, 969, 6

\bibitem[{{Lorimer} {et~al.}(2006){Lorimer}, {Faulkner}, {Lyne}, {Manchester}, {Kramer}, {McLaughlin}, {Hobbs}, {Possenti}, {Stairs}, {Camilo}, {Burgay}, {D'Amico}, {Corongiu}, \& {Crawford}}]{lorimer2006parkes}
{Lorimer}, D.~R., {Faulkner}, A.~J., {Lyne}, A.~G., {et~al.} 2006, \mnras, 372, 777

\bibitem[{{Manchester} {et~al.}(2005){Manchester}, {Hobbs}, {Teoh}, \& {Hobbs}}]{manchester2005australia}
{Manchester}, R.~N., {Hobbs}, G.~B., {Teoh}, A., \& {Hobbs}, M. 2005, \aj, 129, 1993

\bibitem[{{Mattox} {et~al.}(1997){Mattox}, {Schachter}, {Molnar}, {Hartman}, \& {Patnaik}}]{mattox1997identification}
{Mattox}, J.~R., {Schachter}, J., {Molnar}, L., {Hartman}, R.~C., \& {Patnaik}, A.~R. 1997, \apj, 481, 95

\bibitem[{{Mukhopadhyay} \& {Linden}(2022)}]{mukhopadhyay2022self}
{Mukhopadhyay}, P. \& {Linden}, T. 2022, \prd, 105, 123008

\bibitem[{{Murase} {et~al.}(2012){Murase}, {Dermer}, {Takami}, \& {Migliori}}]{murase2012blazars}
{Murase}, K., {Dermer}, C.~D., {Takami}, H., \& {Migliori}, G. 2012, \apj, 749, 63

\bibitem[{{Nakar} {et~al.}(2009){Nakar}, {Ando}, \& {Sari}}]{nakar2009klein}
{Nakar}, E., {Ando}, S., \& {Sari}, R. 2009, \apj, 703, 675

\bibitem[{{Neronov} \& {Semikoz}(2020)}]{neronov2020lhaaso}
{Neronov}, A. \& {Semikoz}, D. 2020, \prd, 102, 043025

\bibitem[{{Popescu} {et~al.}(2017){Popescu}, {Yang}, {Tuffs}, {Natale}, {Rushton}, \& {Aharonian}}]{popescu2017radiation}
{Popescu}, C.~C., {Yang}, R., {Tuffs}, R.~J., {et~al.} 2017, \mnras, 470, 2539

\bibitem[{{Rani} {et~al.}(2023){Rani}, {Moore}, {Eden}, {Rigby}, {Duarte-Cabral}, \& {Lee}}]{rani2023identification}
{Rani}, R., {Moore}, T. J.~T., {Eden}, D.~J., {et~al.} 2023, \mnras, 523, 1832

\bibitem[{{Reid} {et~al.}(2016){Reid}, {Dame}, {Menten}, \& {Brunthaler}}]{reid2016parallax}
{Reid}, M.~J., {Dame}, T.~M., {Menten}, K.~M., \& {Brunthaler}, A. 2016, \apj, 823, 77

\bibitem[{{Reid} {et~al.}(2019){Reid}, {Menten}, {Brunthaler}, {Zheng}, {Dame}, {Xu}, {Li}, {Sakai}, {Wu}, {Immer}, {Zhang}, {Sanna}, {Moscadelli}, {Rygl}, {Bartkiewicz}, {Hu}, {Quiroga-Nu{\~n}ez}, \& {van Langevelde}}]{reid2019trigonometric}
{Reid}, M.~J., {Menten}, K.~M., {Brunthaler}, A., {et~al.} 2019, \apj, 885, 131

\bibitem[{{Sano} {et~al.}(2018){Sano}, {Yamane}, {Tokuda}, {Fujii}, {Tsuge}, {Nagaya}, {Yoshiike}, {Filipovi{\'c}}, {Alsaberi}, {Barnes}, {Onishi}, {Kawamura}, {Minamidani}, {Mizuno}, {Yamamoto}, {Tachihara}, {Maxted}, {Voisin}, {Rowell}, {Yamaguchi}, \& {Fukui}}]{sano2018molecular}
{Sano}, H., {Yamane}, Y., {Tokuda}, K., {et~al.} 2018, \apj, 867, 7

\bibitem[{{Shalchi}(2009)}]{shalchi2009diffusive}
{Shalchi}, A. 2009, Astropart. Phys., 31, 237

\bibitem[{{Slane} {et~al.}(2015){Slane}, {Bykov}, {Ellison}, {Dubner}, \& {Castro}}]{slane2015supernova}
{Slane}, P., {Bykov}, A., {Ellison}, D.~C., {Dubner}, G., \& {Castro}, D. 2015, \ssr, 188, 187

\bibitem[{{Su} {et~al.}(2019){Su}, {Yang}, {Zhang}, {Gong}, {Wang}, {Zhou}, {Wang}, {Chen}, {Sun}, {Chen}, {Xu}, \& {Jiang}}]{su2019milky}
{Su}, Y., {Yang}, J., {Zhang}, S., {et~al.} 2019, \apjs, 240, 9

\bibitem[{{Tam} {et~al.}(2020){Tam}, {Lee}, {Cui}, {Hu}, {Kong}, {Li}, {Tudor}, {He}, \& {Pal}}]{tam2020multiwavelength}
{Tam}, P.-H.~T., {Lee}, K.~K., {Cui}, Y., {et~al.} 2020, \apj, 899, 75

\bibitem[{{Tauber} {et~al.}(2010){Tauber}, {Mandolesi}, {Puget}, {Banos}, {Bersanelli}, {Bouchet}, {Butler}, {Charra}, {Crone}, {Dodsworth}, {Efstathiou}, {Gispert}, {Guyot}, {Gregorio}, {Juillet}, {Lamarre}, {Laureijs}, {Lawrence}, {N{\o}rgaard-Nielsen}, {Passvogel}, {Reix}, {Texier}, {Vibert}, {Zacchei}, {Ade}, {Aghanim}, {Aja}, {Alippi}, {Aloy}, {Armand}, {Arnaud}, {Arondel}, {Arreola-Villanueva}, {Artal}, {Artina}, {Arts}, {Ashdown}, {Aumont}, {Azzaro}, {Bacchetta}, {Baccigalupi}, {Baker}, {Balasini}, {Balbi}, {Banday}, {Barbier}, {Barreiro}, {Bartelmann}, {Battaglia}, {Battaner}, {Benabed}, {Beney}, {Beneyton}, {Bennett}, {Benoit}, {Bernard}, {Bhandari}, {Bhatia}, {Biggi}, {Biggins}, {Billig}, {Blanc}, {Blavot}, {Bock}, {Bonaldi}, {Bond}, {Bonis}, {Borders}, {Borrill}, {Boschini}, {Boulanger}, {Bouvier}, {Bouzit}, {Bowman}, {Br{\'e}elle}, {Bradshaw}, {Braghin}, {Bremer}, {Brienza}, {Broszkiewicz}, {Burigana}, {Burkhalter}, {Cabella}, {Cafferty}, {Cairola}, {Caminade}, {Camus}, {Cantalupo}, {Cappellini},
  {Cardoso}, {Carr}, {Catalano}, {Cay{\'o}n}, {Cesa}, {Chaigneau}, {Challinor}, {Chamballu}, {Chambelland}, {Charra}, {Chiang}, {Chlewicki}, {Christensen}, {Church}, {Ciancietta}, {Cibrario}, {Cizeron}, {Clements}, {Collaudin}, {Colley}, {Colombi}, {Colombo}, {Colombo}, {Corre}, {Couchot}, {Cougrand}, {Coulais}, {Couzin}, {Crane}, {Crill}, {Crook}, {Crumb}, {Cuttaia}, {D{\"o}rl}, {da Silva}, {Daddato}, {Damasio}, {Danese}, {D'Aquino}, {D'Arcangelo}, {Dassas}, {Davies}, {Davies}, {Davis}, {de Bernardis}, {de Chambure}, {de Gasperis}, {de La Fuente}, {de Paco}, {de Rosa}, {de Troia}, {de Zotti}, {Dehamme}, {Delabrouille}, {Delouis}, {D{\'e}sert}, {di Girolamo}, {Dickinson}, {Doelling}, {Dolag}, {Domken}, {Douspis}, {Doyle}, {Du}, {Dubruel}, {Dufour}, {Dumesnil}, {Dupac}, {Duret}, {Eder}, {Elfving}, {En{\ss}lin}, {Eng}, {English}, {Eriksen}, {Estaria}, {Falvella}, {Ferrari}, {Finelli}, {Fishman}, {Fogliani}, {Foley}, {Fonseca}, {Forma}, {Forni}, {Fosalba}, {Fourmond}, {Frailis}, {Franceschet}, {Franceschi},
  {Fran{\c{c}}ois}, {Frerking}, {G{\'o}mez-Re{\~n}asco}, {G{\'o}rski}, {Gaier}, {Galeotta}, {Ganga}, {Garc{\'\i}a L{\'a}zaro}, {Garnica}, {Gaspard}, {Gavila}, {Giard}, {Giardino}, {Gienger}, {Giraud-Heraud}, {Glorian}, {Griffin}, {Gruppuso}, {Guglielmi}, {Guichon}, \& {Guillaume}}]{tauber2010planck}
{Tauber}, J.~A., {Mandolesi}, N., {Puget}, J.~L., {et~al.} 2010, \aap, 520, A1

\bibitem[{Ye {et~al.}(2023)Ye, Hu, Tian, Chang, Chang, Cheng, Gao, Ge, Gong, Guo, Guo, He, Huang, Jiang, Jiang, Jing, Li, Li, Li, Li, Li, Liao, Lin, Lin, Liu, Liu, Liu, Miao, Mo, Morton-Blake, Peng, Sun, Tang, Tang, Tao, Tian, Wang, Wang, Wang, Wei, Wei, Wu, Xian, Xiang, Xu, Xue, Yang, Yang, Yu, Zeng, Zhang, Zhang, Zhang, Zhang, Zhi, Zhong, Zhou, Zhu, \& Zhuang}]{ye2022proposal}
Ye, Z.~P., Hu, F., Tian, W., {et~al.} 2023, Nat. Astron., 7, 1497

\bibitem[{{Yoast-Hull} {et~al.}(2016){Yoast-Hull}, {Gallagher}, \& {Zweibel}}]{yoast2016equipartition}
{Yoast-Hull}, T.~M., {Gallagher}, J.~S., \& {Zweibel}, E.~G. 2016, \mnras, 457, L29

\bibitem[{Yuan \& Chirkin(2024)}]{abbasi2023updated}
Yuan, T. \& Chirkin, D. 2024, in 38th International Cosmic Ray Conference (ICRC2023)

\bibitem[{{Zhang} {et~al.}(2024){Zhang}, {Cui}, {Huang}, {Lin}, {Liu}, {Qiu}, {Shao}, {Shi}, {Xie}, \& {Yang}}]{zhang2024proposed}
{Zhang}, H., {Cui}, Y., {Huang}, Y., {et~al.} 2024, submitted to Astropart. Phys., ArXiv e-prints [\eprint[arXiv]{2408.05122}]

\bibitem[{{Zhang} {et~al.}(2021){Zhang}, {Liu}, {Chen}, \& {Wang}}]{zhang2021morphology}
{Zhang}, Y., {Liu}, R.-Y., {Chen}, S.~Z., \& {Wang}, X.-Y. 2021, \apj, 922, 130

\end{thebibliography}
\end{document}